\documentclass[manuscript]{aastex}
\usepackage{amsmath,bm}
\usepackage{morefloats}
\usepackage{natbib}
\usepackage{threeparttable}

\title{Time-Distance Helioseismology of Two Realistic Sunspot Simulations}
\author{
K. DeGrave,\altaffilmark{1}
J. Jackiewicz,\altaffilmark{1}
M. Rempel\altaffilmark{2}
}

\affil{\altaffilmark{1}New Mexico State University, Department of Astronomy, 1320 Frenger Mall, Las Cruces, NM 88003, USA; degravek@nmsu.edu, jasonj@nmsu.edu}
\affil{\altaffilmark{2}National Center for Atmospheric Research, HAO Division, 3080 Center Green Drive, Boulder, CO 80301, USA; rempel@ucar.edu}

\begin{document}

\begin{abstract}
Linear time-distance helioseismic inversions are carried out using several filtering schemes to determine vector flow velocities within two $\sim100^2\,{\rm Mm^2}\times 20\,{\rm Mm}$ realistic magnetohydrodynamic sunspot simulations of 25~hr. One simulation domain contains a model of a full sunspot (i.e. one with both an umbra and penumbra), while the other contains a pore (i.e. a spot without a penumbra). The goal is to test current helioseismic methods using these state-of-the-art simulations of magnetic structures. We find that horizontal flow correlations between inversion and simulation flow maps are reasonably high ($\sim0.5$--0.8) in the upper 3~Mm at distances exceeding 25--30~Mm from spot center, but are substantially lower at smaller distances and larger depths. Inversions of forward-modeled travel times consistently outperform those of our measured travel times in terms of horizontal flow correlations, suggesting that our inability to recover flow structure near these active regions is largely due to the fact that we are unable to accurately measure travel times near strong magnetic features. In many cases the velocity amplitudes from the inversions underestimate those of the simulations by up to 50\%, possibly indicating nonlinearity of the forward problem. In every case, we find that our inversions are unable to recover the vertical flow structure of the simulations at any depth.
\end{abstract}

\section{Introduction}\label{sec:intro}
Sunspots are a dominant feature in white-light observations of the Sun, and understanding their complex structure is a key goal of solar physics. While key information is obtainable from high-resolution observations at and above the photosphere, there is still no consensus pertaining to the questions regrading their subsurface properties; in particular, their mechanism of formation within the solar interior, how they are assembled and transported through the convection zone, and what their three-dimensional subsurface structure is. The study of sunspots is of particular importance as they are driven by the Sun's magnetic dynamo and are associated with energetic solar events like flares and coronal mass ejections. Recent state-of-the-art numerical simulations is shallow domains have been extremely successful in modeling the photospheric properties of sunspots \citep{rempel2009,rempel2009a,rempel2009b}. To shed more light on some of these outstanding questions in deeper layers, helioseismology can be applied.

Local helioseismology allows for the study of the structure and dynamics of the convection zone over localized patches of the solar surface. The methods of local helioseismology have been useful in gleaning information from these layers regarding subsurface flows and sound-speed perturbations around features like supergranules and active regions. Time-distance helioseismology \citep{duvall1993} (one of the several methods of local helioseismology and the focus of this paper) in particular, relies on the inversion of wave travel-time measurements made at the photosphere to probe the upper convection zone, and has been used along with other local helioseismic methods (i.e. helioseismic holography and ring-diagram analysis) over the past two decades in attempts to study sunspots.

Helioseismic results regarding sunspot structure have generally been mixed \citep{birch2011}, with some pointing to sunspots having a shallow structure, while others point to a deep structure. Here, a ``deep'' sunspot is one exhibiting subsurface properties (i.e. flows or wave-speed perturbations) that are significantly different than the quiet Sun at a depth of more than a few Mm. Some inversion results obtained using time-distance helioseismology and ring-diagram analysis \citep{kosovichev1996,zhaokosovichev2001,basu2004,haber2004,bogart2008,kosovichev2012} suggest that sunspots are deep features, extending down 10--15~Mm or more below the photosphere, while those of forward models \citep{fan1995,crouch2005} and numerical simulations \citep{rempel2009,braun2012} suggest sunspots are quite shallow, with vertical extents of only 2--3~Mm. Additional disagreement regarding sunspot structure was described in \citet{gizon2009}, where wave-speed perturbations inferred though time-distance inversions beneath a sunspot were in stark disagreement with those found via ring-diagram analysis \citep{kosovichev2012}, differing in both magnitude and sign.

In terms of large-scale plasma flows in the extended area surrounding active regions, there seems to be general agreement between time-distance helioseismology and ring-diagram analysis \citep{hindman2004}. These methods show weak converging flows in the near-surface layers of the convection zone, and diverging flows in the depth range of 10--15~Mm with velocities on the order of 50~$\rm{m\,s^{-1}}$ \citep{gizon2001,haber2001,haber2004,zhaokosovichev2004}. On the other hand, \citet{braun2011} have inferred a more complicated combination of both converging and diverging flows towards active regions using holography. In the immediate vicinity of sunspots, helioseismic results nearly always point toward a radial outflow from spot center (the moat flow) \citep{gizon2000, gizon2009, braunlindsey2003} having a near-surface magnitude of a few hundred~$\rm{m\,s^{-1}}$ \citep{gizon2009}. These results are supported through Doppler velocity measurements and the tracking of photospheric bright points \citep{sheeley1969,sheeley1972}. The ring-diagram results of \citet{hindman2009} suggest the moat flow is a shallow feature possessing a depth of only 2~Mm, followed by converging flow that extends more deeply through the convection zone. The time-distance results of \citet{gizon2009}, on the other hand, suggest that the moat flow extends down to at least $\sim5$~Mm without the presence of any flow reversal. In contrast with these findings, time-distance flow results from a study by \citet{zhao2001} actually indicated inflow in the 1--5~Mm directly below a sunspot, with outflow existing thereafter down to a depth of $\sim10$~Mm.

\citet{gizon2009} suggested that disagreement found between various inversions is likely due, at least in part, to the presence of strong magnetic fields in and around sunspots and the associated effects that these have on the wave field that is typically not taken into account. 
The inconsistency between studies, exacerbated (or caused) by these issues, makes it difficult to determine how accurate helioseismic inversions are around active regions. Comparisons between methods can be useful, but agreement between methods does not necessarily guarantee that they are correct. Here, we assess the accuracy of time-distance analysis through the use of realistic solar simulations. Analysis of a model sunspot whose three-dimensional structure is known \textit{a priori} and whose features closely resemble solar sunspots is a valuable tool in assessing the capabilities or limitations of our current helioseismic methods. Not only does artificial data allow us to test the accuracy of our inversion results, but it also allows us to make other comparisons that would otherwise have been impossible if only real solar data were available (i.e. testing the effects of data filter choice on results, tests of kernel performance through forward-modeling, etc.).

The goal of this work is to test current time-distance inversions using some of the most realistic sunspot simulation data available to us today to see how the method performs in this strong perturbation regime. The codes used in this study to measure and invert travel times have been used previously by \citet{degrave2014} to study the flow structure in the upper 5~Mm of two $\sim100^{2}~\rm{Mm^{2}} \times 20~\rm{Mm}$ quiet-Sun simulation domains (Rempel 2014, in preparation). In practice, all measurement and inversion procedures described in this work are identical to those outlined in \citet{degrave2014}. Thus, we can quantitatively compare the helioseismic findings from very different simulation data. In Section~\ref{sec:sim}, we describe these data and their overall properties, and in Section~\ref{sec:filt} the data filtering process is briefly explained. Comparisons are made between two travel-time definitions in Section~\ref{sec:tt}, and measured travel times are compared with forward-modeled ones in Section~\ref{sec:fmtt}. A brief overview of the inversion procedure is given in Section~\ref{sec:results} along with the inverted flows. Finally, these results are discussed in Section~\ref{sec:dis}. For a more complete discussion regarding the data filtering, sensitivity kernel computation, and inversion procedure implemented in this work, we again refer the reader to \citet{degrave2014} where these topics are discussed in more detail.

\section{Sunspot Simulation Data}\label{sec:sim}




Our analysis is based on two sunspot simulations of domain size $98.304\times98.304\times18.432~\rm{Mm}^3$ in the horizontal and vertical directions. The simulations are started from a snapshot of a quiet-Sun simulation (including a mixed polarity field maintained through a small-scale dynamo) after insertion of a self-similar axisymmetric field structure as described in \citet[][Appendix A]{rempel2012}. Boundaries in the horizontal directions are periodic. At the open bottom boundary, the magnetic field is symmetric (i.e. both horizontal and vertical field components are allowed). At the closed top boundary, the magnetic field is computed following the same procedure as \citet{rempel2012} in which the field inclination is increased by about a factor of two compared to a potential field extrapolation. This was found to lead to the formation of a penumbra for a sufficiently high numerical resolution.

We consider here two setups that differ in terms of initial flux and field strength, as well as resolution. The high-resolution case is $0.048\times0.048\times0.024$~Mm, which is a compromise between being able to resolve a penumbra and being able to run the simulation for a time scale of a few days. The initial flux of the spot is $9\times10^{21}$~Mx, and the initial field strength varies from 20~kG at the bottom boundary to 3~kG at the top. The second simulation uses a lower resolution of $0.128\times0.128\times0.48$~Mm, with an initial flux and field strength 10\% larger than the first. For the first six hours, we evolved both simulations with a strong damping term on all three velocity components in a cylinder of radius 15~Mm surrounding the spot, and a closed bottom boundary condition at the foot point of the spot (radius of 5~Mm) in order to suppress convective motions and allow them to re-grow in a fashion that is consistent with the presence of the spot. After this initialization phase, the damping was switched off and an open bottom boundary condition was used everywhere in the domain.

While the high-resolution case (hereafter abbreviated HRes) maintains a penumbra through the entire duration of the simulation, the lower resolution model (hereafter LRes) loses its penumbra (and associated Evershed flow) after a few hours. This leads to different flow structures in the proximity of the spot as detailed below. Furthermore, the low-resolution spot decays more quickly compared to the high-resolution setup. We analyze a time series that begins 12.5 hours after the start of these simulations (6.5 hours after the initialization phase has ended).

A Doppler velocity time series was obtained from each simulation at the $\tau=0.01$ level, and we define the data cubes as $v_z (\bm{r}, z=0, t)$, where $\bm{r}=(x,y)$ is the horizontal coordinate and $z$ is the vertical coordinate. The series each span 25~hr, and both are sampled with a time cadence $h_t= 45$~s. The data cubes were interpolated onto grids with horizontal spacing $h_x=h_y=1$~Mm before helioseismic analysis was performed. 

\begin{figure}
\begin{center}$
\begin{array}{cc}
\includegraphics[width=0.5\linewidth,clip=]{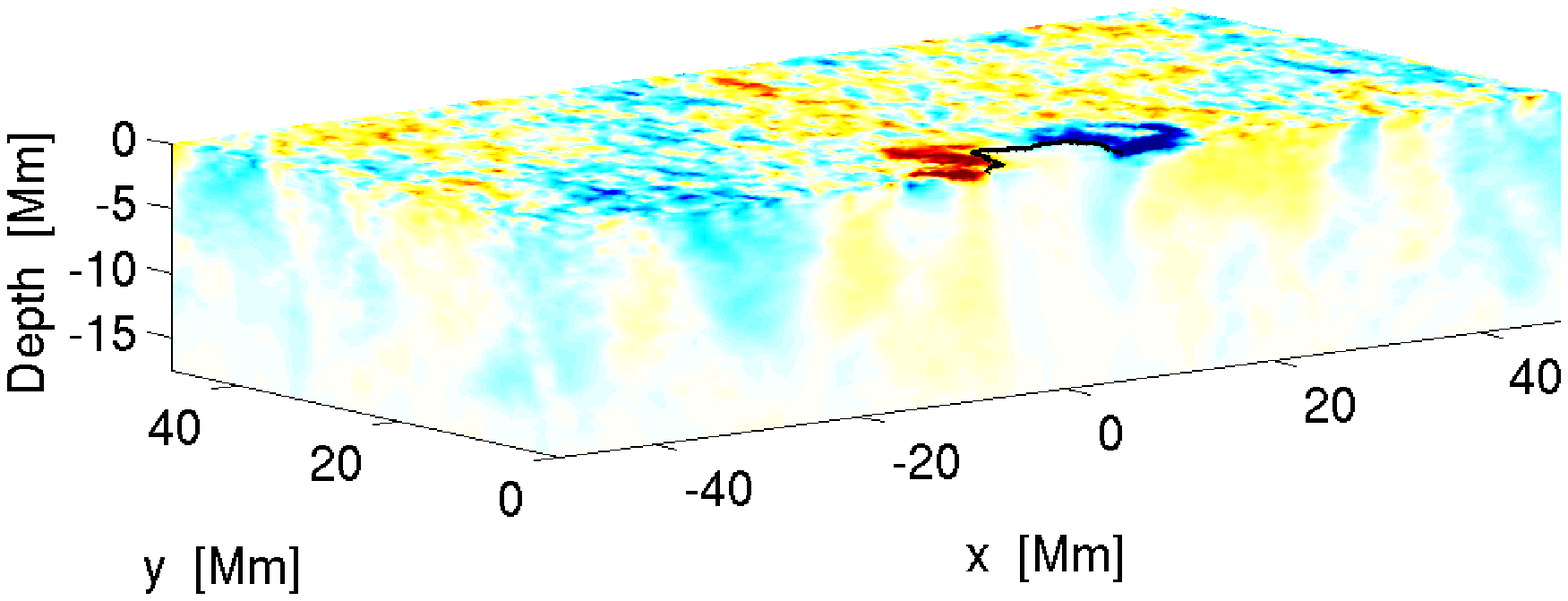} &
\includegraphics[width=0.5\linewidth,clip=]{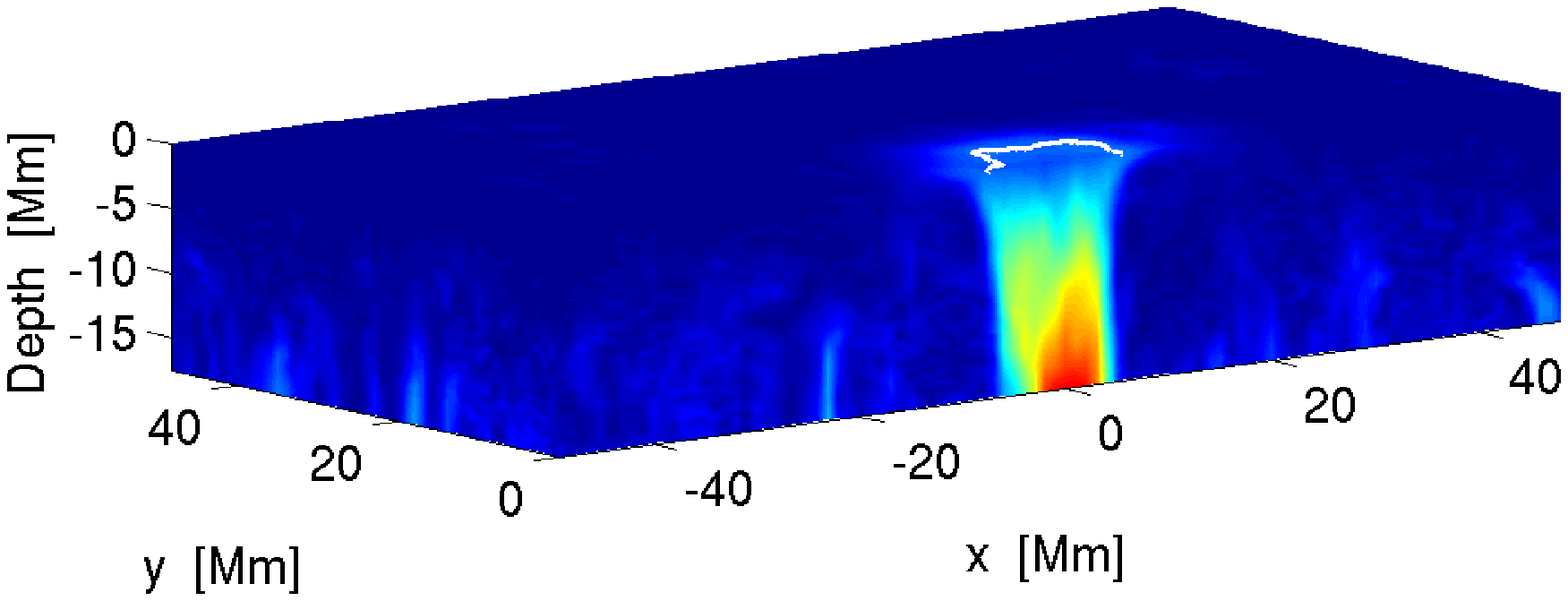} \\
\includegraphics[width=0.5\linewidth,clip=]{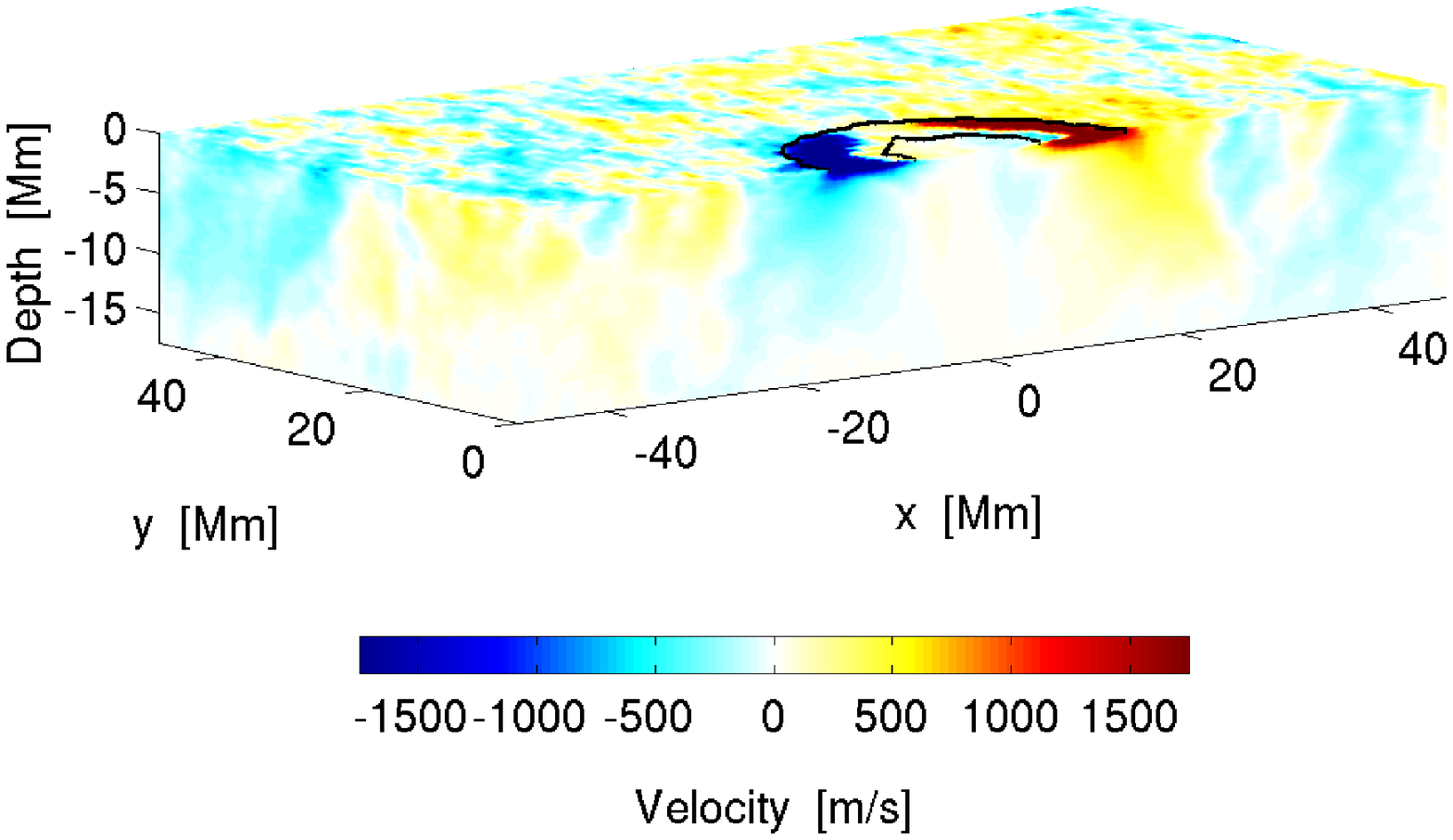} &
\includegraphics[width=0.5\linewidth,clip=]{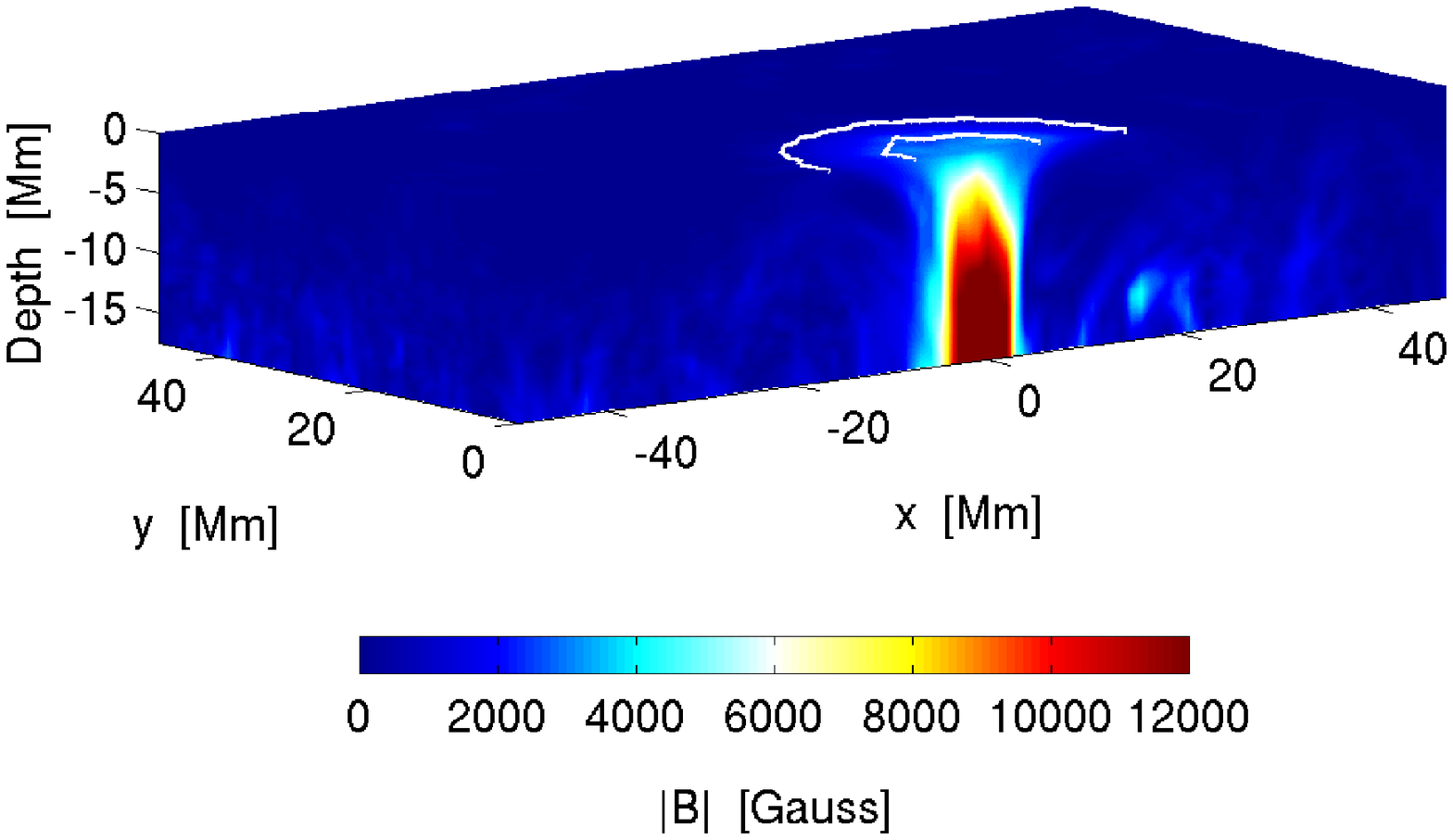}
\end{array}$
\end{center}
\caption{The two simulations used in this study. Shown are time averages (over $25$~hr) of the $v_x$ velocity (left column) and the B-field magnitude ($|B|=\sqrt{B_x^2+B_y^2+B_z^2}$, right column) for half of the computation domain ($y\ge 0$) for LRes (top row) and HRes (bottom row). The horizontal slice at the top is taken at the $\tau_{500} =0.01$ ($z=0$) level. The contours at the surface mark the boundaries of the spot umbra and penumbra (in the case of HRes) as defined in the text. Figures in a given column share the same color bar.}
\label{fig:slices}
\end{figure}

Figure~\ref{fig:slices} shows time-averaged horizontal flows and magnetic field strength in the LRes and HRes simulations. Power spectra computed from both velocity series show a rich spectrum of acoustic modes, similar to the example in Figure~2 of \citet{degrave2014}.

\section{Measurements}\label{sec:tt}

\subsection{Mode Filtering}\label{sec:filt}
The simulation data were filtered using a series of ridge and phase-speed filters identical to those discussed in \citet{degrave2014}. The ridge filters were chosen to isolate ridges $f $--$p_3$, and we implemented the first five lowest time-distance phase-speed filters defined in \citet{couvidat2006}, referred to as $\rm{td_1}$--$\rm{td_5}$ hereafter. Each of these filters was constructed to keep the signal from waves whose lower turning points are above a depth of $\sim12$~Mm (i.e. waves with phase-speeds less than roughly 40~$\rm{km\,s^{-1}}$) to avoid wave reflections at the bottom boundary of the simulation domains. All filters are confined within the frequency range of 2.5--5.3~mHz.

We filter the Doppler time series, $v_z(x,y,z=0,t) = \phi(x, y, t)$, by multiplying the Fourier transform of the data cube with the square root of each filter $F_m (\bm{k},\omega)$ as

\begin{equation}
 \phi_m (\bm{k},\omega) = \phi (\bm{k},\omega) \sqrt{F_m (\bm{k},\omega)}.
\end{equation}
Here, $\phi_m (\bm{k},\omega)$ is the filtered cube in Fourier space containing wave signal isolated using a filter with index $m$. The filtered data are then transformed back to real space, giving $\phi_m(x,y,t)$.

Temporal cross-covariances were then computed from the filtered data $\phi_m (x,y,t)$ in the center-to-annulus and quadrant configurations \citep[e.g.,][]{duvall1997} for a range of skip distances $\Delta$ depending on filter. For the ridge-filtered data, $\Delta = 11-27$~Mm in increments of 4~Mm, for a total of five distances per ridge. For the phase-speed filtered data, the valid $\Delta$ values over which the cross-correlations can be computed depend on central phase-speed as discussed in \citet{couvidat2006}. The phase-speed $\Delta$ values used here are 5--9, 7--11, 9--15, 15--19, 19--29~Mm in increments of 2~Mm for filters $\rm{td_1}$--$\rm{td_5}$ respectively.

\subsection{Comparison of Two Travel-Time Measurement Methods}\label{sec:ttcomp}
Travel-time differences were computed from the measured cross-covariances in the `oi' (out minus in), `we' (west - east), and `ns' (north - south) geometries following the two methods outlined in \citet{gb02, gb04} (hereafter referred to as GB02 and GB04). To implement these definitions, we calculate the difference between the measured cross-covariances at each point and a symmetric reference cross-covariance function found by spatially averaging the oi cross-covariances in the quiet regions of the simulation domain. In this context we define ``quiet" to mean regions where the surface $z=0$ magnetic field magnitude $|B|=\sqrt{B_x^2+B_y^2+B_z^2} \le 250~\rm{G}$.

\begin{figure}
\begin{center}$
\begin{array}{cc}
\includegraphics[width=0.5\linewidth,clip=]{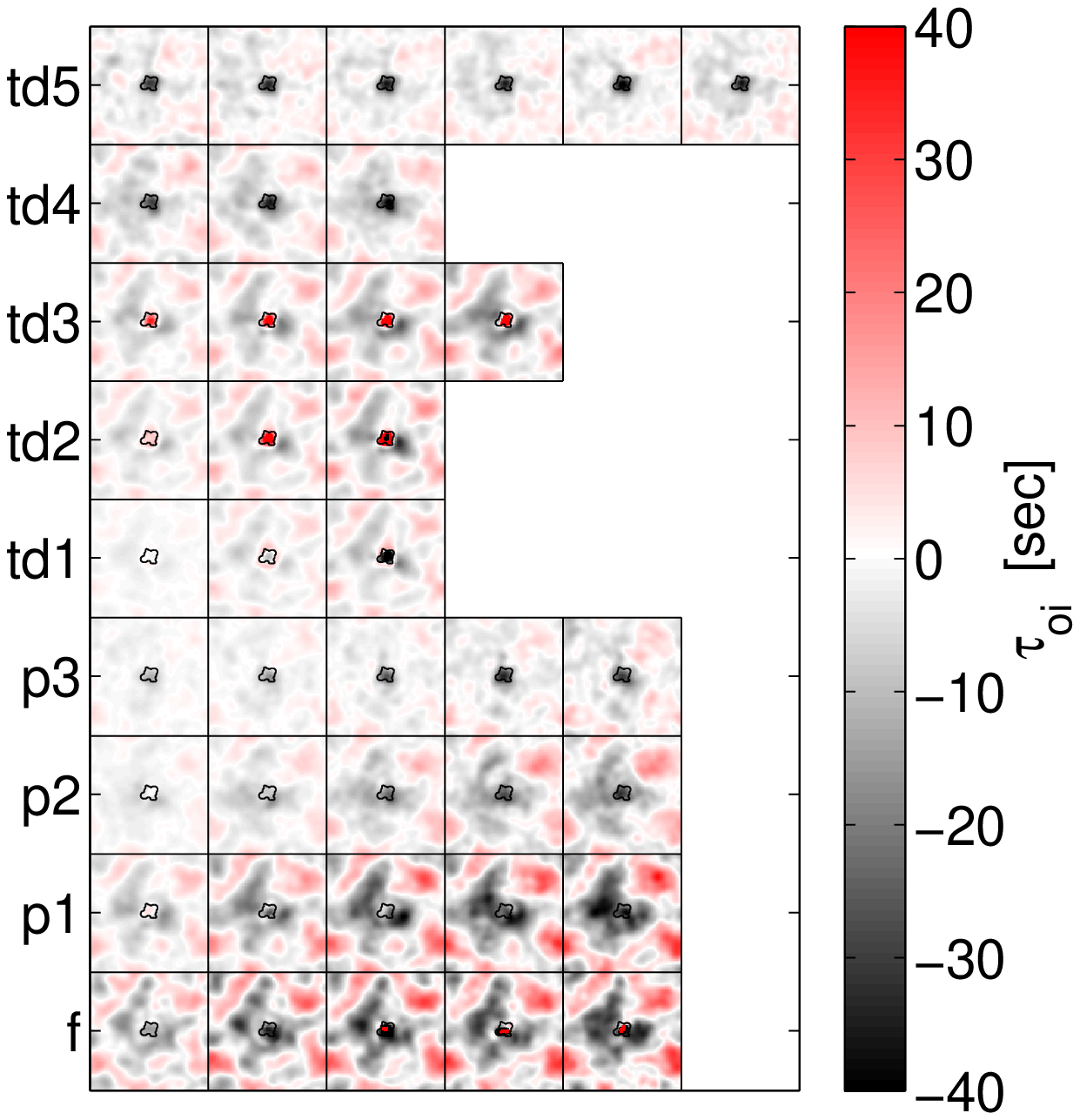} &
\includegraphics[width=0.5\linewidth,clip=]{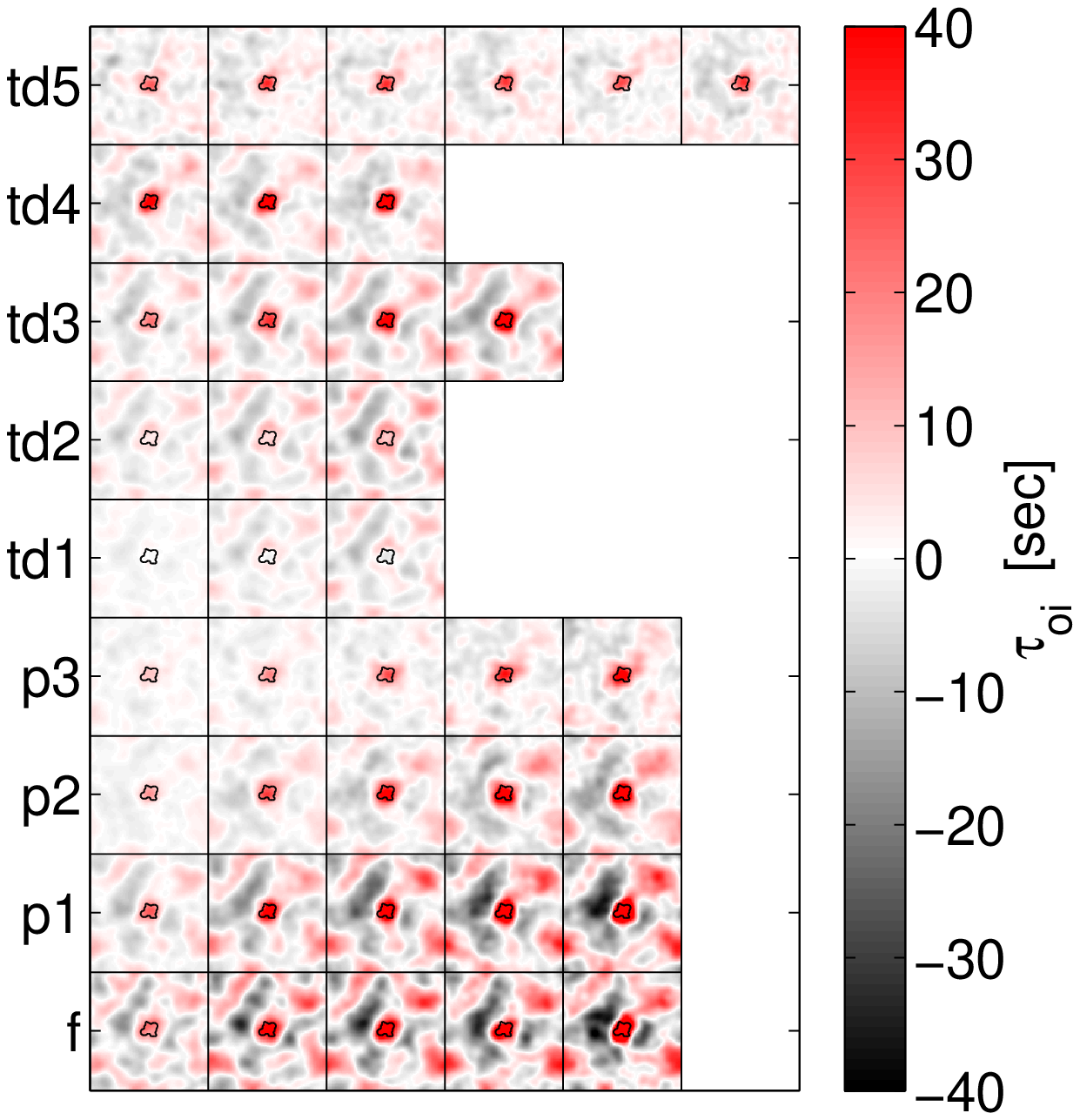} \\
\includegraphics[width=0.5\linewidth,clip=]{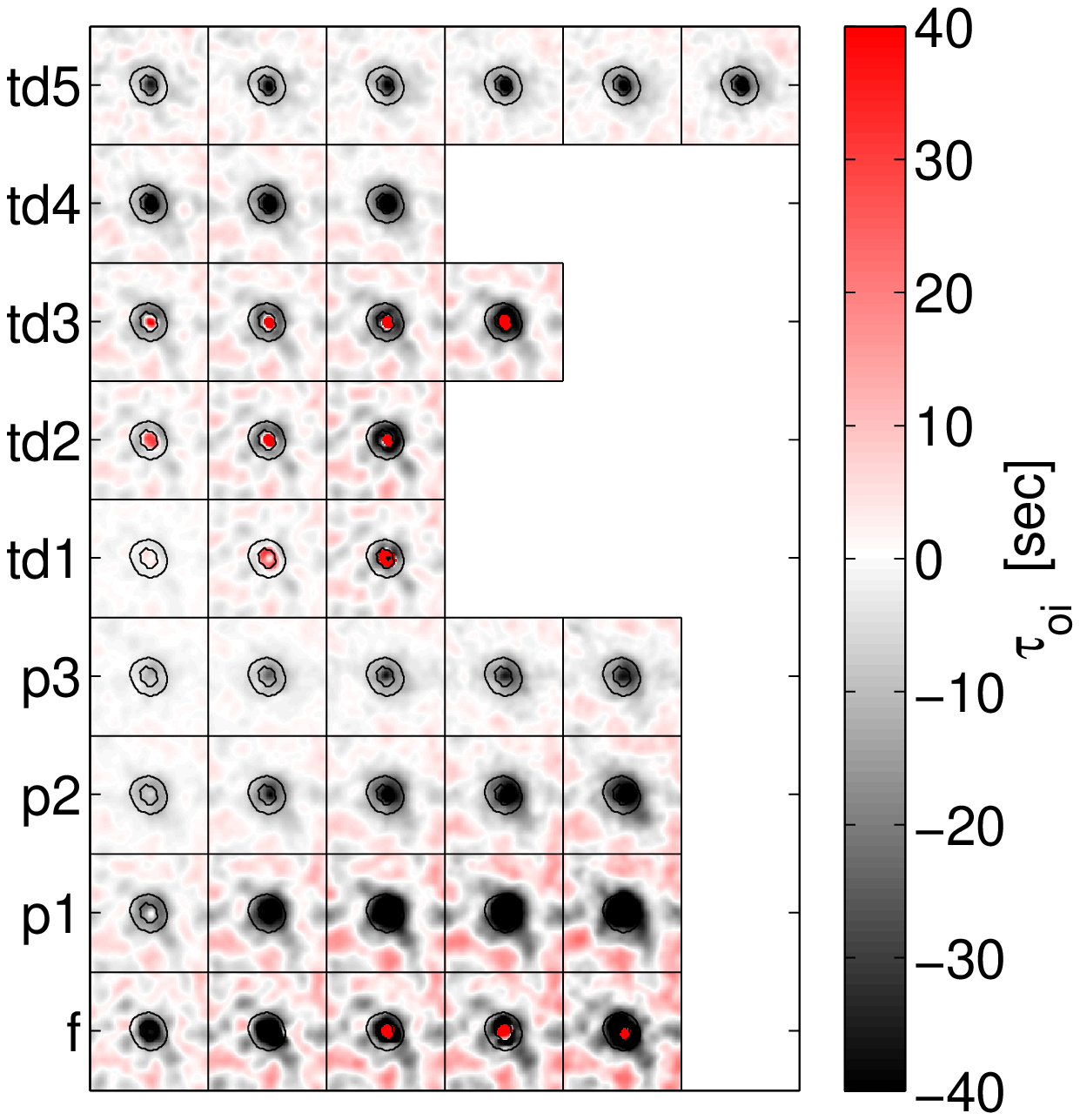} &
\includegraphics[width=0.5\linewidth,clip=]{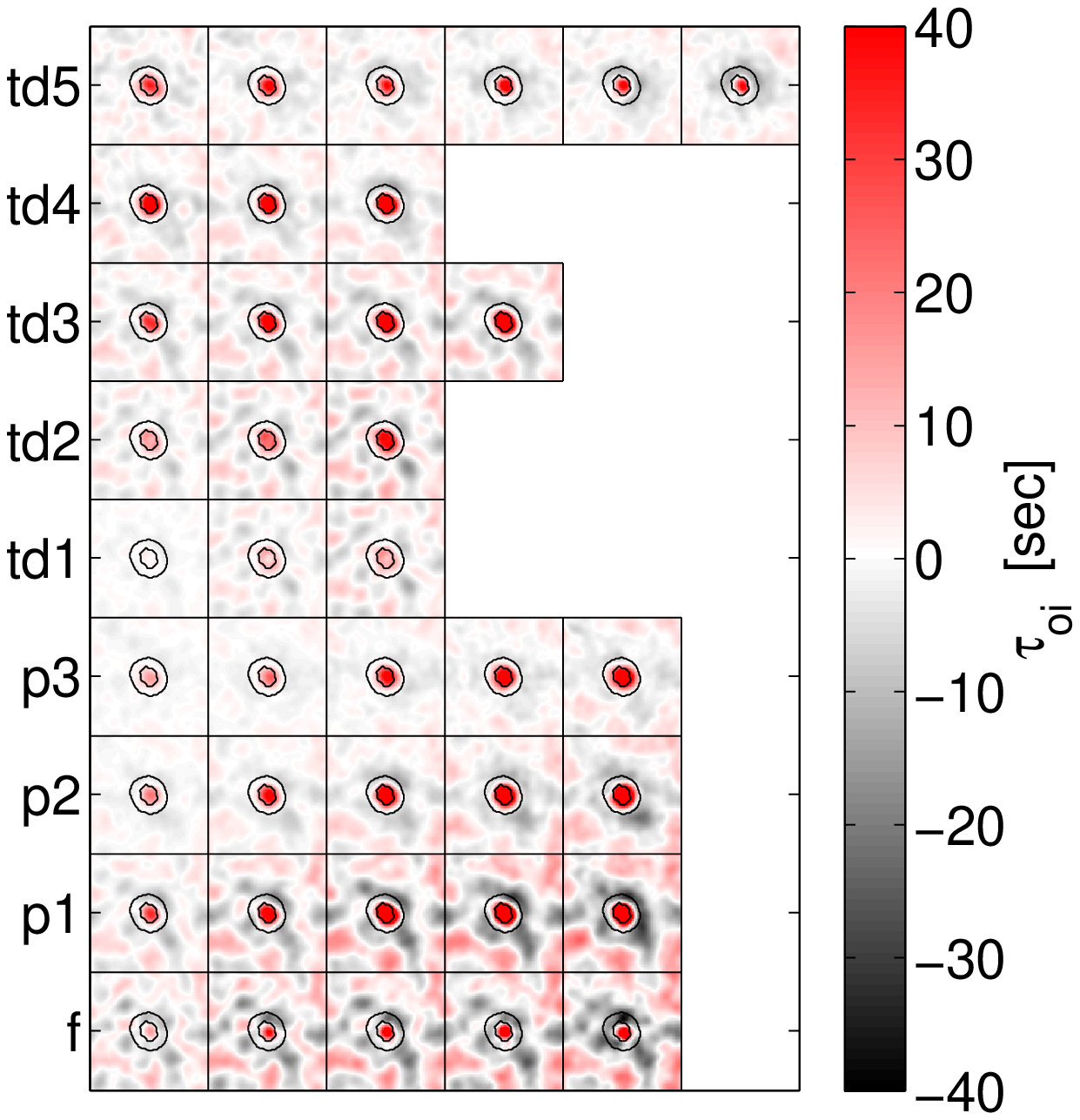}
\end{array}$
\end{center}
\caption{All LRes (top row) and HRes (bottom row) oi travel-time maps measured using GB02 (left column) and GB04 (right column). The $\Delta$ for which each map is computed increases from left to right in each plot, covering the appropriate range of values depending on filter type as defined in the text. The $f$-mode travel times have been divided by a factor of two for easier comparison. The contours mark the boundaries of the simulation umbra and penumbra.}
\label{fig:ttoi}
\end{figure}

Figure~\ref{fig:ttoi} shows each of the measured GB02 and GB04 oi travel-time maps for LRes and HRes for every filter over the appropriate range of $\Delta$ values. The spot umbra and penumbra (in the case of HRes) are marked by the black contour lines shown in each panel. The boundaries of these regions were determined from the simulation time-averaged continuum intensity. Following \citet{braun2012}, we define the umbra to be the region where the continuum intensity is less than half of the surrounding quiet-Sun intensity, while the penumbra is defined to be the location where the intensity is 0.5 to 0.9 times that of the quiet Sun. The color scales have been clipped to more easily see the lower-amplitude features away from the spots. A negative oi travel time denotes a signal from a diverging flow (moat flow), which is clearly exhibited by HRes in Figure~\ref{fig:slices}. We find that in the region outside of the spots, the two travel-time definitions generally agree quite well.

Significant differences do exist between the two methods, however, for travel times within a spot. First of all, travel times measured using GB02 exhibit sign changes which depend on data filter and even $\Delta$ in a few cases, most notably for the $f$ mode. A similar sign change with $\Delta$ was also observed by \citet{couvidat2012} \citep[and in the case of holography,][]{braun2008} when $p$-mode travel times were measured around a real solar active region. While this may be interpreted as a signature of varying flow structure at the different depths for which the modes are sensitive, the models do not exhibit such variation, and its cause must be explained by other means.

Secondly, the GB04 travel-time differences are almost uniformly the same sign (positive, suggesting an inflow) in all strong-field regions in both simulations. The disagreement with the GB02 measurements and the signatures anticipated from the simulations is not unexpected, as pointed out elsewhere \citep[e.g.,][]{couvidat2012}. The reason is that the GB04 definition is a linearized version of the GB02 one and only reduces to it when no noise is present and when the cross-covariances do not have strong amplitude variations. This is not the case in the strong magnetic regions of the model. Indeed, a correction is routinely applied to the cross-covariance functions before using GB04 which normalizes their amplitudes at each spatial position by dividing by the maximum value \citep{rajaguru2006,couvidat2012}. Without doing so, the overall amplitudes of the travel times would be underestimated by as much as a factor of two or so.

\begin{figure}
  \centering
  \includegraphics[width=\textwidth,clip=]{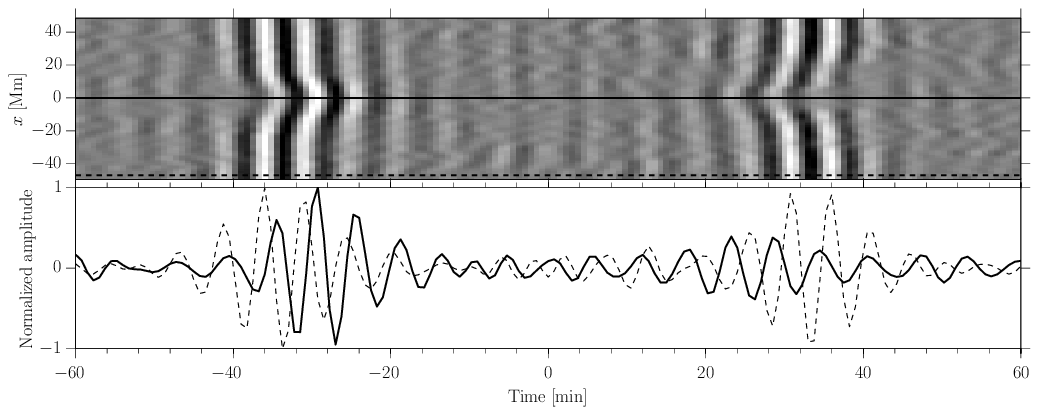}\\
  \includegraphics[width=\textwidth,clip=]{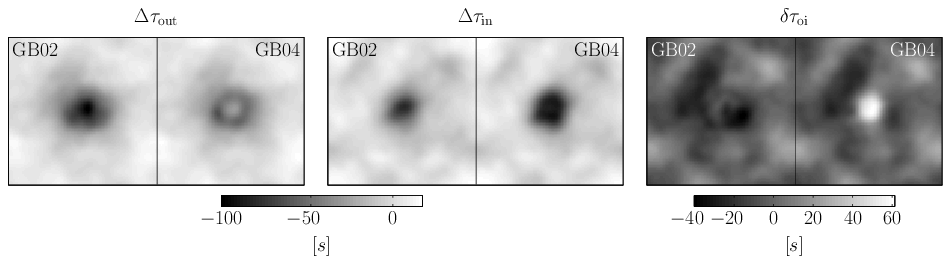}
  \caption{Effects of magnetic field on cross-covariances and travel-time measurements. The top panel shows a slice along $y=0$ of the annulus cross-covariance function (amplitude corrected) measured using the $p_1$ ridge at a travel distance $\Delta=19$Mm from the LRes model. Two functions from individual pixels from quiet (dashed line) and pore (solid line) regions are shown below to illustrate the amplitude variations on each branch of the time lag. The bottom row are the corresponding GB02 and GB04 travel-time maps measured from the positive and negative time branches, as well as the final `out-in' map.}
  \label{fig:ttcomp}
\end{figure}

The fact that the GB04 travel-time perturbations are always positive is also easy to see. Consider Figure~\ref{fig:ttcomp} where we show example cross-covariances measured from the LRes simulation using an annulus of radius $19$~Mm and $p_1$ modes. The cross-covariances at each spatial position have been scaled by their maximal value. Far from the pore, the positive and negative time lag branches have similar amplitudes (dashed line), and GB04 applied to these works reasonably well. However, within the pore structure (solid line), the positive branch of the covariance function, denoting the ``outgoing'' waves from the pore center, are noticeably lower in amplitude than the incoming waves on the negative time branch. While helioseismic techniques are designed to measure the phase of the wave packets, the linear GB04 definition is also influenced by the overall amplitude, as stated previously. Since we know acoustic modes are absorbed in sunspots \citep{braun1987}, the outgoing amplitude is reduced compared to the ingoing (although both branches would be greatly reduced in amplitude compared to the quiet regions of the model, if not for the normalization). Therefore, since $\delta\tau_{\rm oi} = \Delta\tau_{\rm out}-\Delta\tau_{\rm in}$ and $\Delta\tau_{\rm out}$ is smaller (less negative) than $\Delta\tau_{\rm in}$, the resulting measurement gives a positive travel-time difference in all cases. Note that both $\Delta\tau_{\rm out}$ and $\Delta\tau_{\rm in}$ are both negative themselves in the magnetic region, as acoustic waves travel faster in spots than in quiet Sun \citep{gizon2009,moradi2010}.

While developing more accurate measurements within strong magnetic regions is beyond the scope of this paper, we point out that dividing by the maximum value in each (positive and negative) branch separately would be a better strategy than just using one normalization factor at each pixel. Initial tests show this to be the case. In the analysis and inversions that follow, besides demonstrating that the GB02 and GB04 results agree quite well in the quiet Sun, we consider only the GB04 measurements using the ``standard'' normalization procedure \citep{couvidat2012}.

\begin{figure}
\begin{center}$
\begin{array}{cc}
\includegraphics[width=0.5\linewidth,clip=]{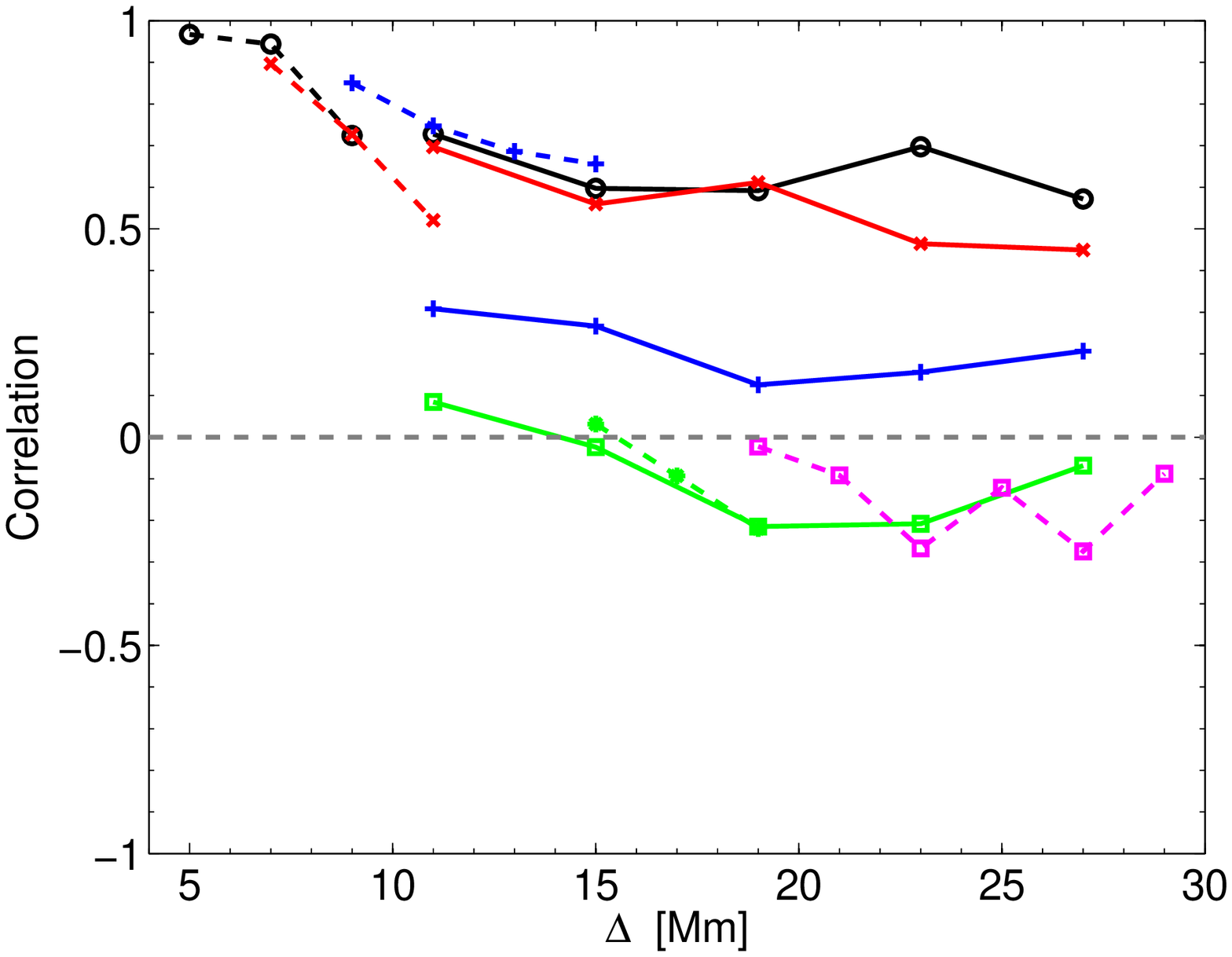} &
\includegraphics[width=0.5\linewidth,clip=]{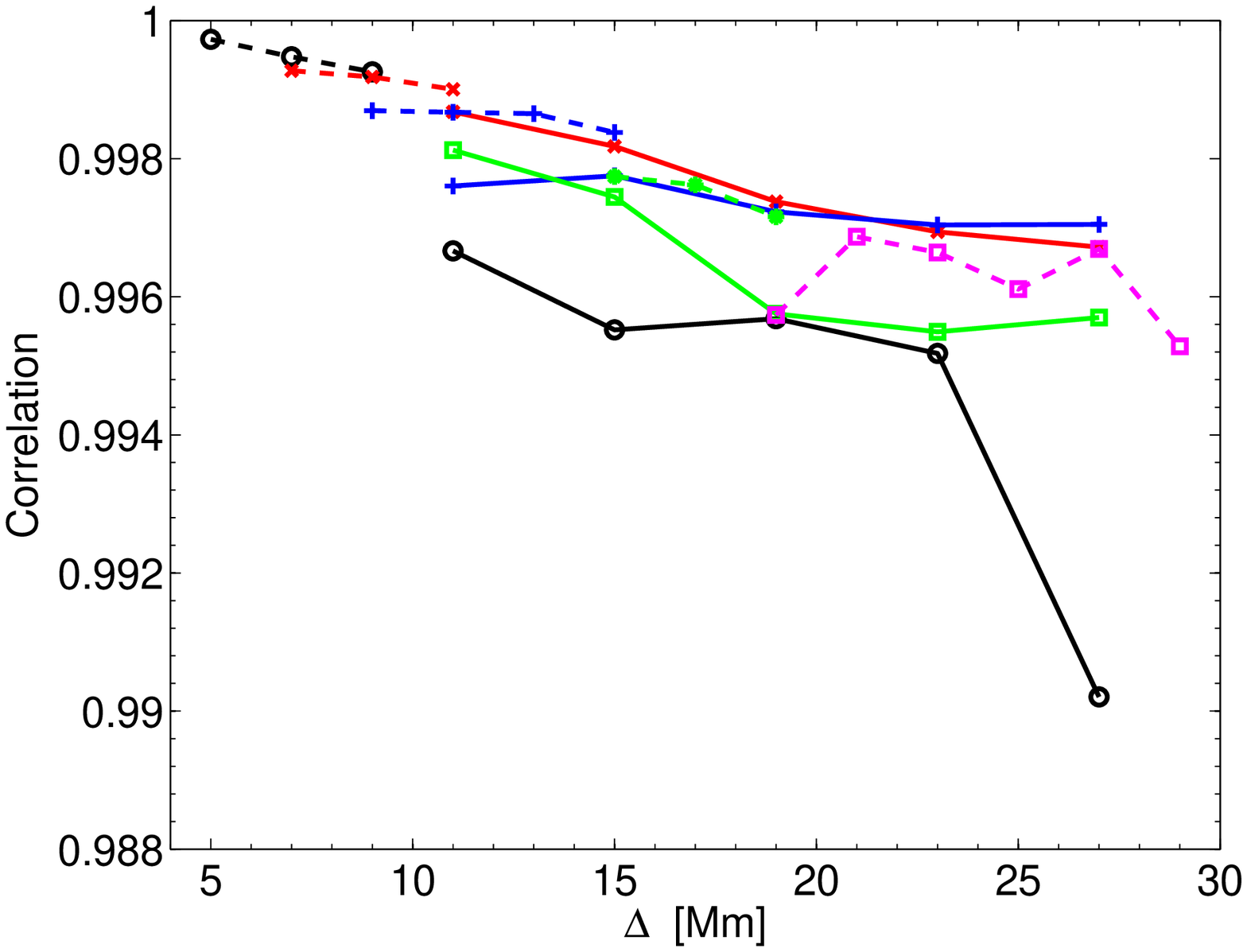} \\
\includegraphics[width=0.5\linewidth,clip=]{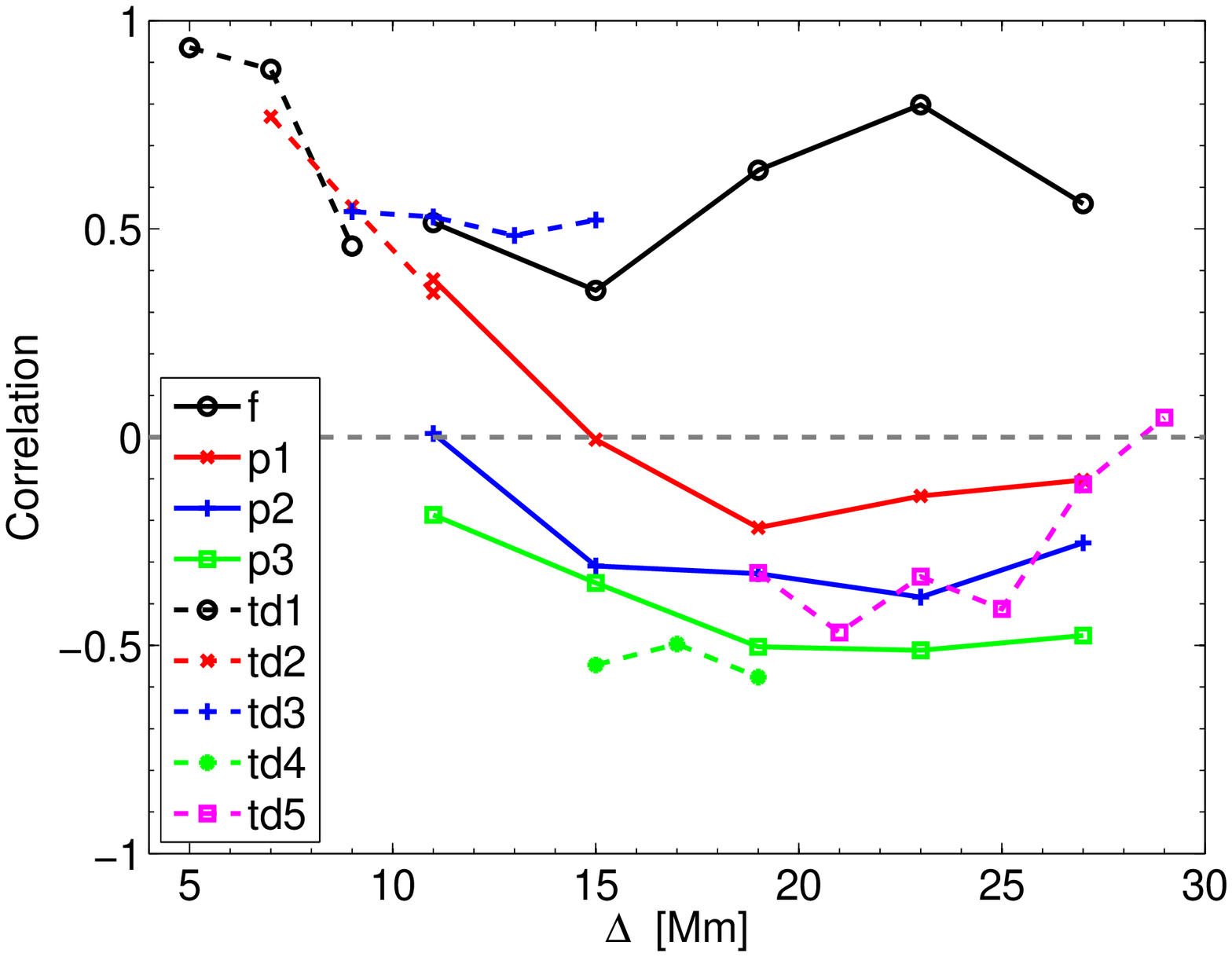} &
\includegraphics[width=0.5\linewidth,clip=]{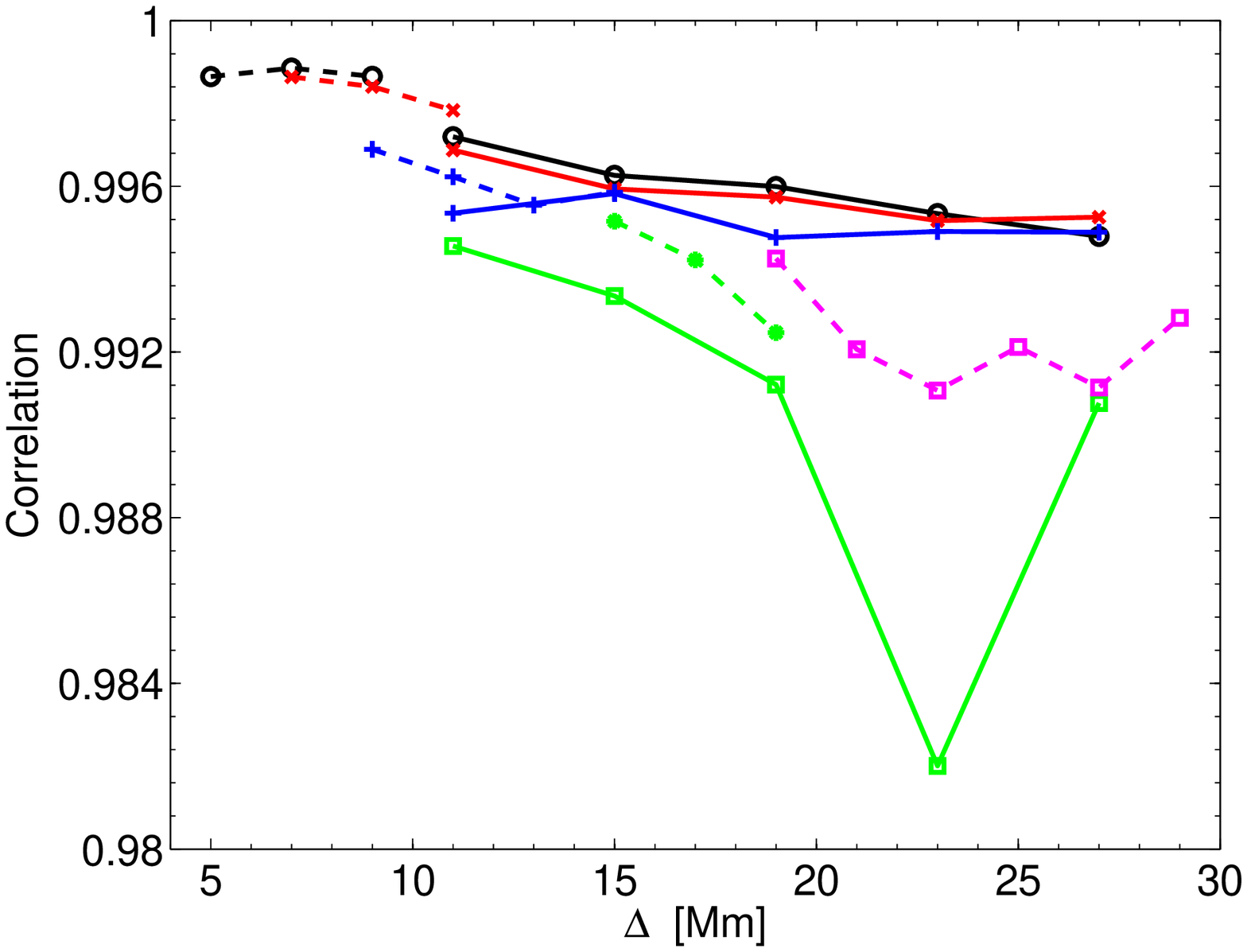}
\end{array}$
\end{center}
\caption{The 2D spatial correlation between GB02 and GB04 oi travel-time maps versus $\Delta$ for simulations LRes (top tow) and HRes (bottom row) before (left column) and after (right column) applying a circular mask of radius 25~Mm to each map to eliminate the spot and immediate surrounding area.}
\label{fig:ttoicorr}
\end{figure}

\begin{figure}
\begin{center}$
\begin{array}{cc}
\includegraphics[width=0.5\linewidth,clip=]{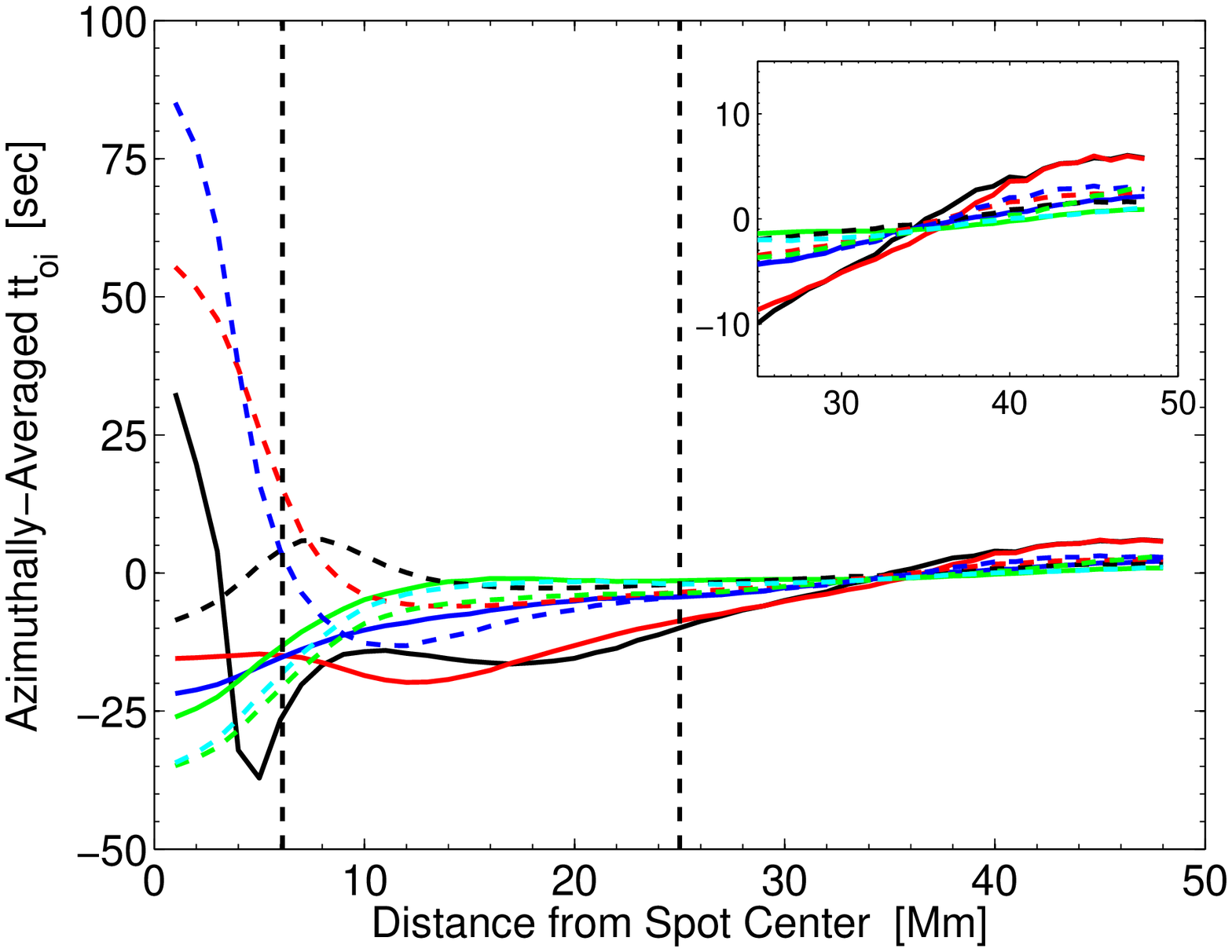} &
\includegraphics[width=0.5\linewidth,clip=]{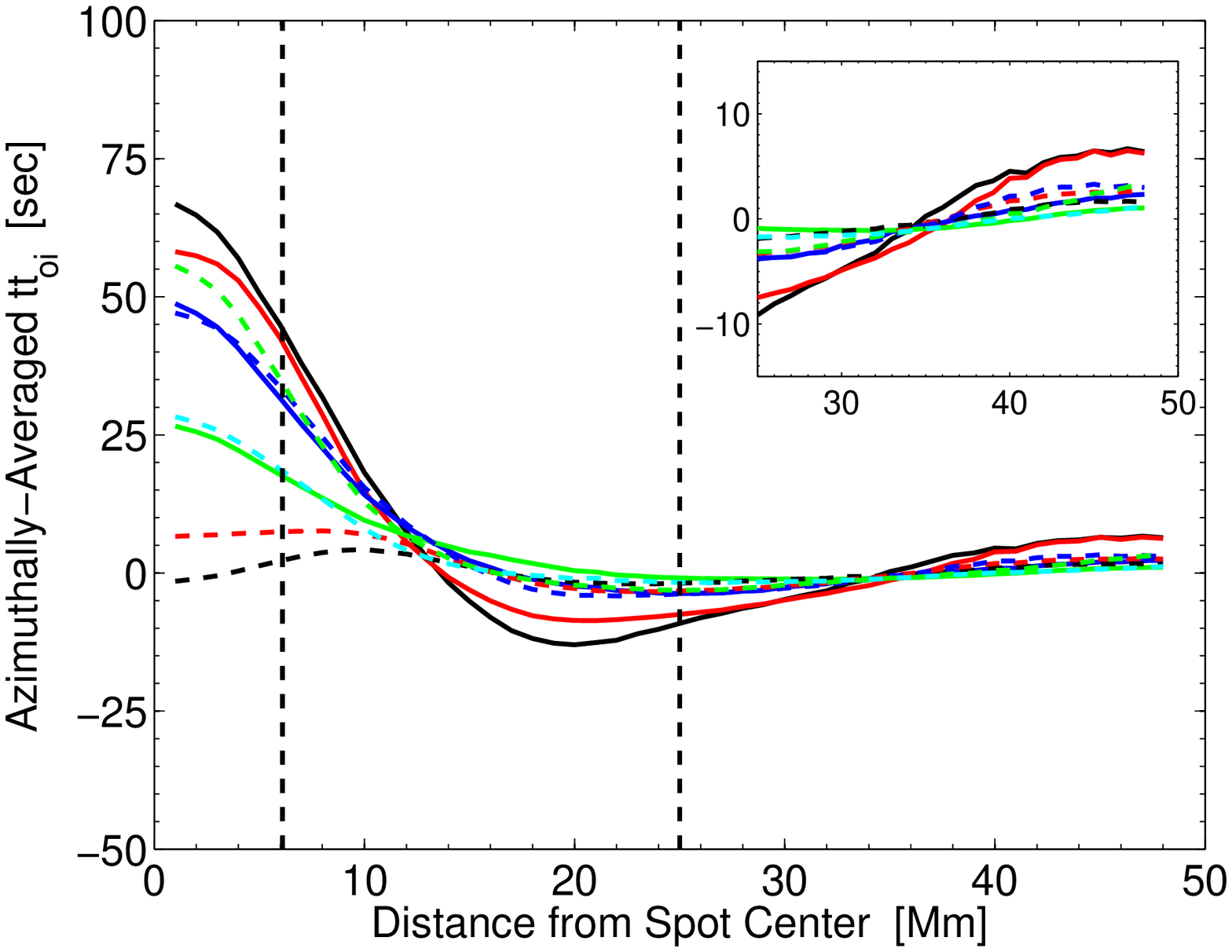} \\
\includegraphics[width=0.5\linewidth,clip=]{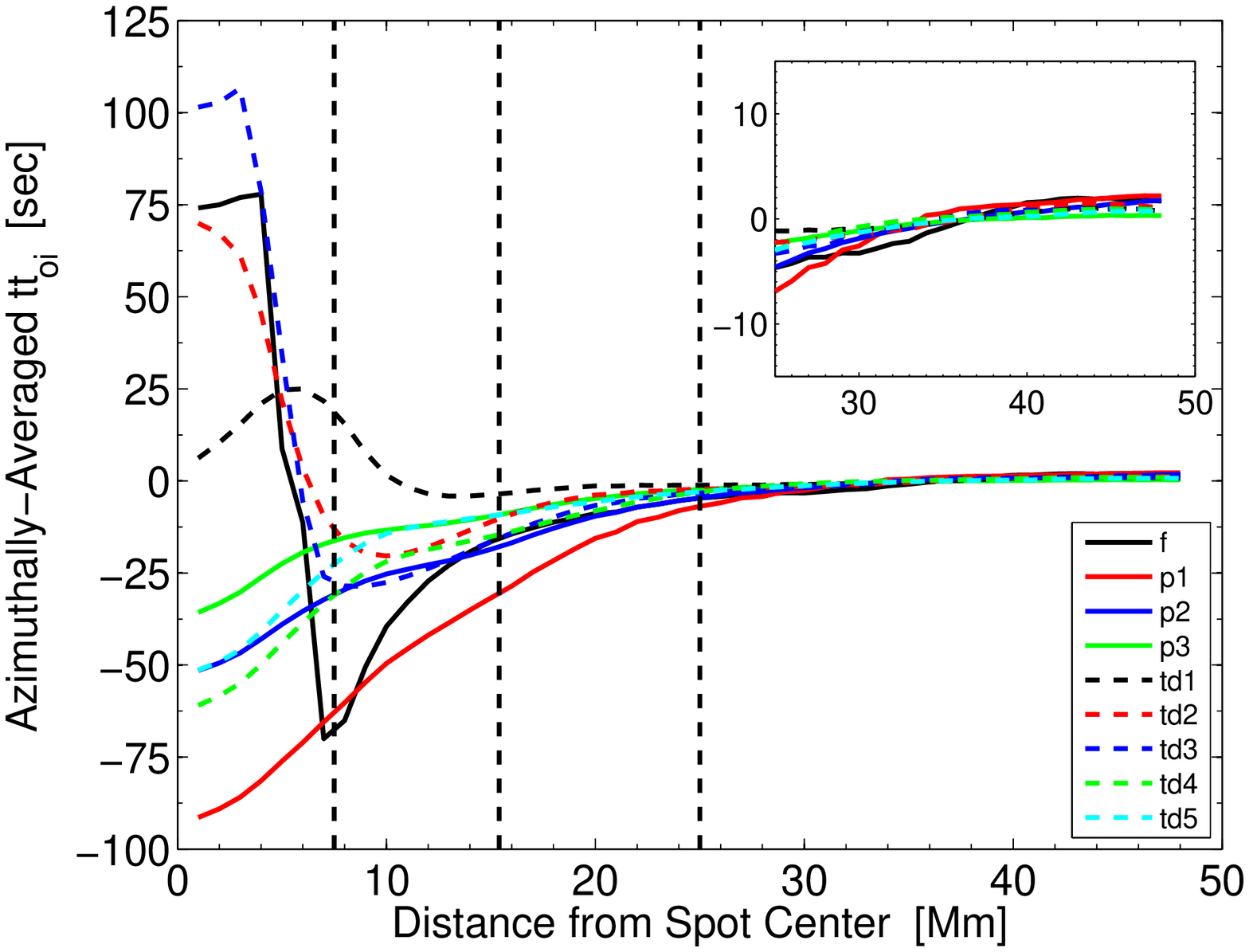} &
\includegraphics[width=0.5\linewidth,clip=]{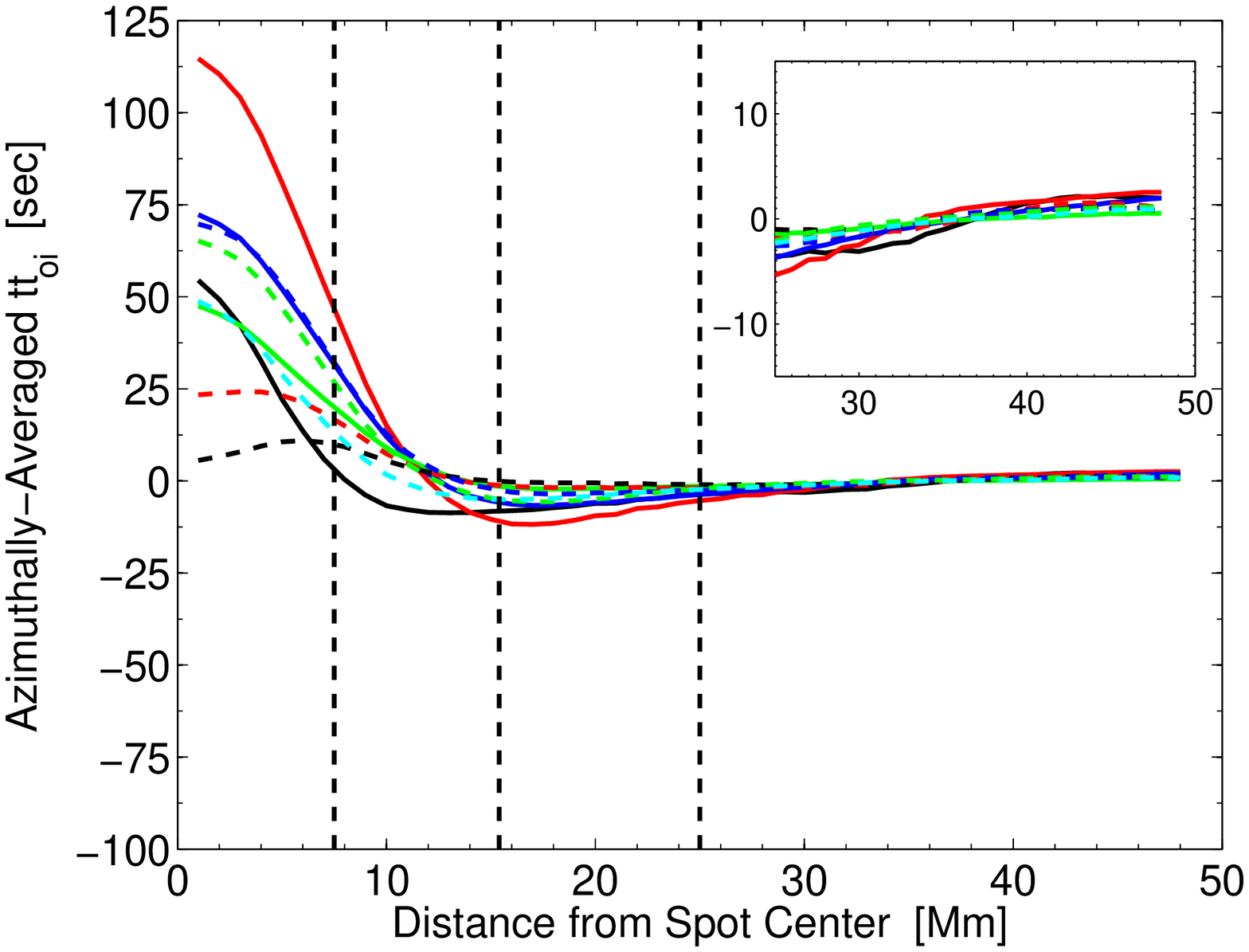}
\end{array}$
\end{center}
\caption{Azimuthally-averaged GB02 (left column) and GB04 (right column) oi travel times for LRes (top row) and HRes (bottom row). The $f$-mode travel times have been divided by a factor of two before averaging for easier comparison. The vertical lines represent the boundaries of the simulation umbra, penumbra, and the circular mask. The travel-time distances for which these profiles are computed correspond to the mid-range $\Delta$ value for each filter. The figure inset is a zoom-in of the profiles at a radius $\ge 25~\rm{Mm}$.}
\label{fig:ttoiazim}
\end{figure}

Figure~\ref{fig:ttoicorr} shows the Pearson correlation between GB02 and GB04 travel times before and after a circular mask of radius 25~Mm was applied to each map to remove the central spot region and immediate surrounding area. As expected, after masking, the correlation is significantly improved (note the vertical axis limits are different), reaching values $> 0.98$, similar to what was found by \citet{couvidat2012} and \citet{degrave2014} using real and simulated quiet-Sun data. Plots of the azimuthally-averaged oi travel times for each filter are shown in Figure~\ref{fig:ttoiazim}, with the boundaries of the umbra, penumbra, and circular mask marked for reference. It is clear that the influence of the magnetic field at the center of the model domain has a significant impact on the magnitude and sign of measured travel times. Travel-time definitions begin to show agreement at a radius of $\sim30$~Mm from the spots, with the travel-time averages converging to small values around zero near the edge of the simulation domains.

\begin{figure}
\begin{center}$
\begin{array}{cc}
\includegraphics[width=0.5\linewidth,clip=]{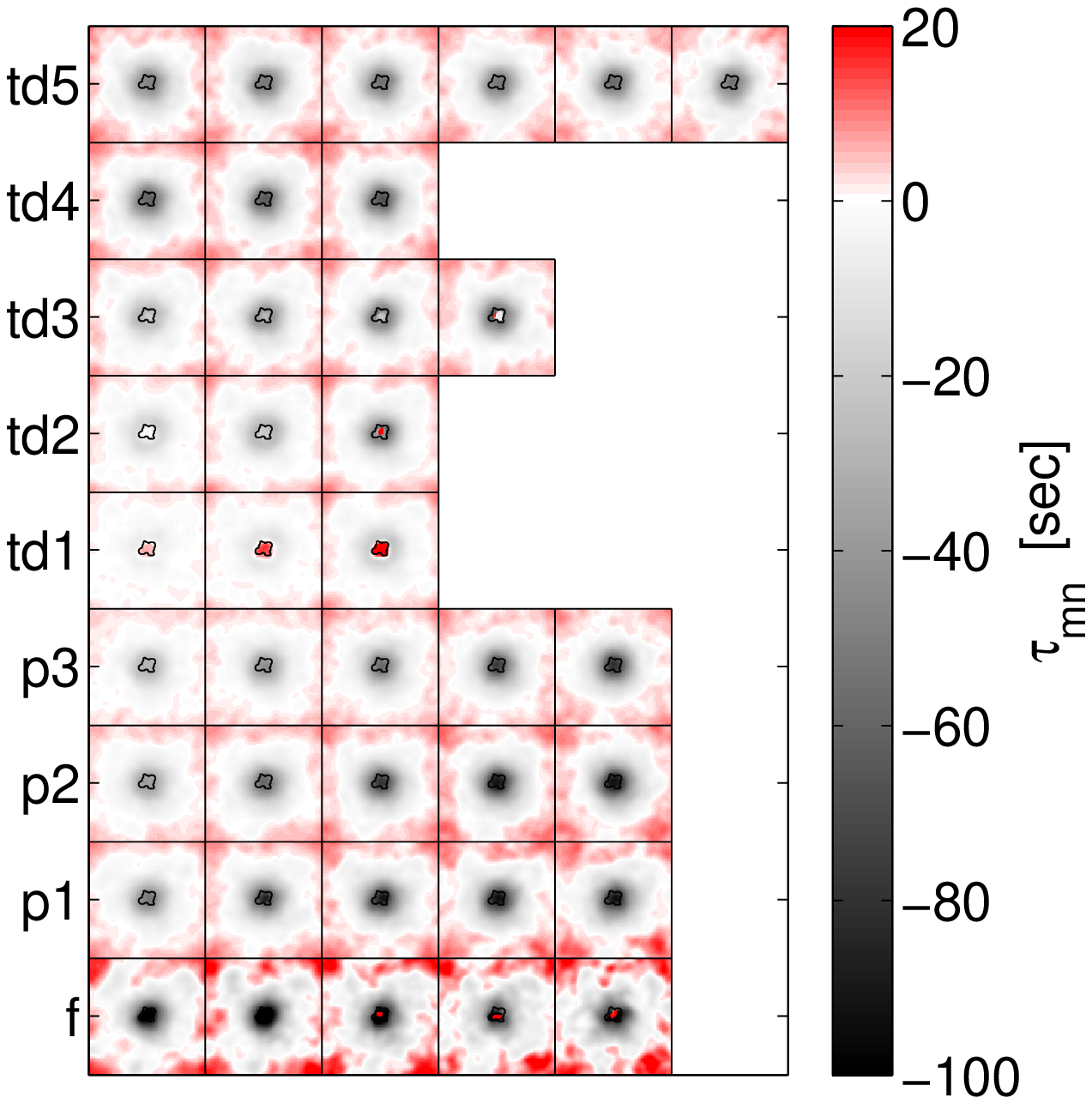} &
\includegraphics[width=0.5\linewidth,clip=]{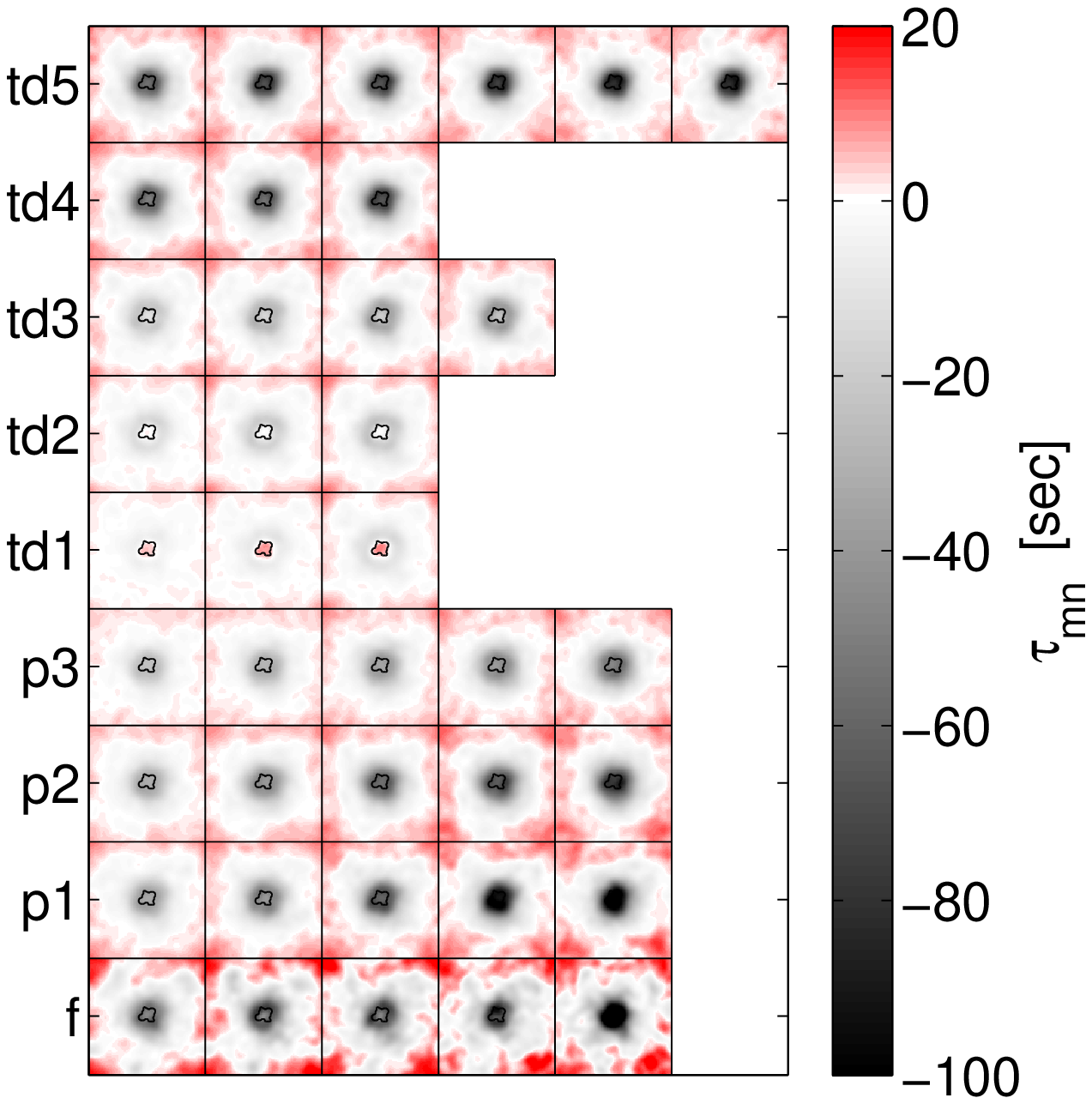} \\
\includegraphics[width=0.5\linewidth,clip=]{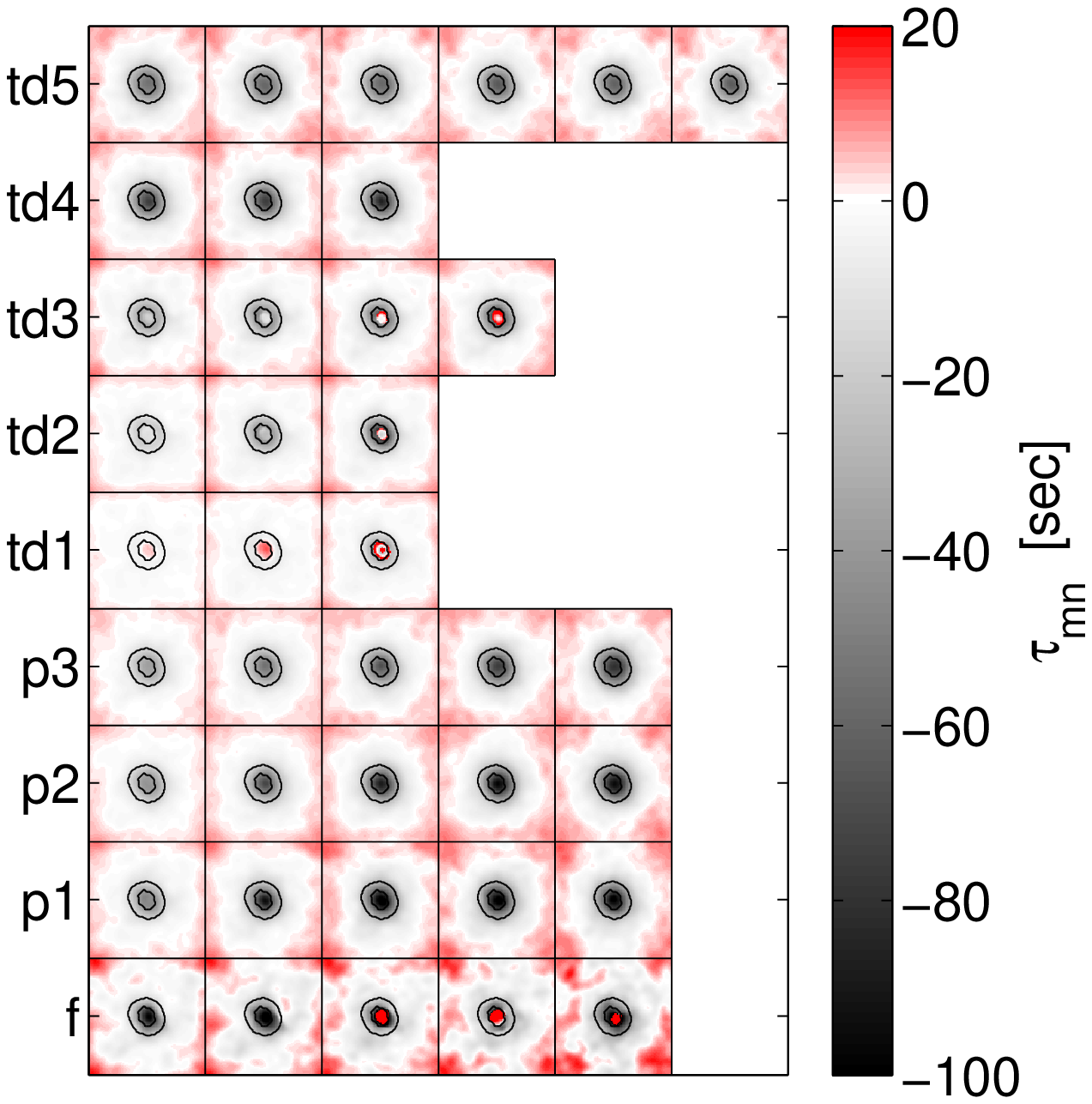} &
\includegraphics[width=0.5\linewidth,clip=]{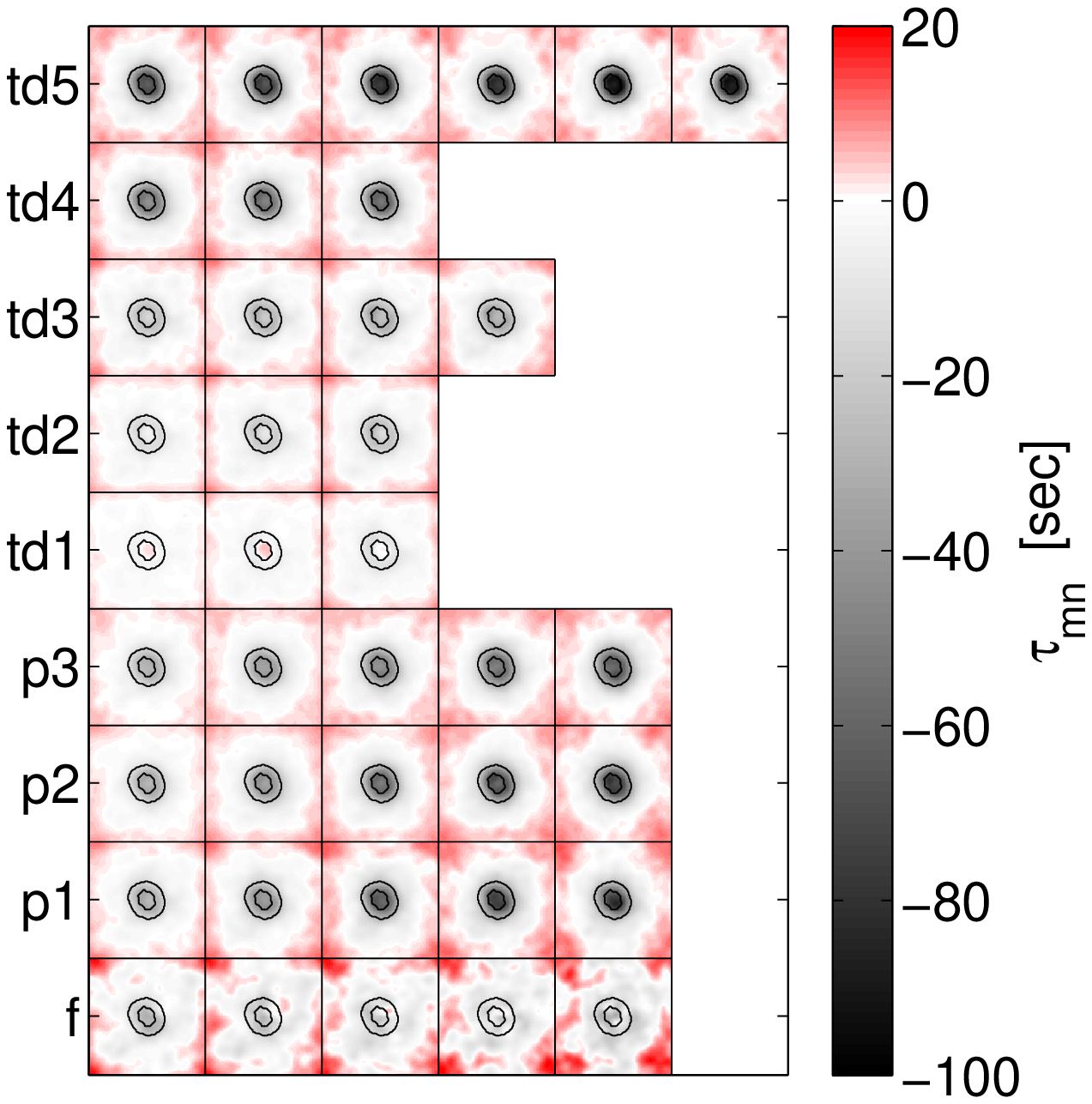}
\end{array}$
\end{center}
\caption{All LRes (top row) and HRes (bottom row) mean travel-time maps measured using GB02 (left column) and GB04 (right column). The $\Delta$ for which each map is computed increases from left to right in each plot, covering the appropriate range of values depending on filter type as defined in the text. The contours mark the boundaries of the simulation umbra and penumbra.}
\label{fig:ttmn}
\end{figure}

For completeness, mean travel-time maps were also computed by averaging the incoming and outgoing travel-time shifts relative to the quiet Sun. These are shown in Figure~\ref{fig:ttmn}. Overall, the two definitions generally agree well with one another and therefore correlate well spatially. Some differences do exists, however, most notably in the $f$-mode measurements made inside the umbra, depending on the value of $\Delta$. Both methods show a dependence of mean travel time on filter type inside the spot, with the low phase-speed measurements differing from those made using the $p$-modes and the higher phase-speed filters. Similar arguments as above concerning the amplitudes of the cross-covariance functions can be made to explain many of the observed differences.

\subsection{Comparison With Forward-Modeled Travel Times}\label{sec:fmtt}
Time-distance helioseismology typically assumes a linear relationship between wave travel times and subsurface perturbations to the wave field. This relationship is often given in the form of an integral equation

\begin{align}
 \delta\tau^{a}(\bm{r}) = h_rh_z\sum_{ij} \bm{K}^{a} (\bm{r}_i - \bm{r},z_j) \cdot \bm{v} (\bm{r}_i,z_j) + N^{a} (\bm{r})
 \label{eq:dt}
\end{align}

\noindent where $h_r=h_xh_y$ and $h_z$ is the vertical grid spacing, $\bm{K}^a$ are three-dimensional vector-valued kernels describing the sensitivity of wave travel times to flows for each particular measurement geometry, filter, and $\Delta$ captured in the $a$ index. $N^a$ represents the noise for travel-time measurement $a$.

A set of kernels $\{K_{v_x}, K_{v_y}, K_{v_z}\}$ was computed by \citet{degrave2014} for flows in the $+\hat{\bm{x}}$ direction in the single scattering Born approximation \citep{birch2007} for use in flow inversions of quiet-Sun simulation data. The calculation of such kernels relies on the accurate modeling of the data power spectrum. The power spectra of the two spot simulations were inspected and compared to those of the quiet-Sun simulations presented in \citet{degrave2014}. These were found to be essentially indistinguishable from one another, in the sense that they both matched the model power equally well. Such a match therefore allows us to reuse the set of kernels from \citet{degrave2014} for the current work.

The use of simulated data gives us an opportunity to test the accuracy of the sensitivity kernels through forward-modeling that would otherwise have been impossible if real solar data were used. A set of forward-modeled travel-time maps was computed by convolving the sensitivity kernels with the known flow fields taken directly from the two spot simulations (Eq.~\ref{eq:dt}). If the kernels were perfect and the measurements free of noise, each modeled travel-time map would match its measured counterpart exactly. This is not the case, as the measured travel times always contain some level of noise, and perturbations due to the presence of magnetic fields are neglected in the sensitivity kernel computation. This is a reasonable approximation in the quiet Sun where such perturbations are relatively small, but could pose a problem for inversions near active regions where the field strength becomes large. We therefore expect some (possibly significant) mismatch between measured and forward-modeled travel-time maps.

\begin{figure}
\begin{center}$
\begin{array}{cc}
\includegraphics[width=0.5\linewidth,clip=]{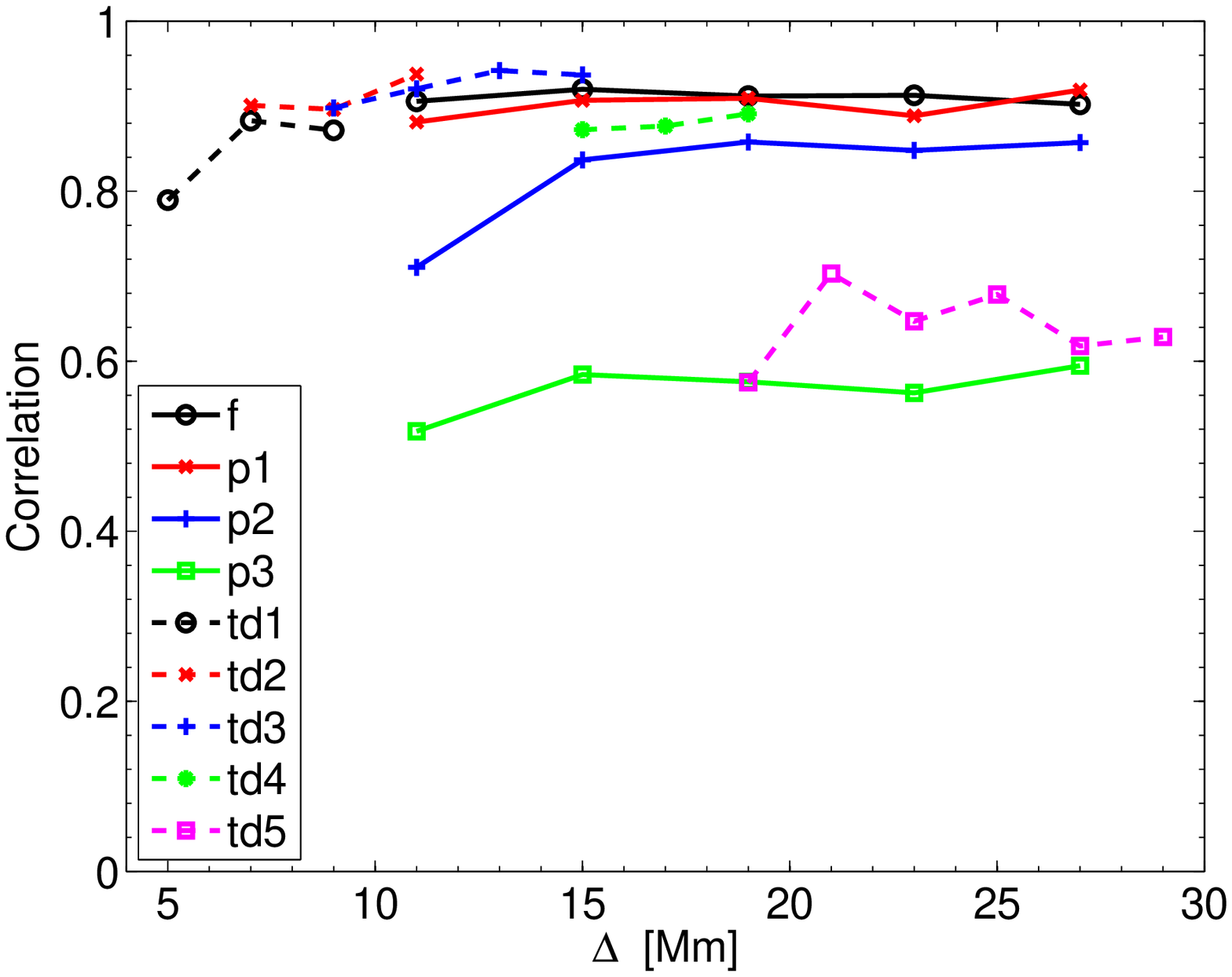} &
\includegraphics[width=0.5\linewidth,clip=]{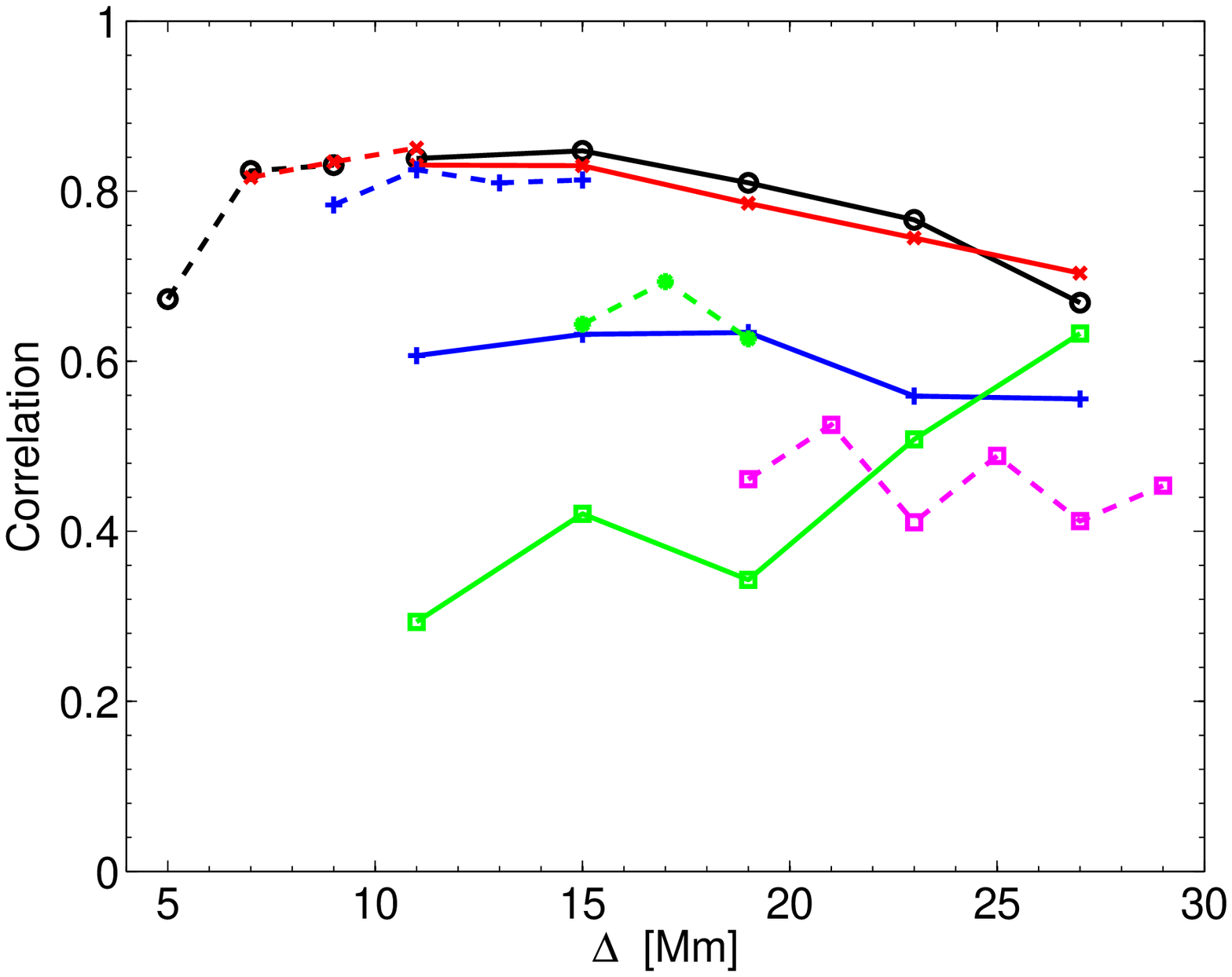}
\end{array}$
\end{center}
\caption{The 2D correlation between measured and forward modeled LRes (left) and HRes (right) oi travel-time maps using the GB02 definition after applying the circular mask to each map to eliminate the spot and immediate surrounding area.}
\label{fig:ttmeasmod}
\end{figure}

Figure~\ref{fig:ttmeasmod} shows the correlation of the measured and forward-modeled oi travel times for every filter for LRes and HRes respectively. Only data from the quiet Sun are shown after applying the circular mask to remove the strong-field regions. We show the results for the GB04 definition as both sets of travel-time maps give indistinguishable results in the quiet Sun. The correlation with the modeled travel times show reasonably high values in the range of $\sim0.5$--0.9. Comparing these figures to Figure~6 of \citet{degrave2014}, there appears to be some consistency with correlation values found outside of the spots. We find that on average the LRes correlations are somewhat higher than those of HRes even after masking, and both simulations show lower masked correlations for filters $p_3$ and $\rm{td_5}$. This was also found by \citet{degrave2014} for filters $p_3$ and $\rm{td_5}$. The disagreement indicates that there are likely effects from strong flows present in the quiet regions of the simulations that the linear kernels are not adequately capturing.

\subsection{Travel-Time Noise}\label{sec:ttnoise}

The noise produced by the stochastic convective motions that excite solar oscillations induces correlations in the travel-time measurements. In \citet{degrave2014}, noise covariances were computed based on the model of \citet{gb04}. Since the simulations studied here, particularly the convective properties, are very similar to the simulations analyzed in \citet{degrave2014}, we estimate the noise from those matrices already computed. 


\section{Inversion Results}\label{sec:results}
Time-distance inversions were carried out according to the inversion scheme outlined by \citet{degrave2014}. The goal is to recover all three vector velocity flow components $(v_x, v_y, v_z)$ at depths of 1, 3, and 5~Mm below the surface of both simulation domains. To do this, we employ the Subtractive Optimally Localized Averaging (SOLA) method \citep{pijpers1992,svanda2011,jackiewicz2012}. Given a set of sensitivity kernels and noise covariance matrices, the SOLA method searches for a set of inversion weights that, when linearly combined with the travel-time measurements, gives an estimate of flow component $\alpha=\{x,y,z\}$ at a targeted location centered at a depth $z_0$ within the simulation domain
\begin{equation}
 v_{\alpha}^{\rm inv} (\bm{r}; z_0) = \sum_i \sum_{a=1}^M w_a^{\alpha} (\bm{r}_i - \bm{r}; z_0) \delta \tau^a(\bm{r}_i),
 \label{vinv}
 \end{equation}
where M represents the number of travel-time maps used in a particular inversion. The weights are used to linearly combine the sensitivity kernels to create an averaging kernel. Ideally, if the weights are suitable, the averaging kernel will be well-localized in space and will closely match a pre-defined 3D Gaussian-shaped `target function' $T$.

Because the sensitivity kernels and noise covariance matrices for this work are identical to those of the \citet{degrave2014} quiet-Sun simulation inversions, we use those pre-computed weights ($w^\alpha_a$) in conjunction with the new spot travel-time maps ($\delta\tau^a$) to recover flows in the LRes and HRes domains. Using these old weights means that all parameters (i.e. resolution, noise, etc.) for subsequent inversions presented in this paper are identical to those listed in Table~1 of \citet{degrave2014}. We show only the results obtained using the weights from inverting the ``QS1'' simulation in \citet{degrave2014}.

We quantify the inversion results by comparing our recovered flows denoted by $v_\alpha^{\rm inv}$, to the flows from the numerical simulations. To facilitate direct comparisons, we smooth the artificial data to the expected resolution of the inverted flows by convolving the inversion target function $T$ with the raw simulation flow field to obtain the ``targeted'' answer $v_\alpha^{\rm tgt}$ for flow component $\alpha$
\begin{equation}
 v_\alpha^{\rm tgt}(\bm{r};z_0) = h_r h_z\sum_{ij} T_\alpha(\bm{r}_i - \bm{r},z_j;z_0)v_\alpha(\bm{r}_i,z_j).
 \label{valph}
\end{equation}
The sum over index $i$ represents a horizontal convolution, while the sum of the products over $j$ takes place at the same depth slices.

We note here that because the magnetic perturbation in the model sunspots introduces many problems as described above, we do not expect to accurately recover the flow structure there. However, we show the inversion results in these regions anyway for the sake of completeness.


\subsection{Horizontal Flow Inversions}\label{sec:hinv}
\begin{figure}
\begin{center}$
\begin{array}{ccc}
\includegraphics[width=0.2\linewidth,clip=]{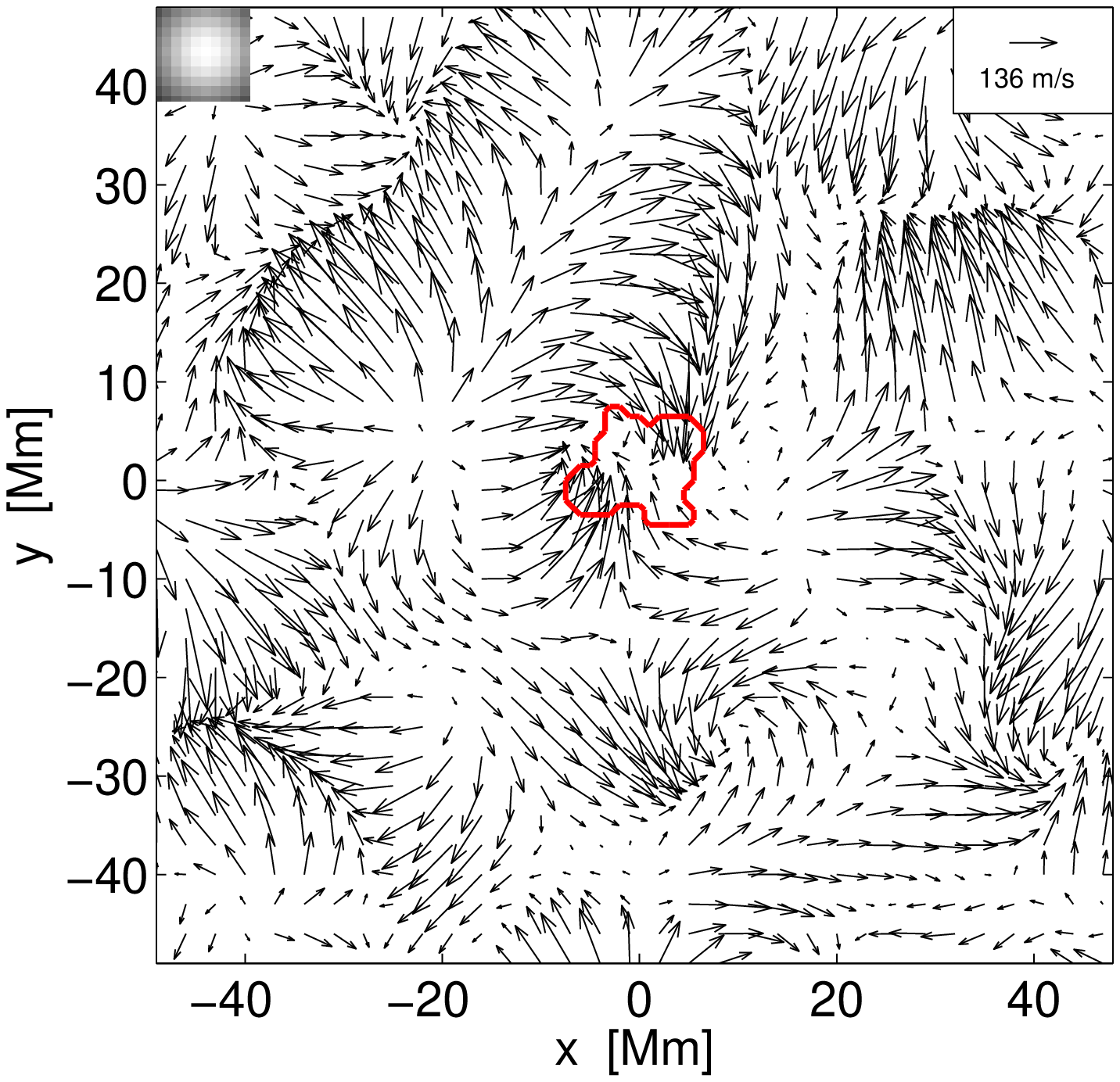} &
\includegraphics[width=0.2\linewidth,clip=]{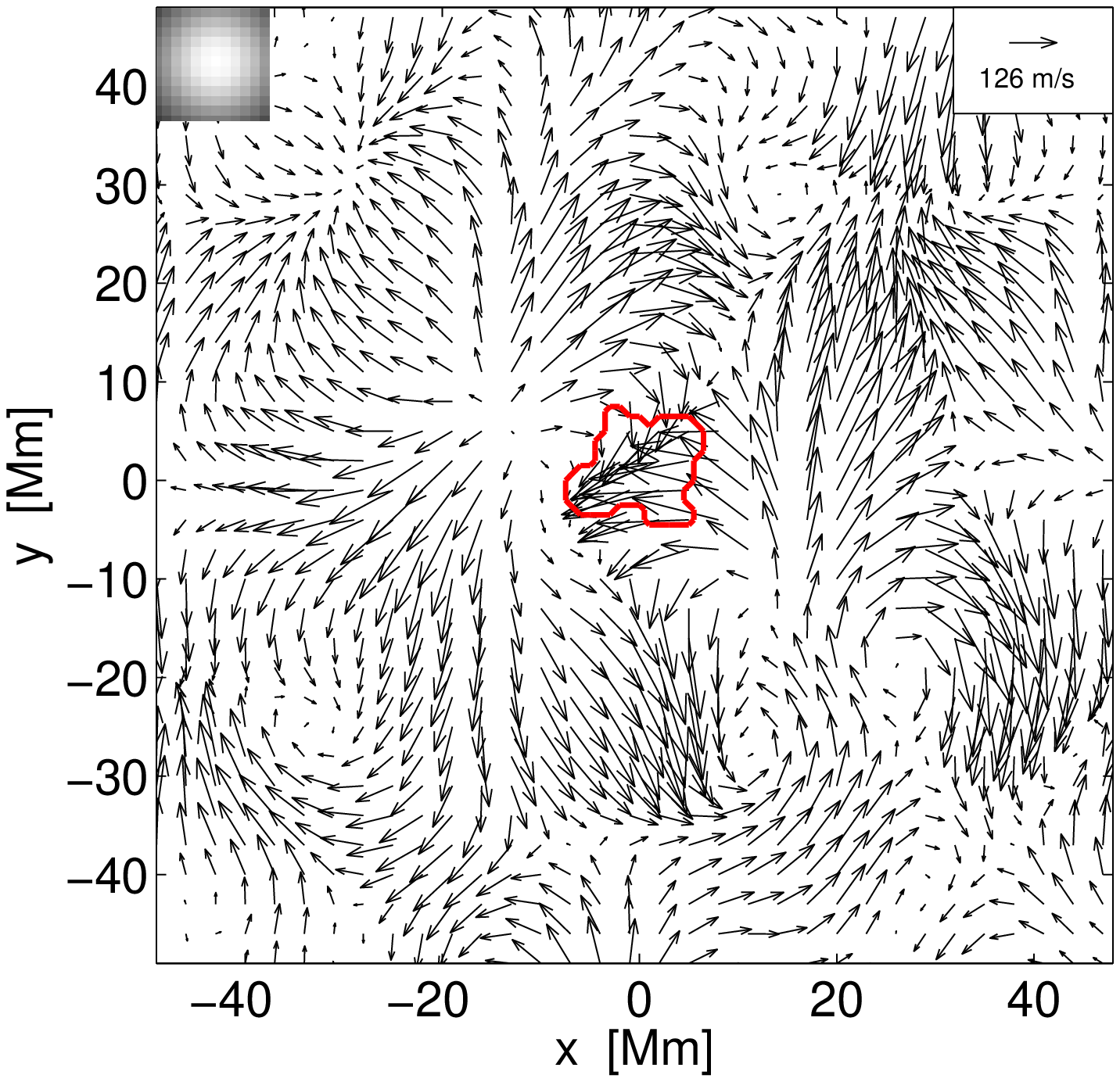} &
\includegraphics[width=0.2\linewidth,clip=]{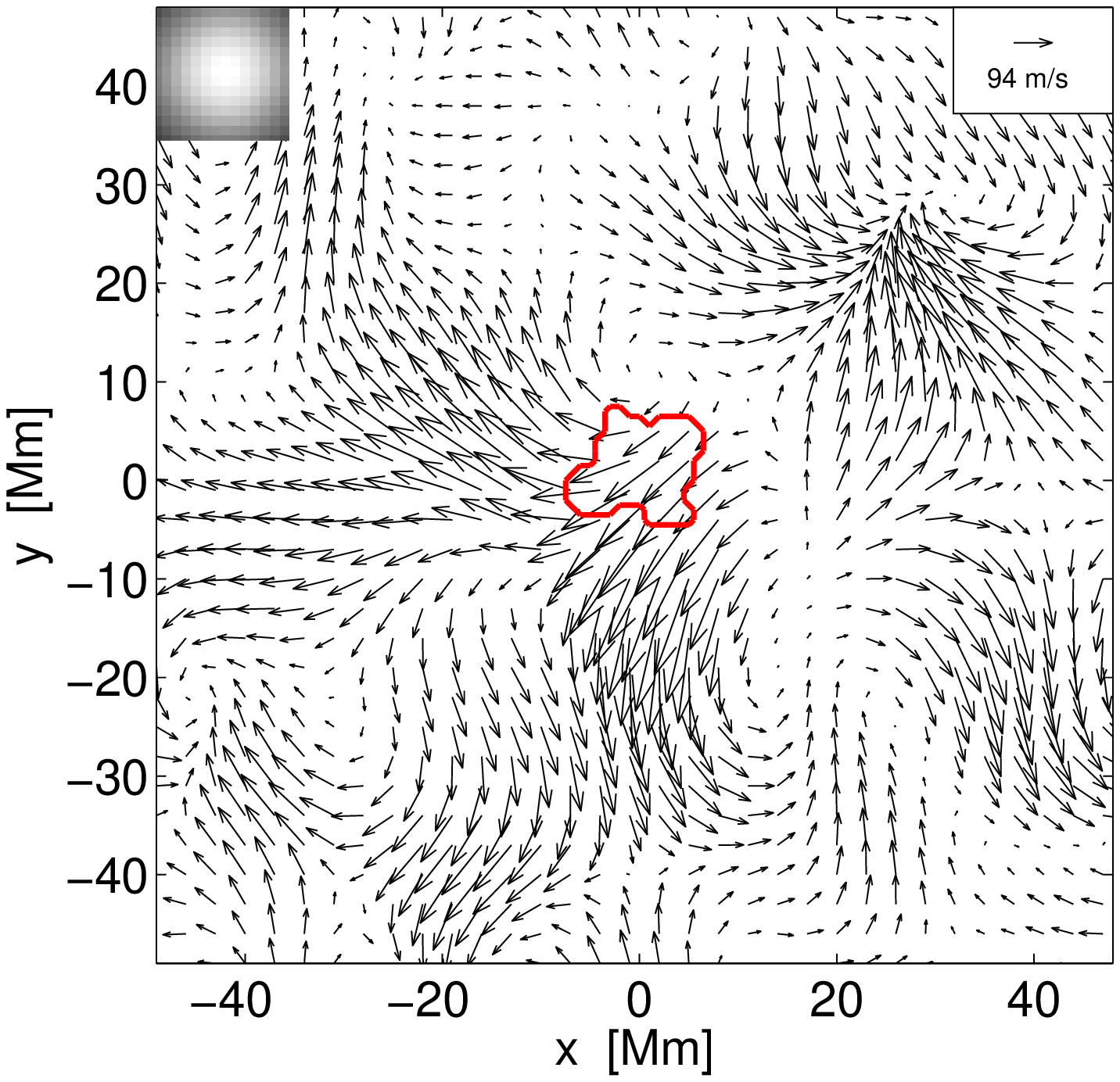} \\
\includegraphics[width=0.2\linewidth,clip=]{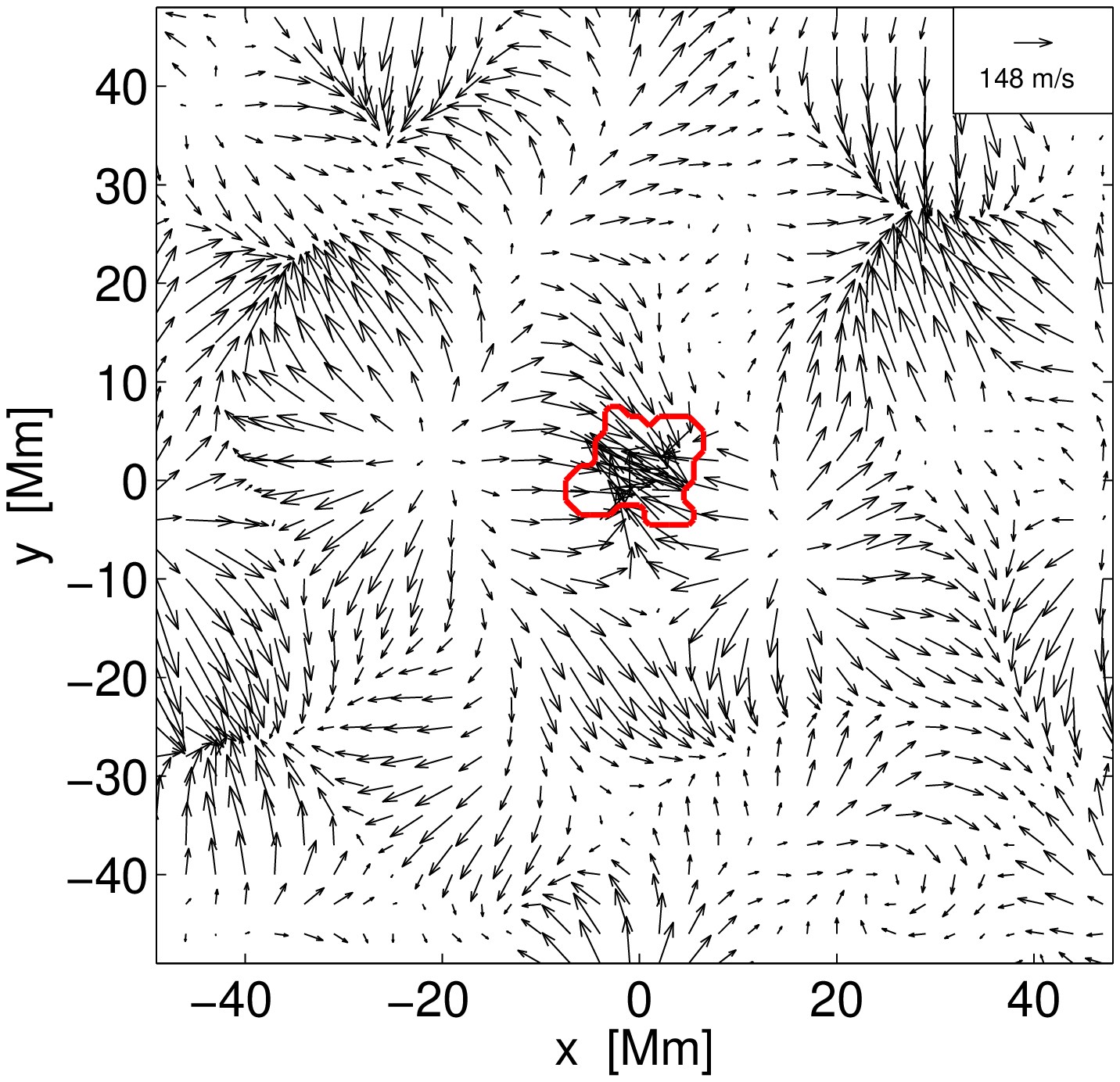} &
\includegraphics[width=0.2\linewidth,clip=]{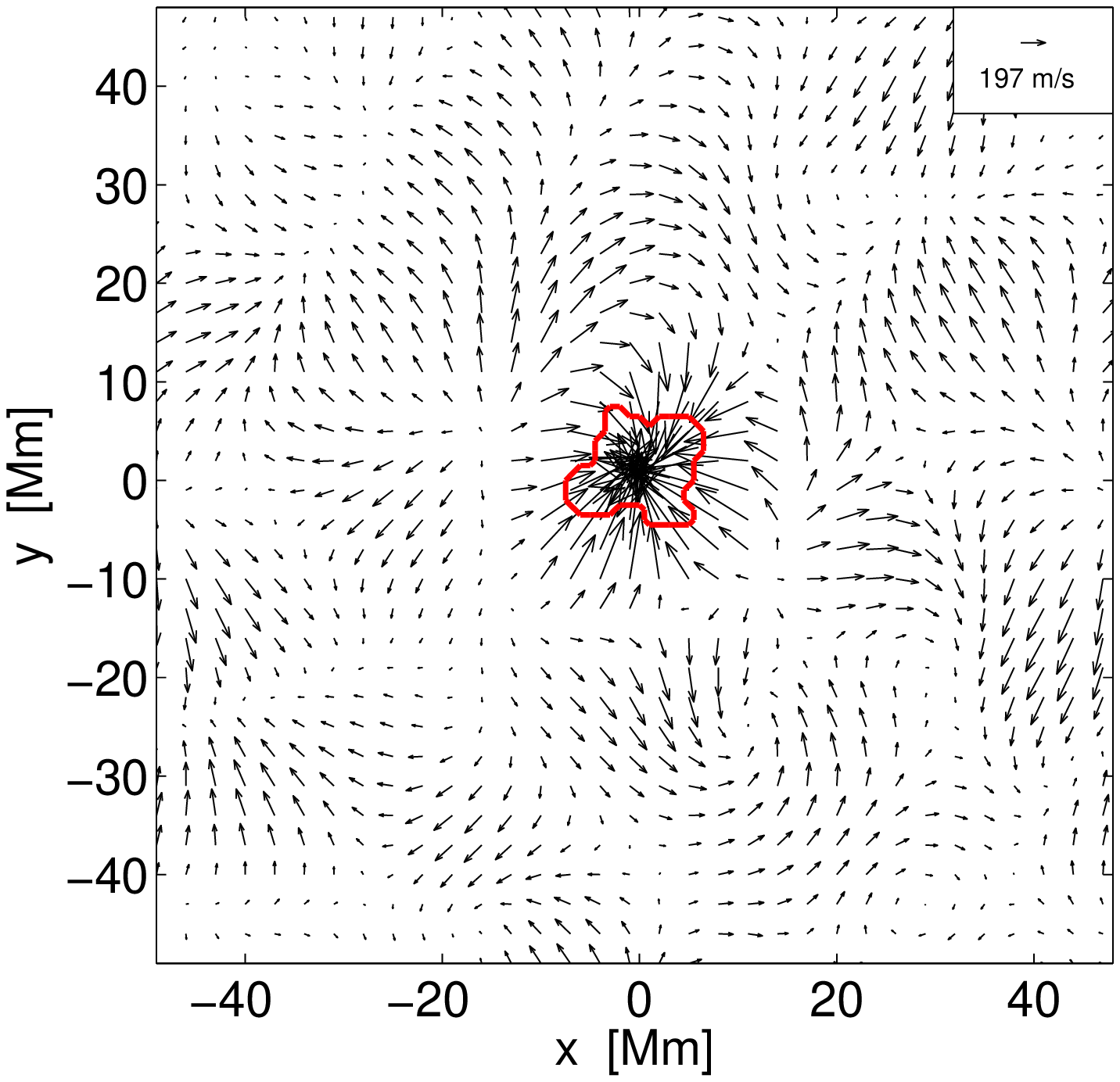} &
\includegraphics[width=0.2\linewidth,clip=]{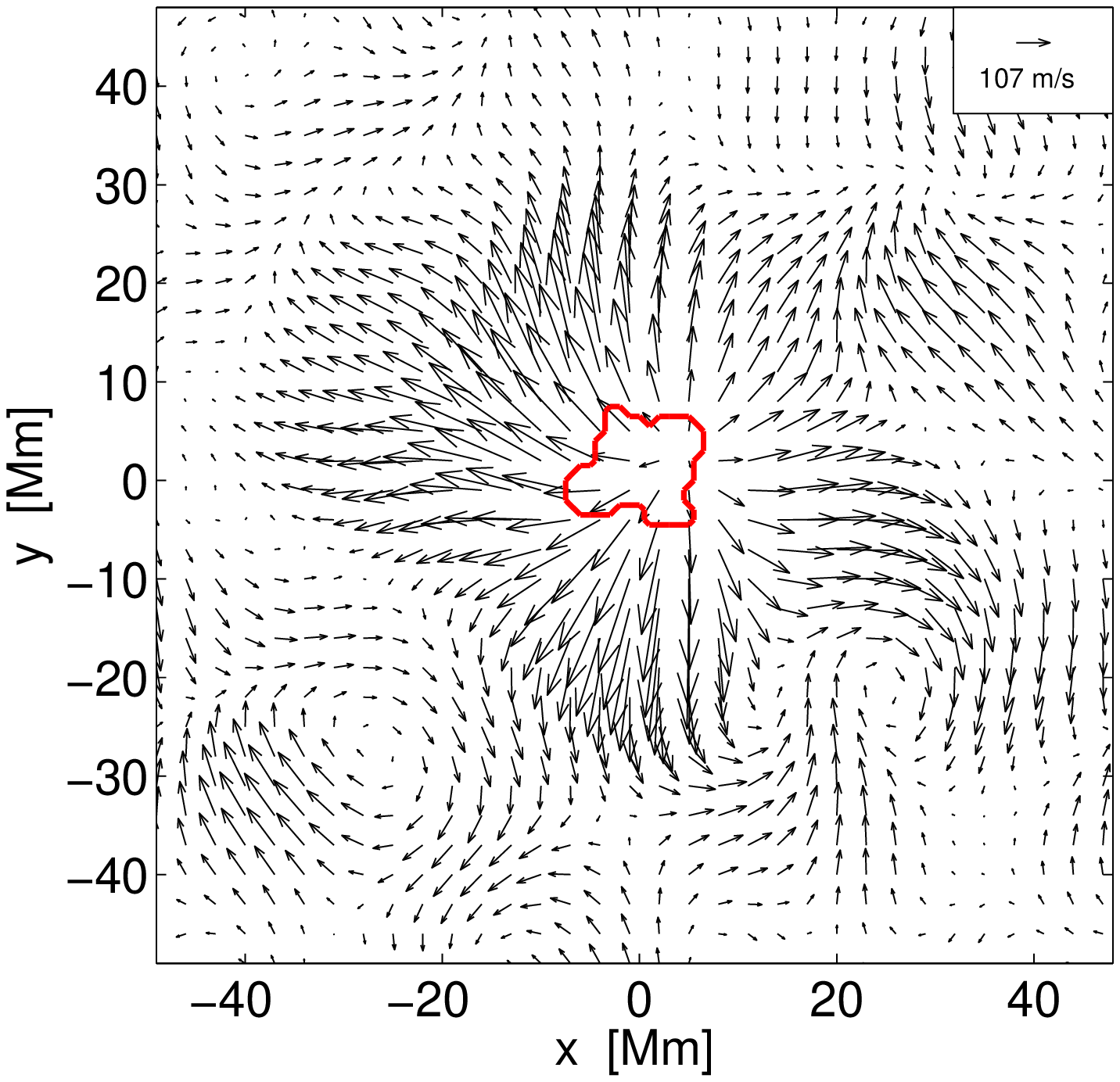} \\
\includegraphics[width=0.2\linewidth,clip=]{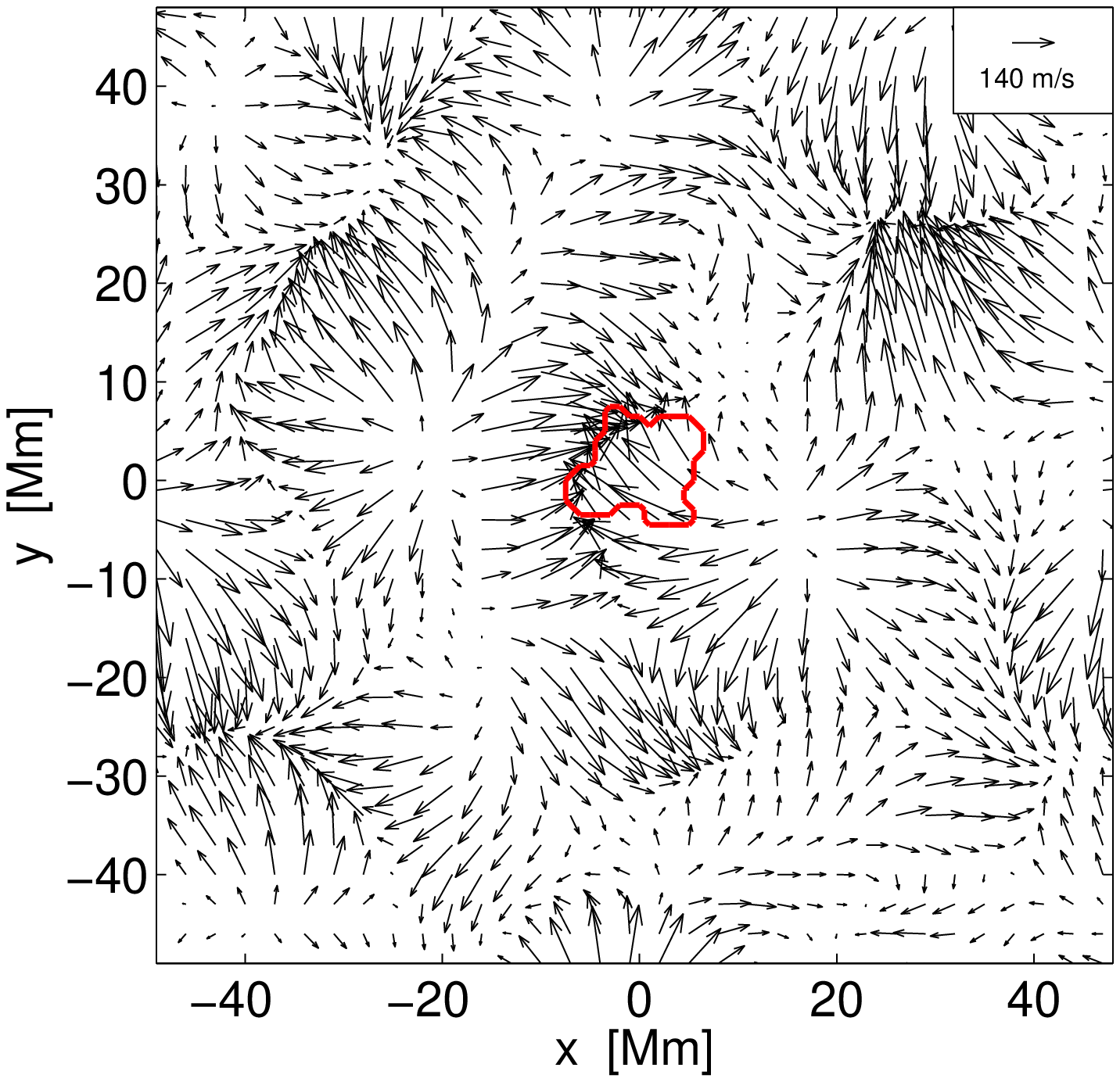} &
\includegraphics[width=0.2\linewidth,clip=]{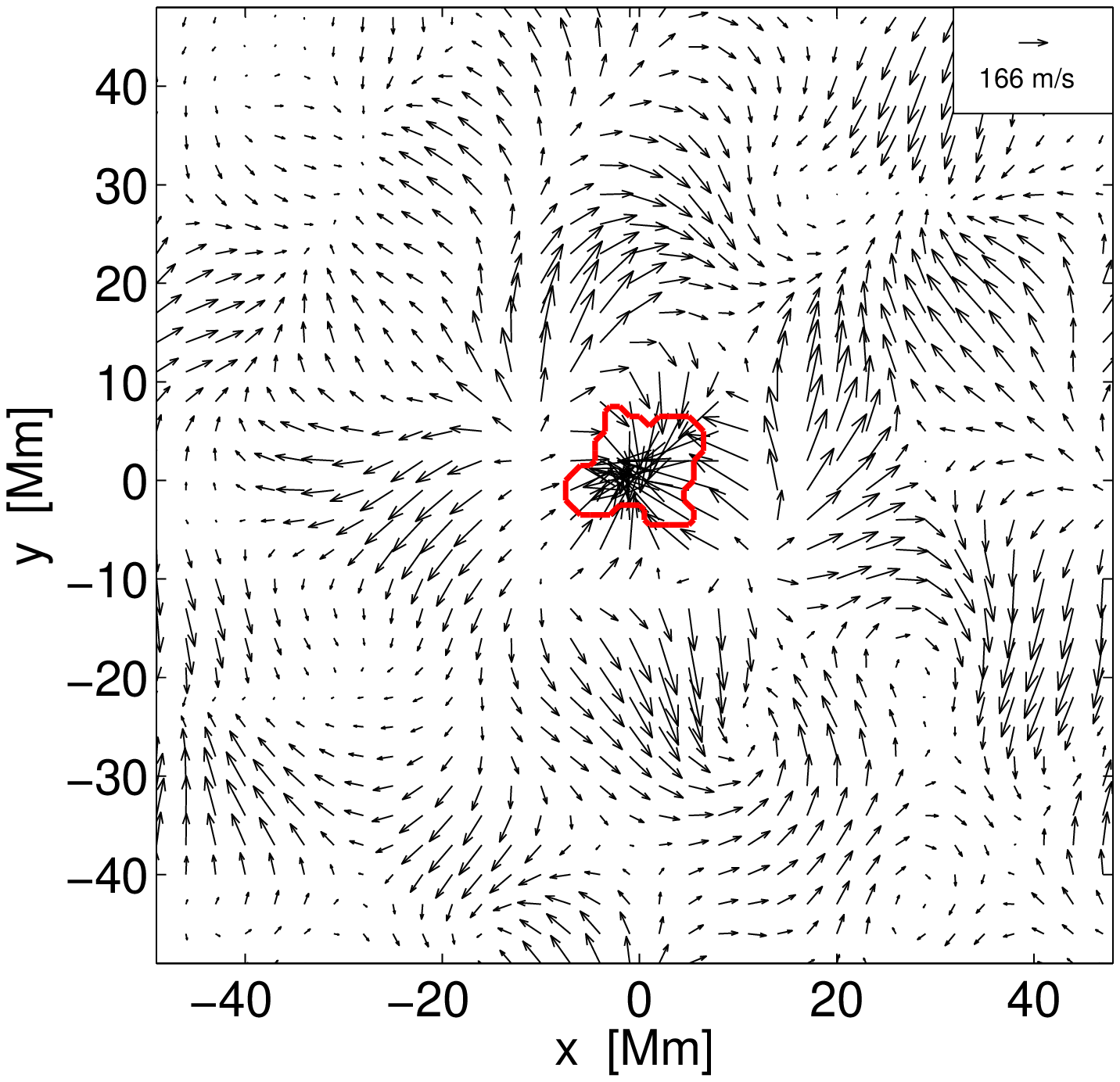} &
\includegraphics[width=0.2\linewidth,clip=]{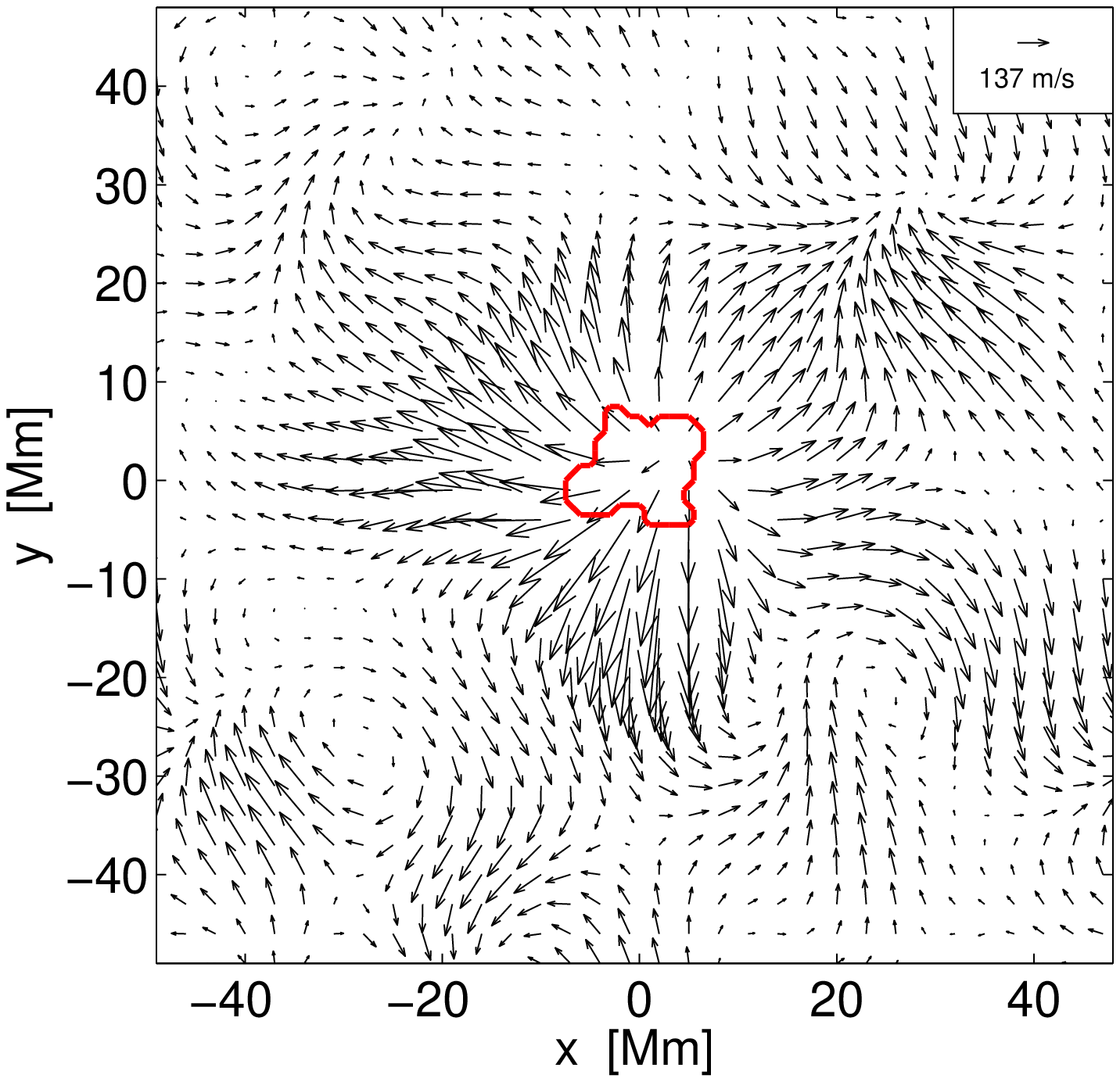} \\
\includegraphics[width=0.2\linewidth,clip=]{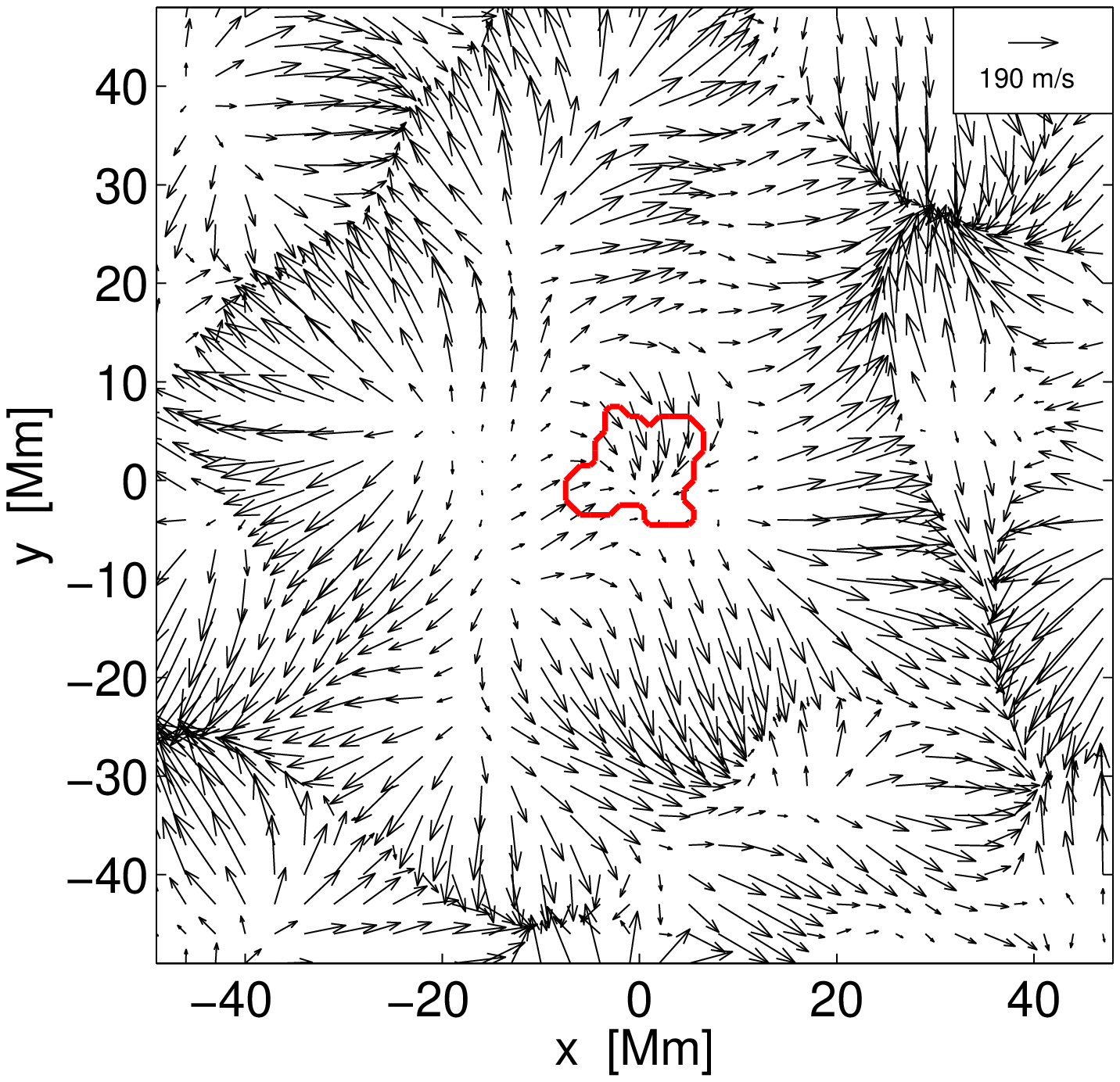} &
\includegraphics[width=0.2\linewidth,clip=]{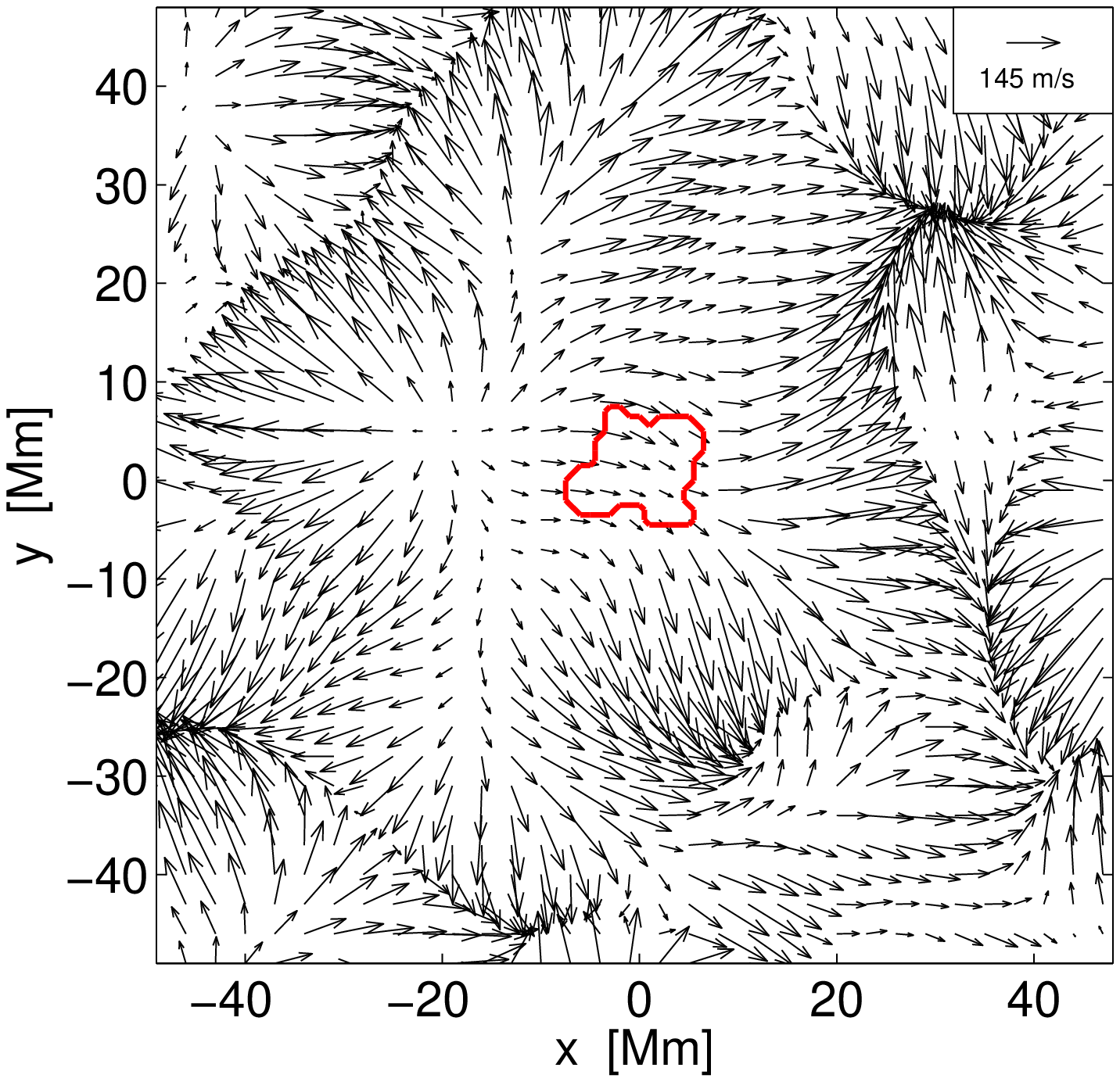} &
\includegraphics[width=0.2\linewidth,clip=]{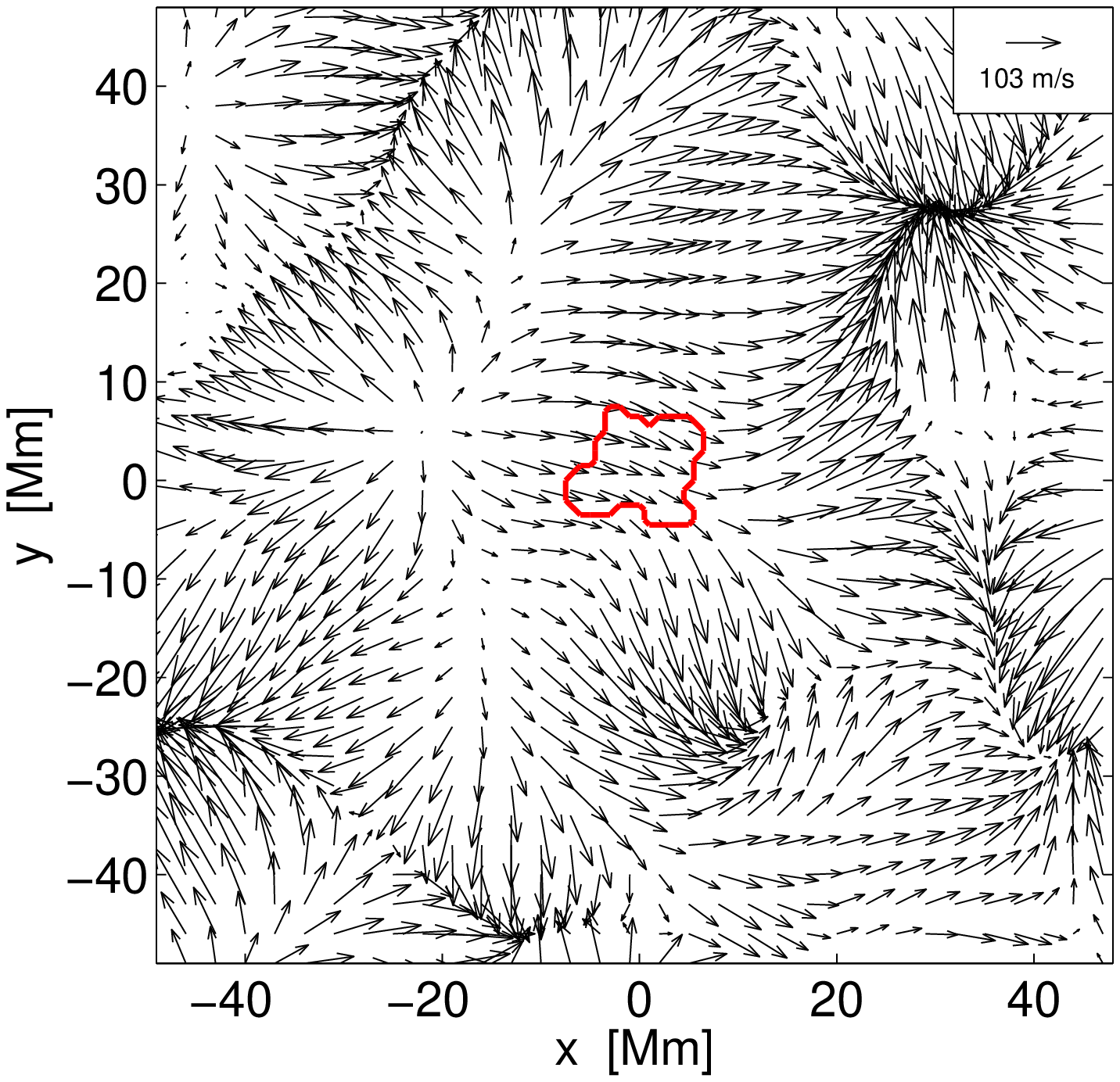}
\end{array}$
\end{center}
\caption{LRes GB02 horizontal $(v_x, v_y)$ inversion flow maps for the ridge (first row), phase-speed (second row), and ridge+phase-speed (third row) travel-time differences for depths (left to right) 1, 3 and 5~Mm. The smoothed simulation flow maps (i.e. $v_{x,y}^{\rm tgt}$) at these depths are shown in the bottom row. The noise for each inversion is $\sim35~\rm{ms^{-1}}$ and the reference arrows represent the RMS velocity corresponding to each flow map. The 2D target function at each depth is shown in the upper lefthand corner of the first row figures. The width of the box corresponds to the horizontal FWHM of each target function and represents the approximate spatial resolution of each flow map. All maps in the same column have identical horizontal resolution. The contour marking the boundary of the spot umbra has been overplotted.}
\label{fig:vxy02SC}
\end{figure}

\begin{figure}
\begin{center}$
\begin{array}{ccc}
\includegraphics[width=0.2\linewidth,clip=]{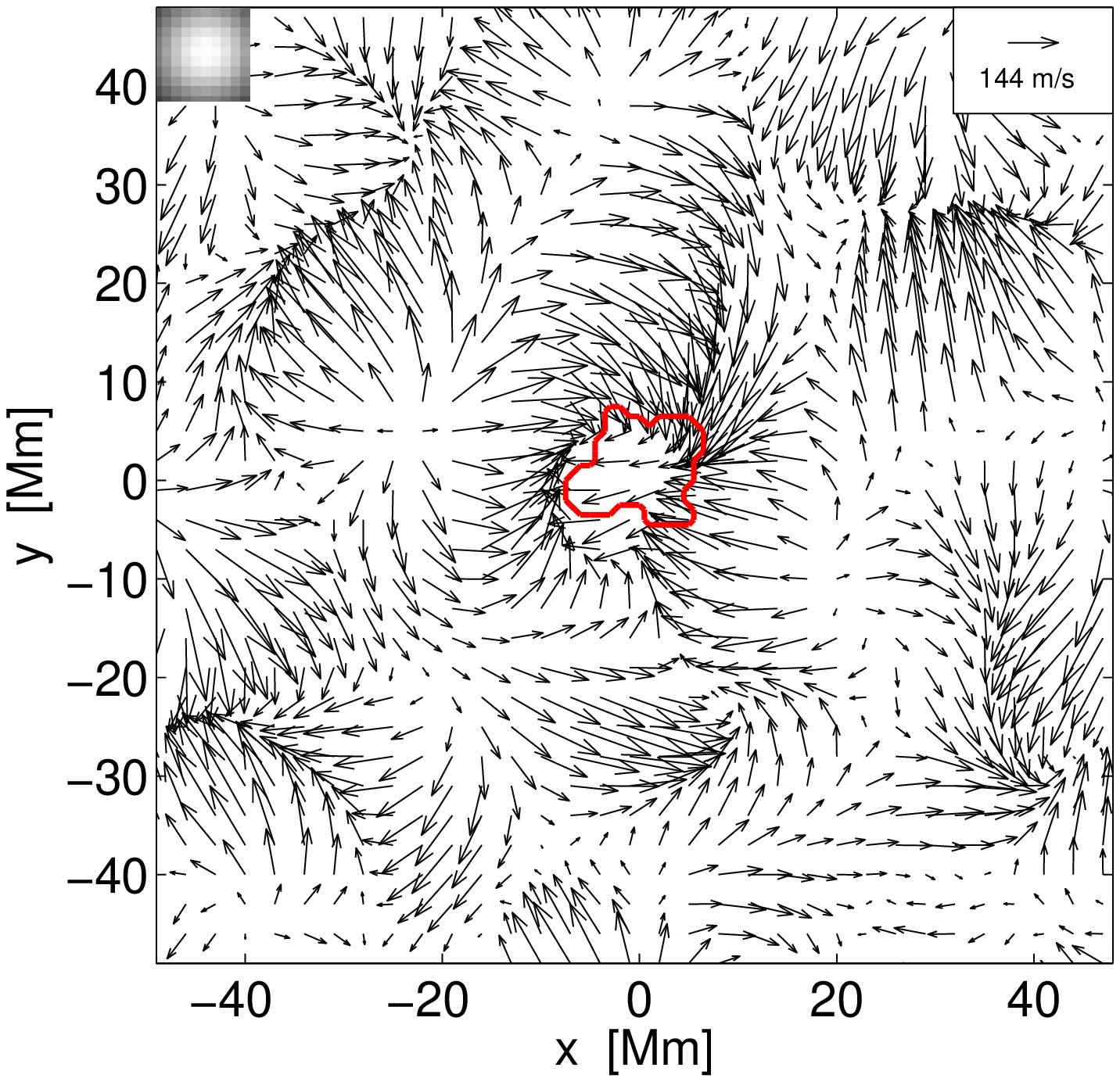} &
\includegraphics[width=0.2\linewidth,clip=]{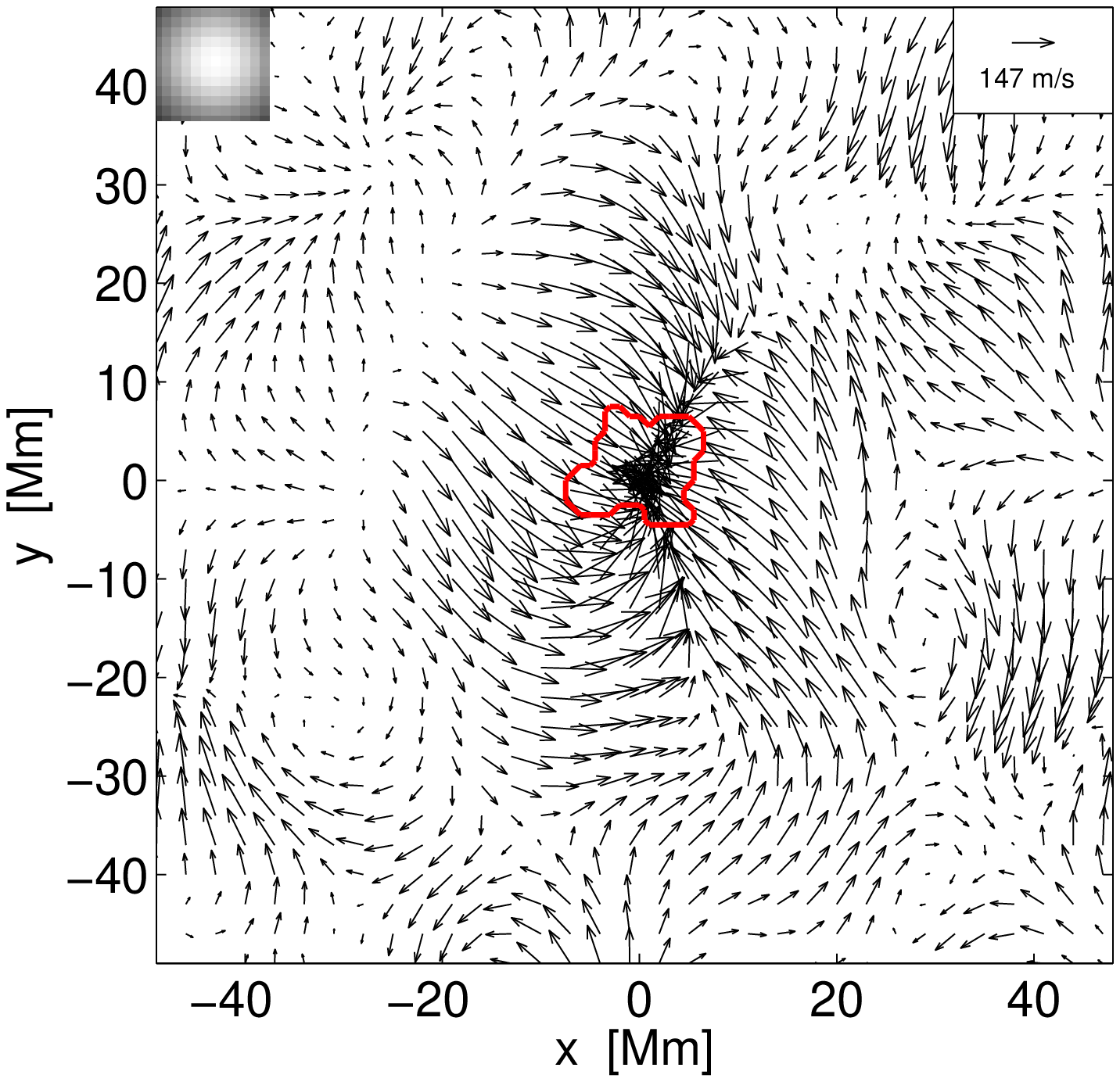} &
\includegraphics[width=0.2\linewidth,clip=]{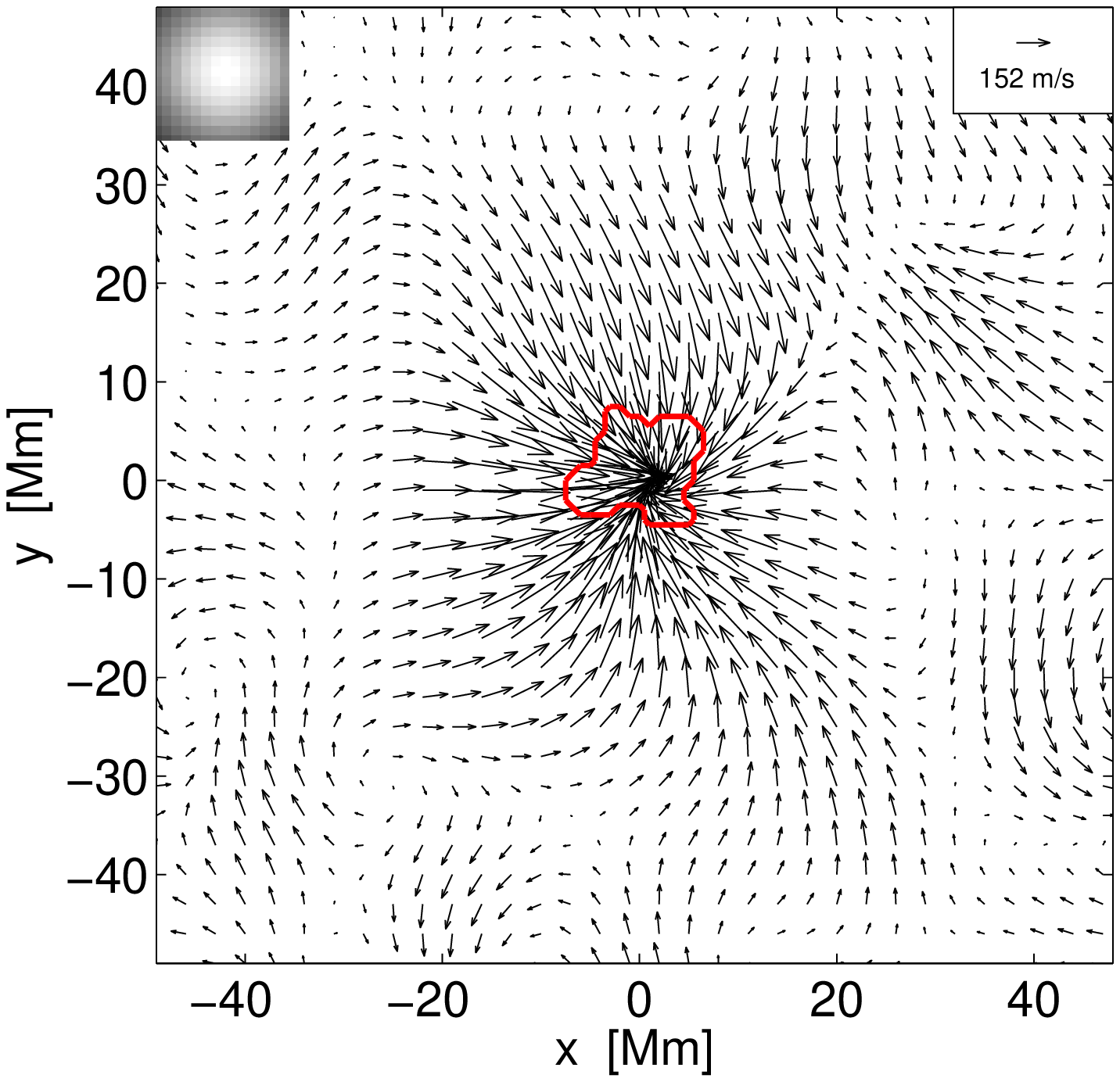} \\
\includegraphics[width=0.2\linewidth,clip=]{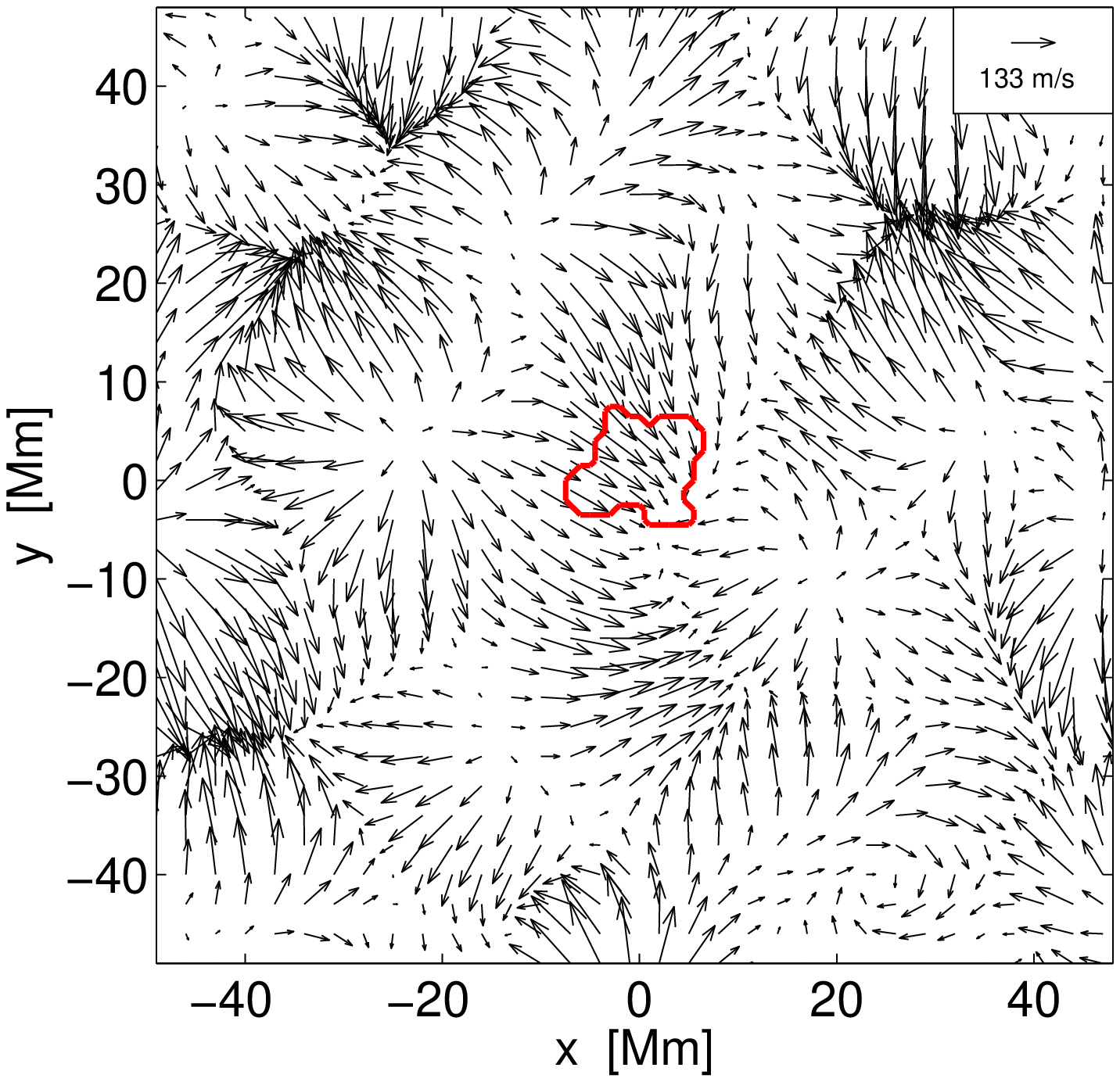} &
\includegraphics[width=0.2\linewidth,clip=]{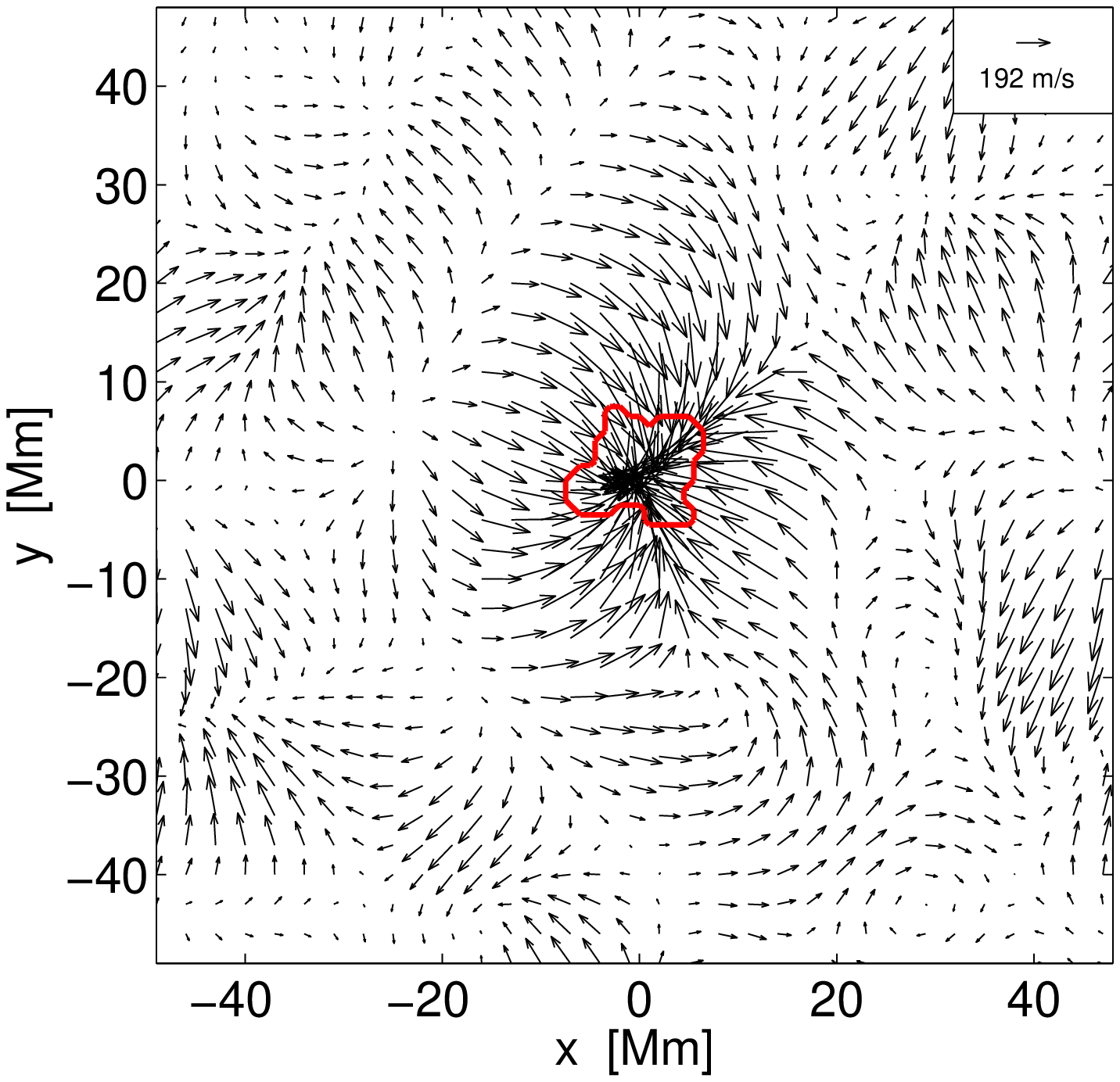} &
\includegraphics[width=0.2\linewidth,clip=]{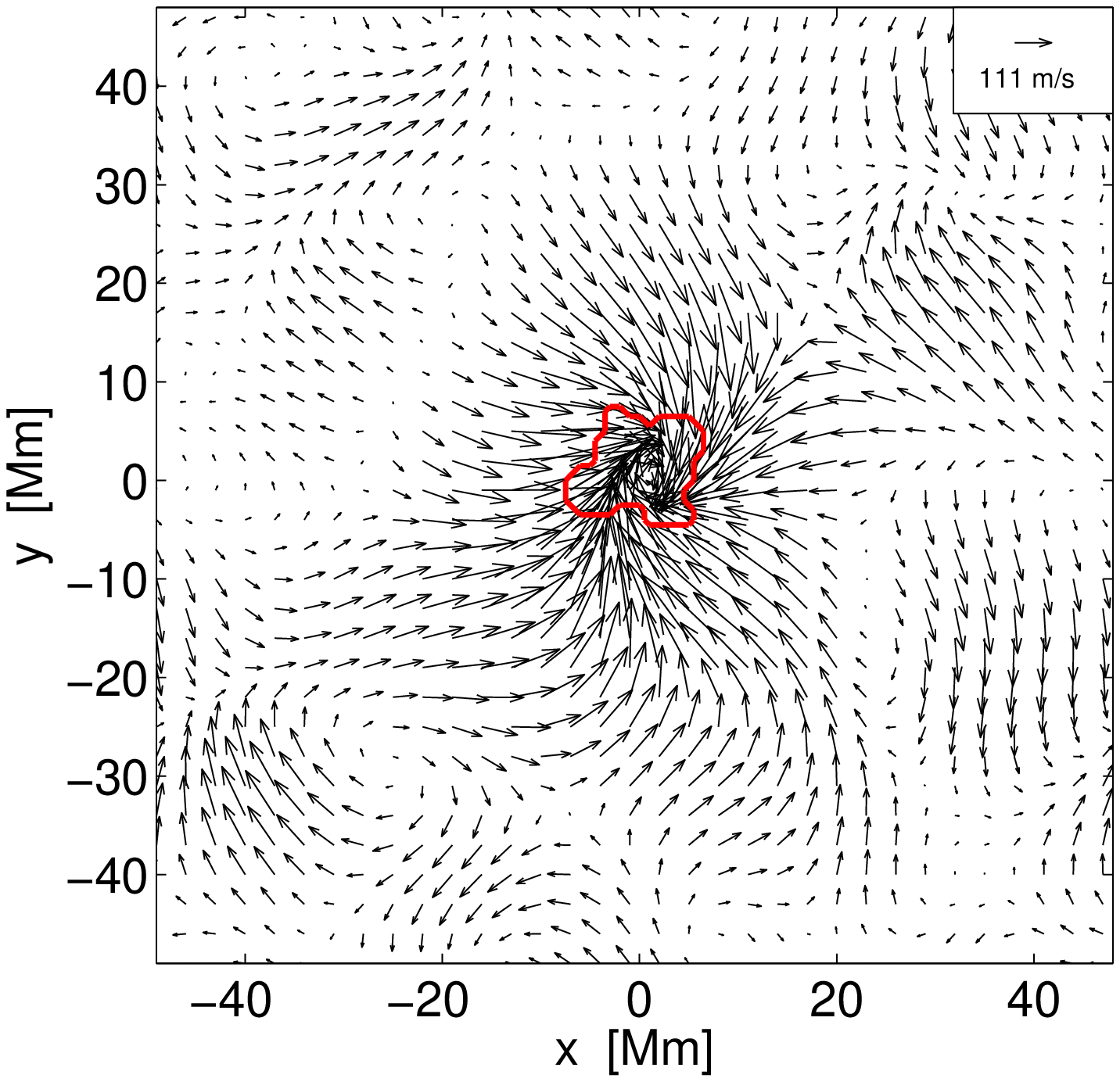} \\
\includegraphics[width=0.2\linewidth,clip=]{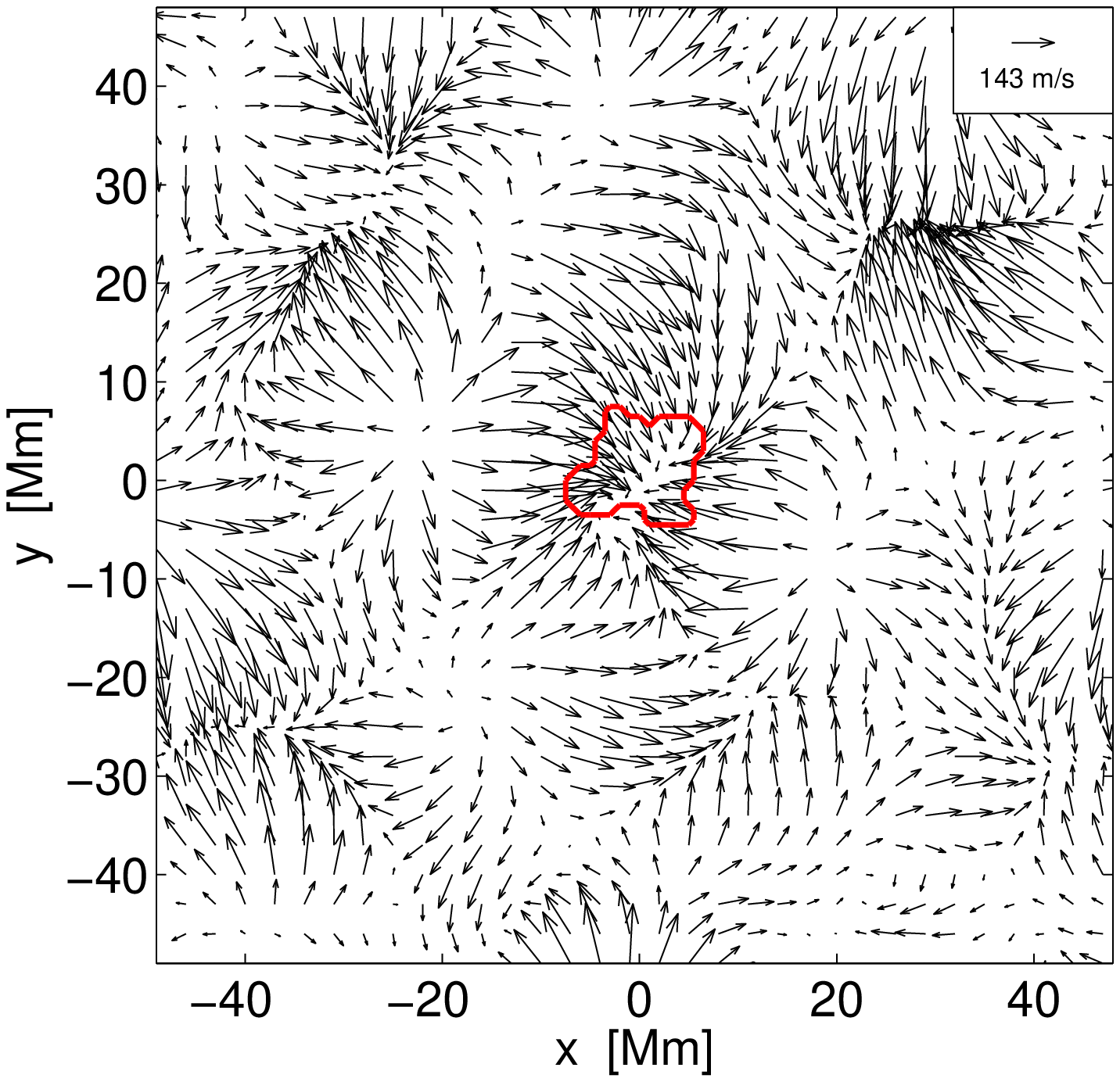} &
\includegraphics[width=0.2\linewidth,clip=]{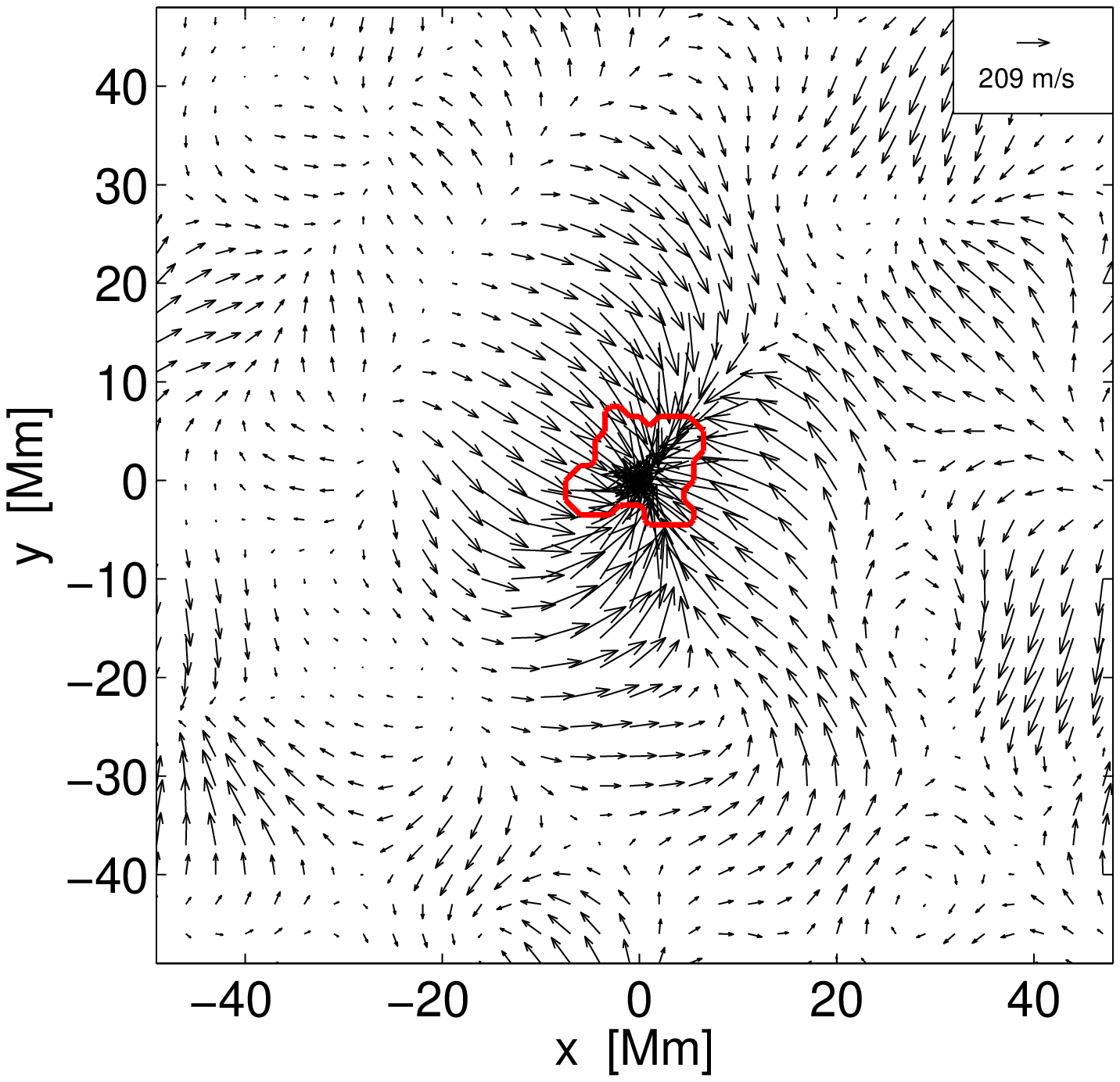} &
\includegraphics[width=0.2\linewidth,clip=]{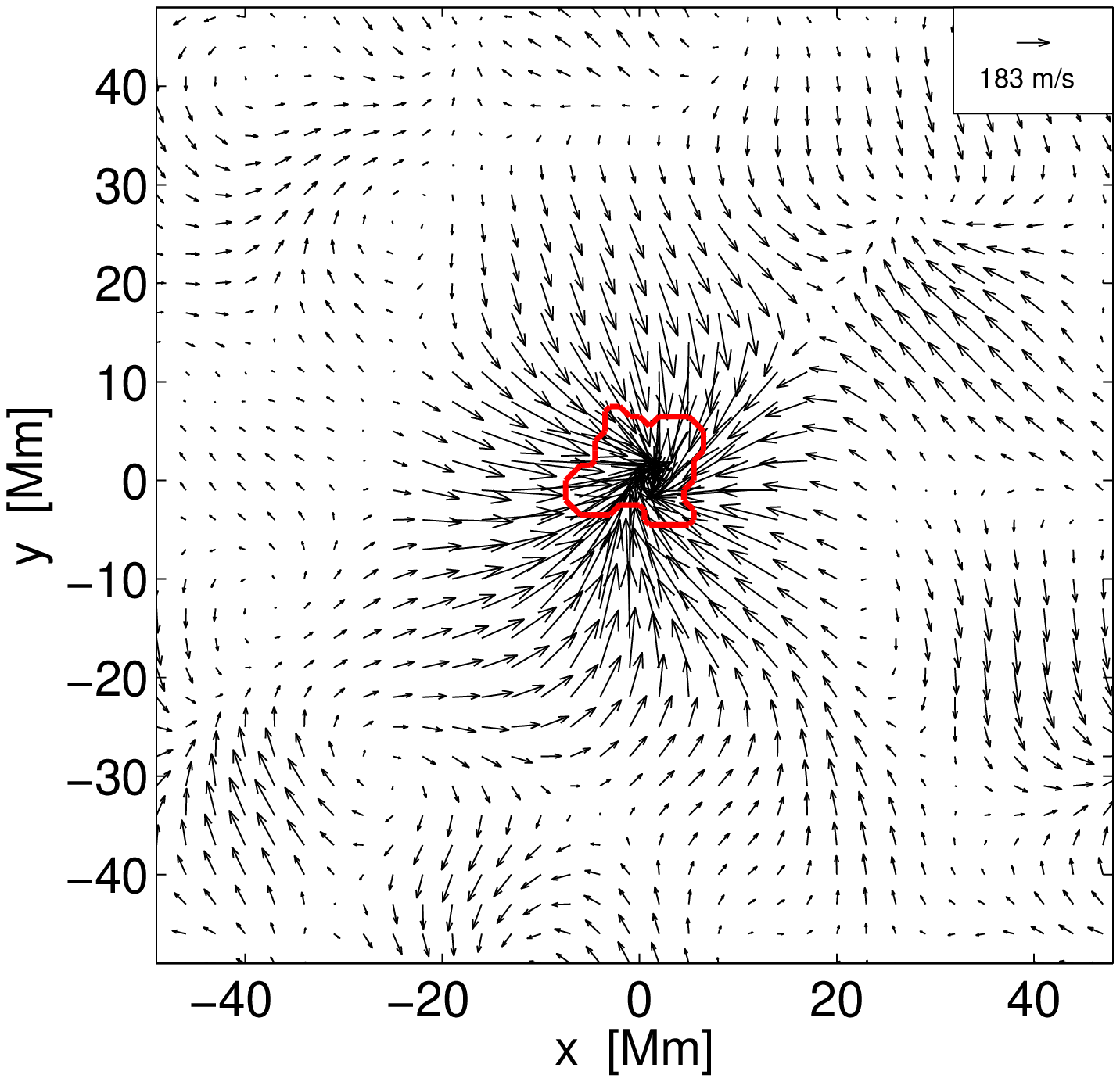} \\
\includegraphics[width=0.2\linewidth,clip=]{f8j.eps} &
\includegraphics[width=0.2\linewidth,clip=]{f8k.eps} &
\includegraphics[width=0.2\linewidth,clip=]{f8l.eps}
\end{array}$
\end{center}
\caption{LRes GB04 horizontal $(v_x, v_y)$ inversion flow maps for the ridge (first row), phase-speed (second row), and ridge+phase-speed (third row) travel-time differences for depths (left to right) 1, 3 and 5~Mm. The smoothed simulation flow maps (i.e. $v_{x,y}^{\rm tgt}$) at these depths are shown in the bottom row. The noise for each inversion is $\sim35~\rm{ms^{-1}}$ and the reference arrows represent the RMS velocity corresponding to each flow map. The 2D target function at each depth is shown in the upper lefthand corner of the first row figures. The width of the box corresponds to the horizontal FWHM of each target function and represents the approximate spatial resolution of each flow map. All maps in the same column have identical horizontal resolution. The contour marking the boundary of the spot umbra has been overplotted.}
\label{fig:vxy04SC}
\end{figure}

Figures~\ref{fig:vxy02SC} and \ref{fig:vxy04SC} are the inverted $v_{x,y}^{\rm inv}$ horizontal flow maps for LRes at each depth using the GB02 and GB04 travel-time definitions respectively. Each row represents the separate ridge and phase-speed filter inversions (rows 1 and 2 respectively), and the combined ridge+phase-speed filter inversions (row 3). The noise for each is $\sim35~\rm{m\,s^{-1}}$. For comparison, the simulation flows, $v_{x,y}^{\rm tgt}$ (see Eq.~\ref{valph}), at each corresponding depth are shown in row 4. These represent the best case scenario that we can hope to accomplish with our inversions. For reference, the boundary of the spot umbra is marked by a single contour line. The approximate horizontal and vertical resolutions of each inversion (i.e. the horizontal and vertical FWHM of the target function) along with the SOLA regularization parameters used are given in Table~1, inversion set~1 of \citet{degrave2014}. All averaging kernels can also be found in the online supplement of that paper. The smoothed LRes simulation flows do not show a remarkable moat flow from the spot, yet there is evidence of quiet-Sun flow structures at each layer. At a depth of 1~Mm, we find that both sets of inversions agree well with one another away from the spot and are able to capture much of the overall large-scale flow structure present in the simulation. Here the inversion flow fields appear to be influenced by the spot only within the umbra itself, having little effect on the surrounding areas. At 3~Mm, some of this structure is still visible, but the flows have been washed out by the large-amplitude ones directly in and around the spot. The phase-speed and ridge+phase-speed filter inversions begin to show increased inflow through the spot, with the spot's influence no longer constrained only to within the umbra. By 5~Mm the GB02 inversions show a sign change around the spot, showing an outflow in this region rather than inflow. Such strong inflow and outflow around the spot suggested by the inversions is not observed in the simulation $v_{x,y}^{\rm tgt}$ flow maps at any depth, and likely arises as a consequence of the strong magnetic field in this region coupled with decreased wave sensitivity in the deeper layers.

\begin{figure}
\begin{center}$
\begin{array}{ccc}
\includegraphics[width=0.2\linewidth,clip=]{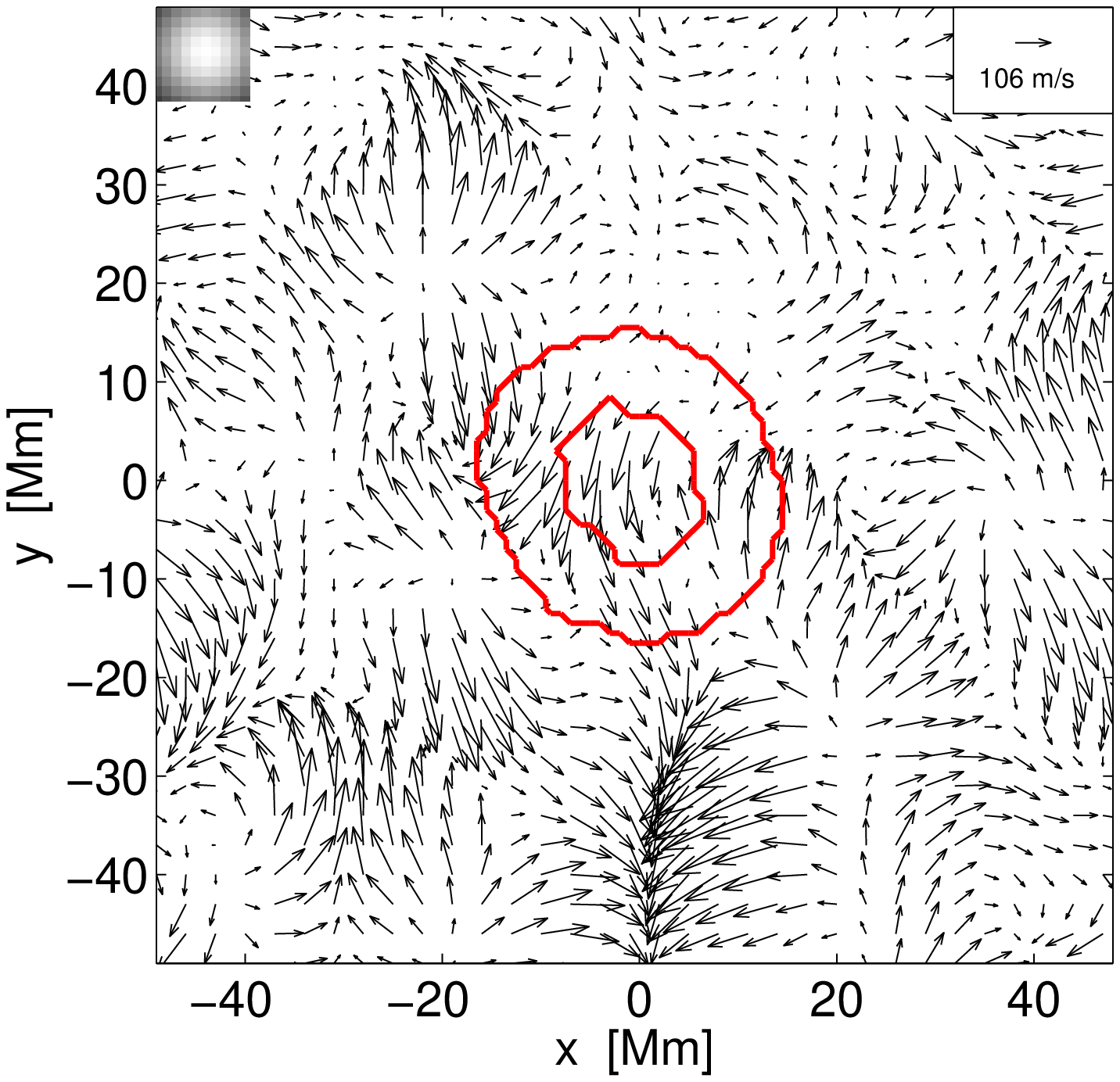} &
\includegraphics[width=0.2\linewidth,clip=]{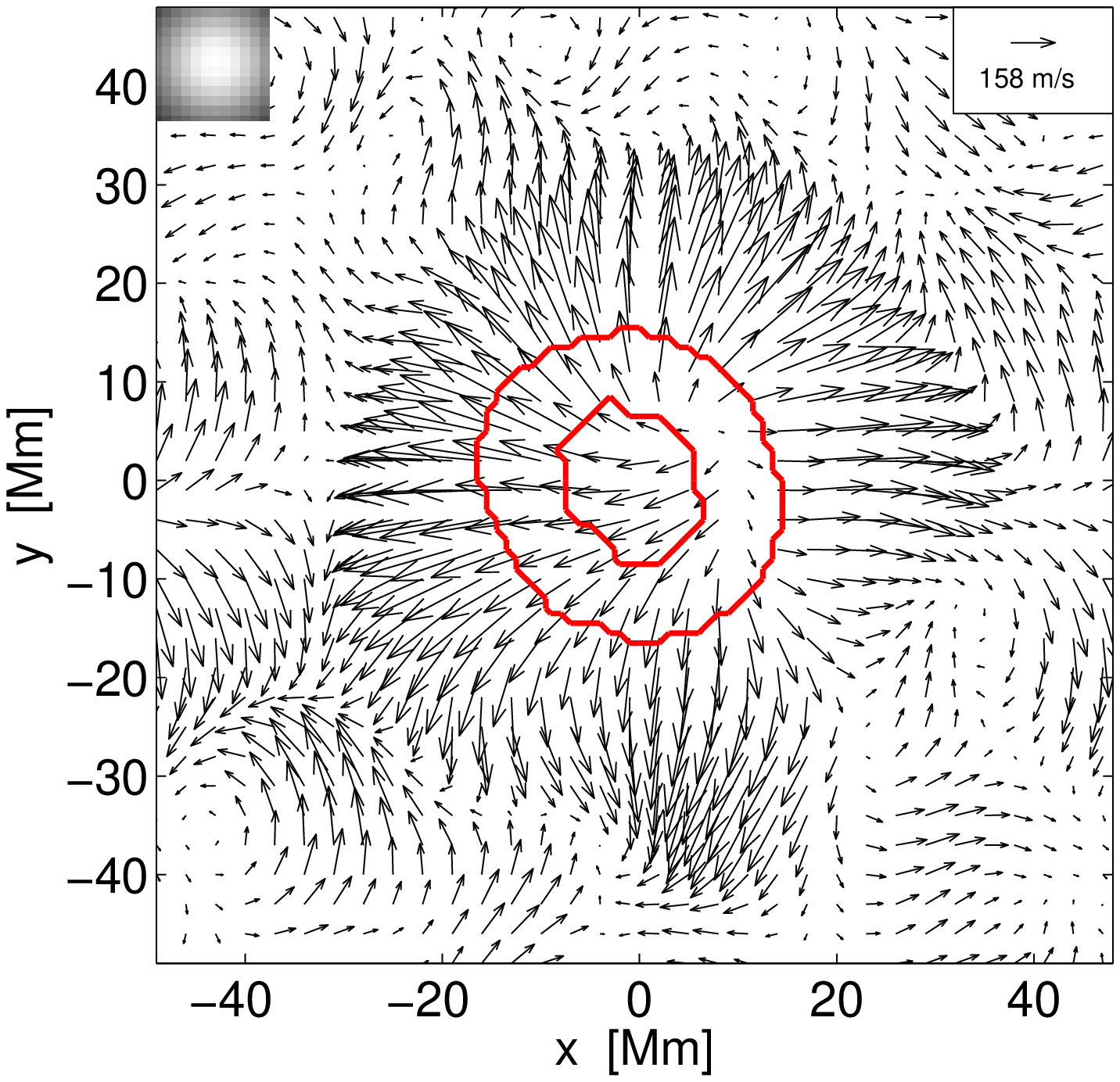} &
\includegraphics[width=0.2\linewidth,clip=]{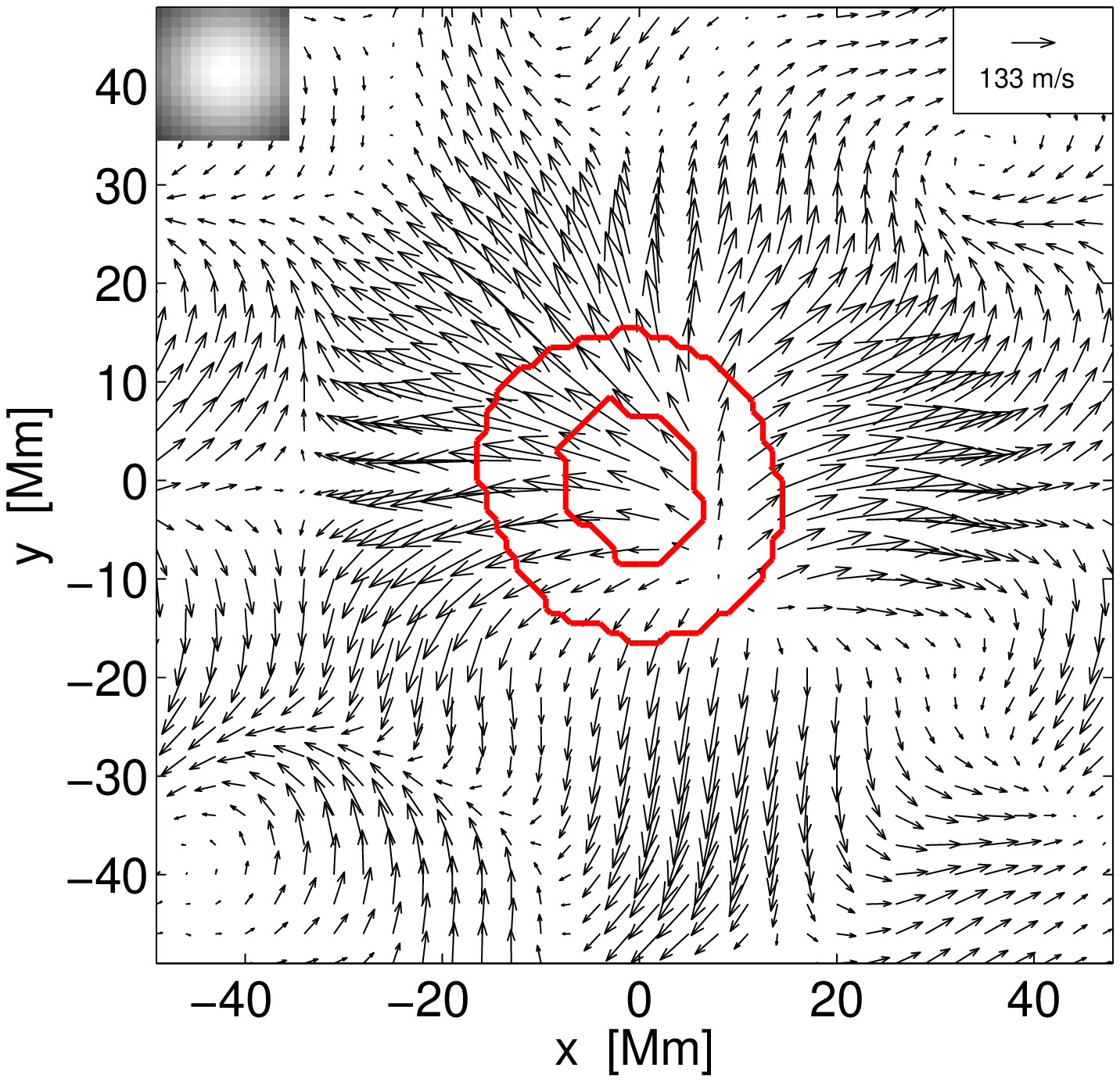} \\
\includegraphics[width=0.2\linewidth,clip=]{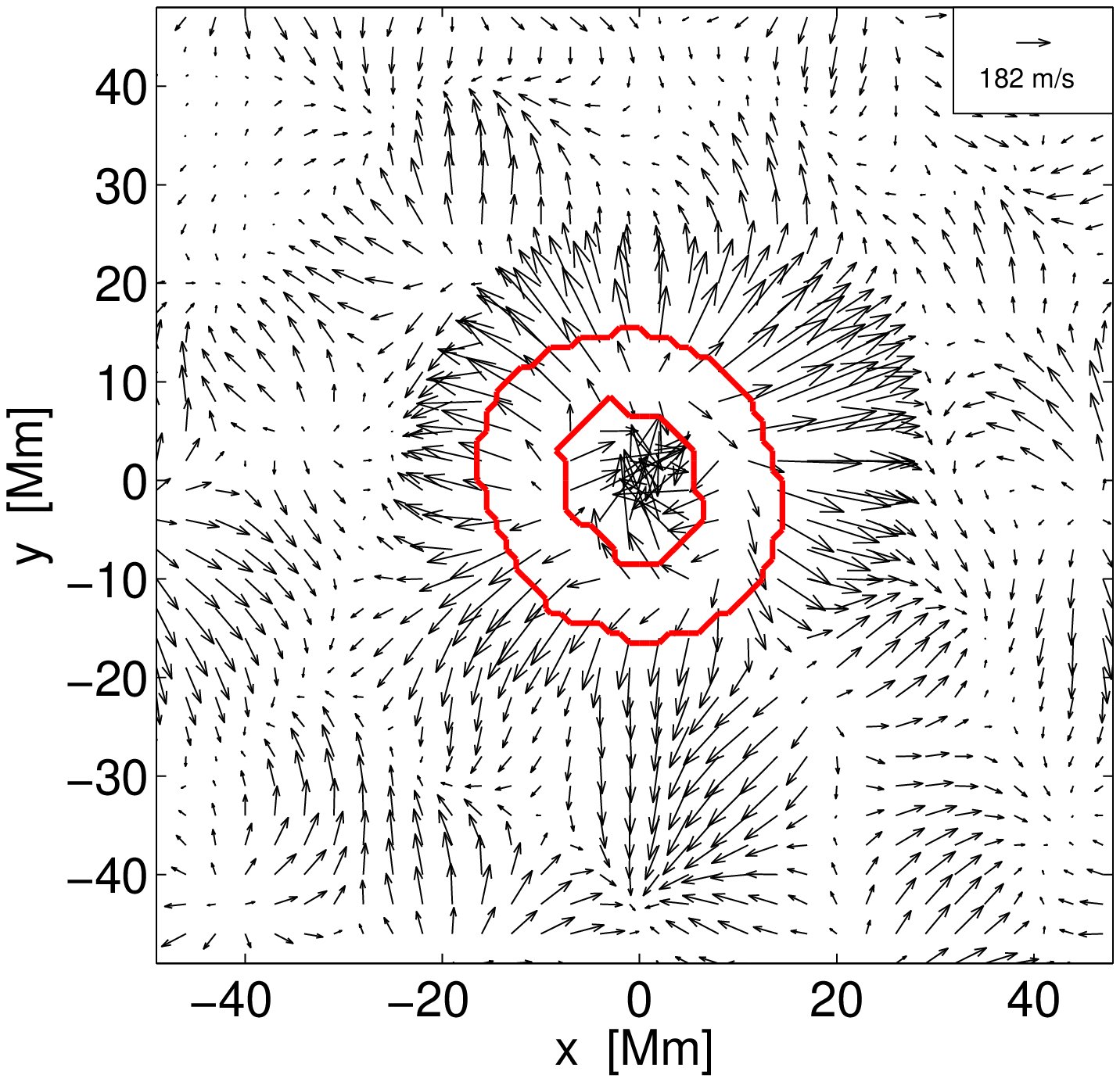} &
\includegraphics[width=0.2\linewidth,clip=]{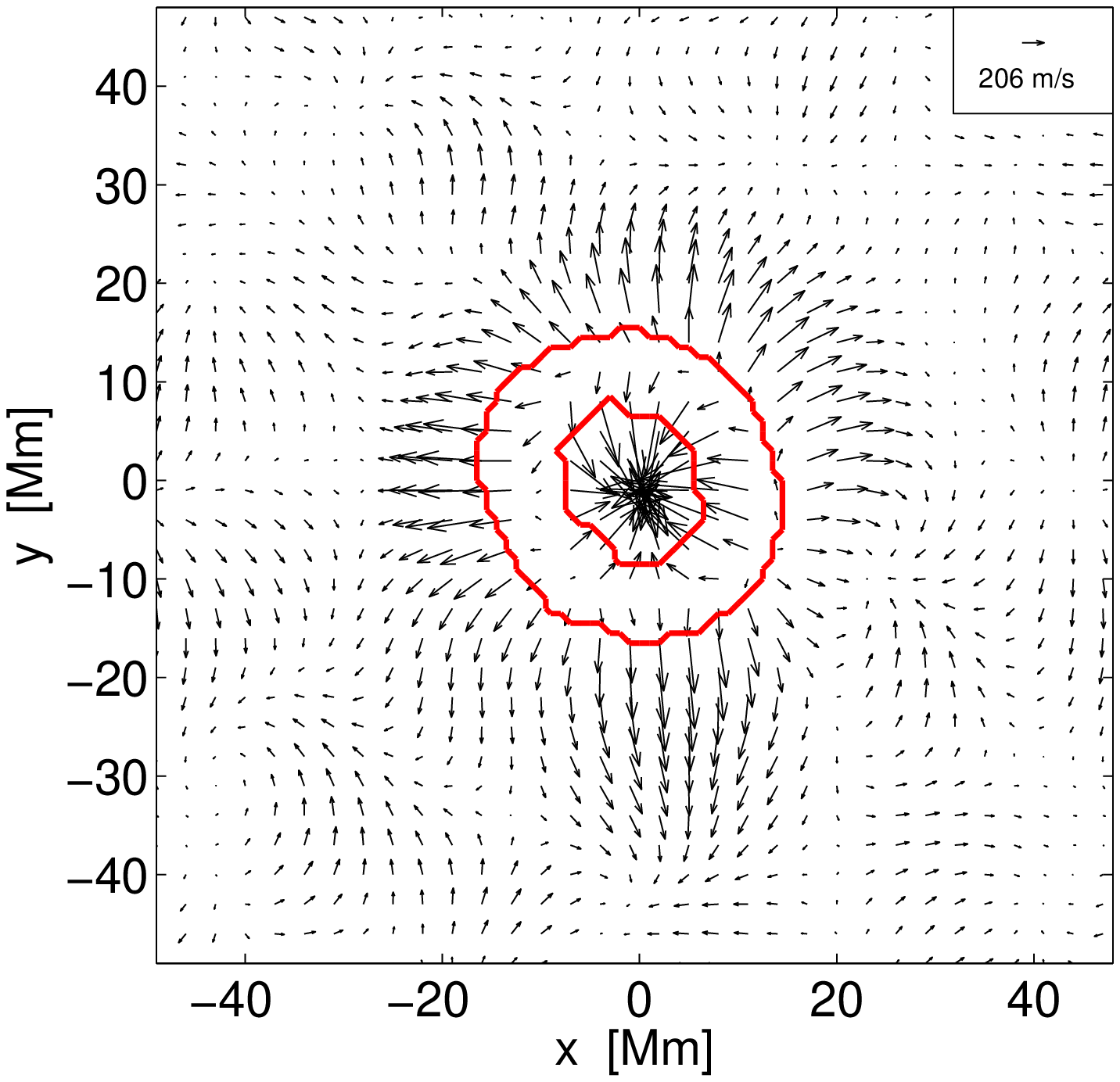} &
\includegraphics[width=0.2\linewidth,clip=]{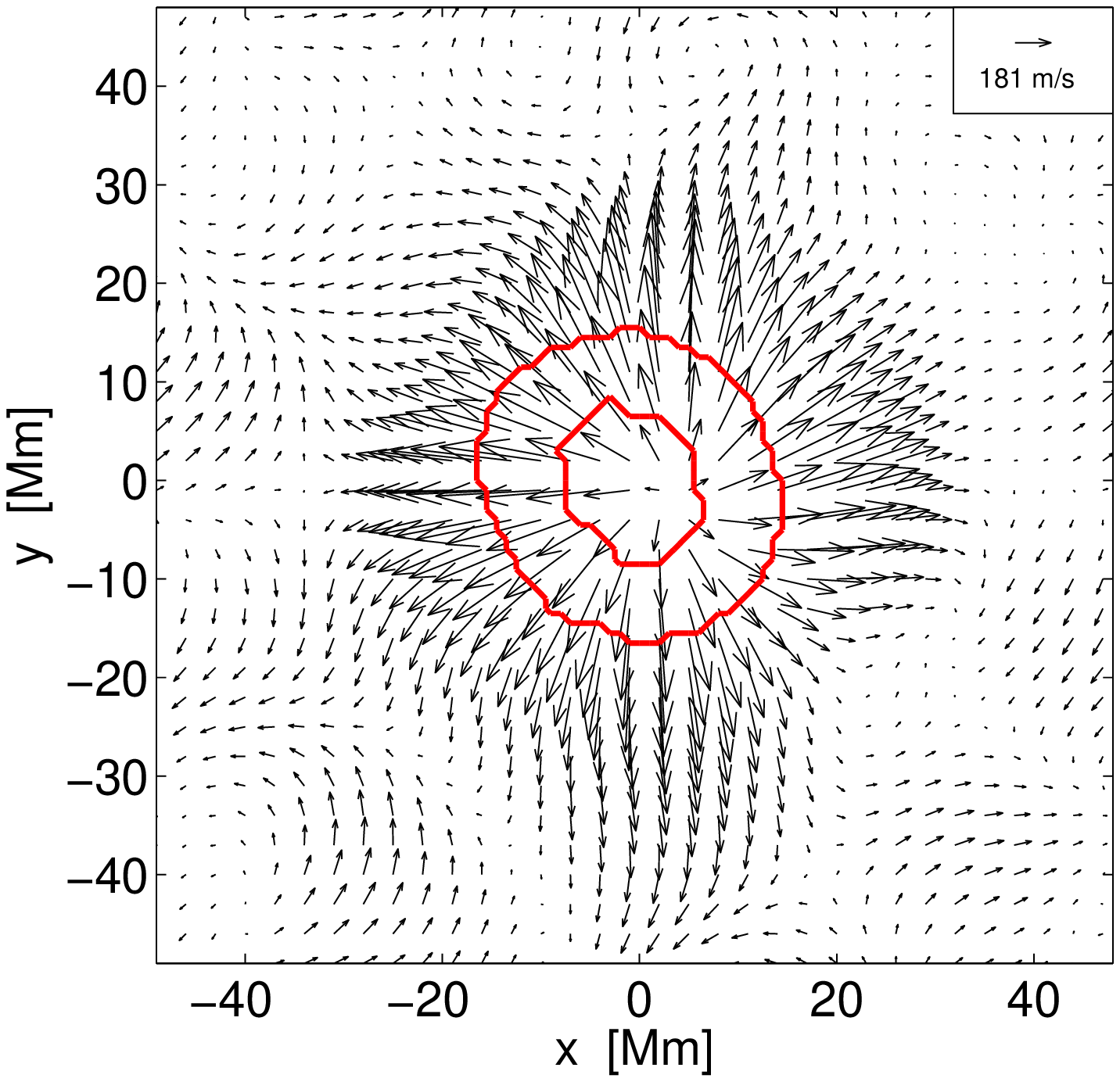} \\
\includegraphics[width=0.2\linewidth,clip=]{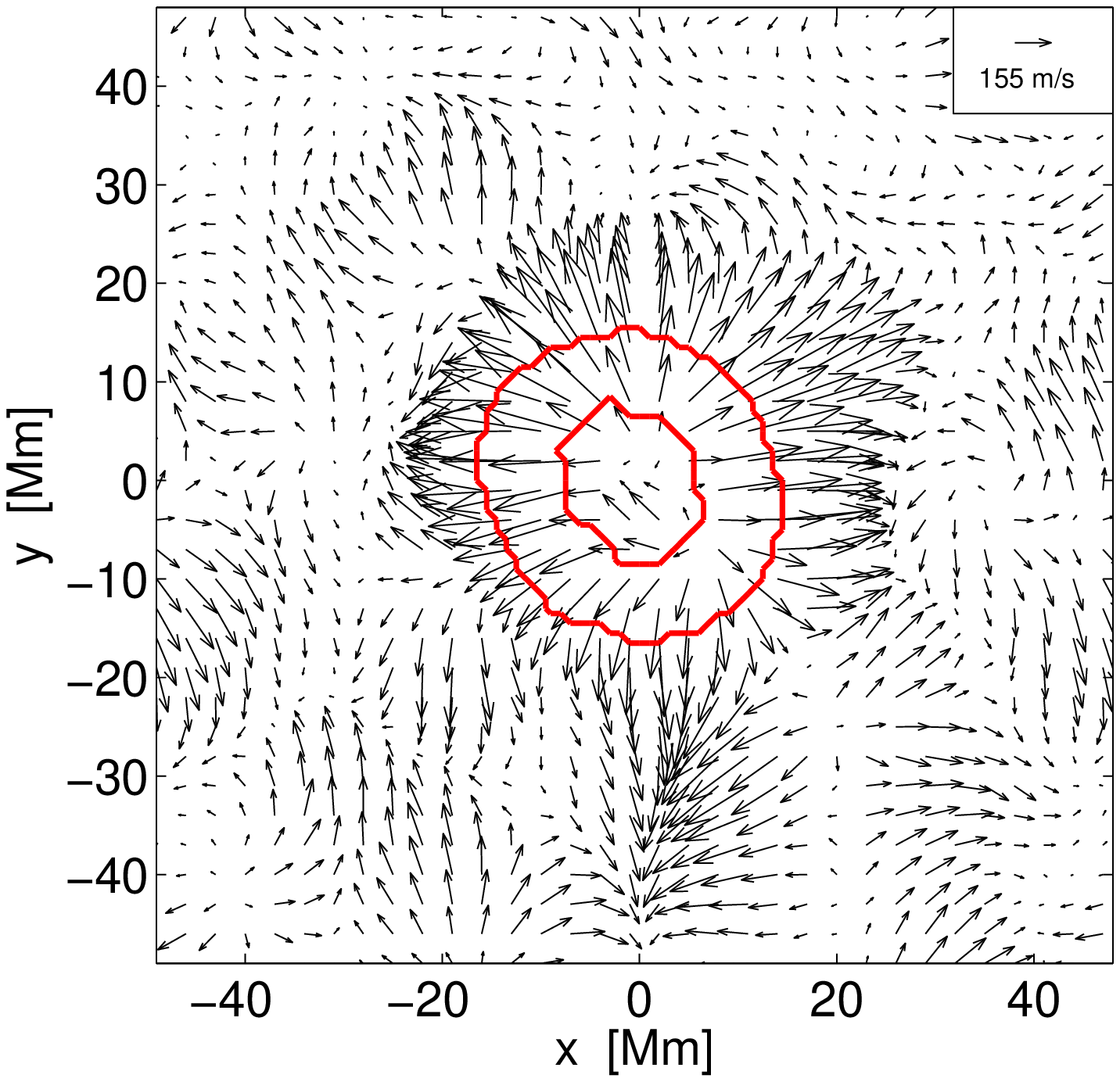} &
\includegraphics[width=0.2\linewidth,clip=]{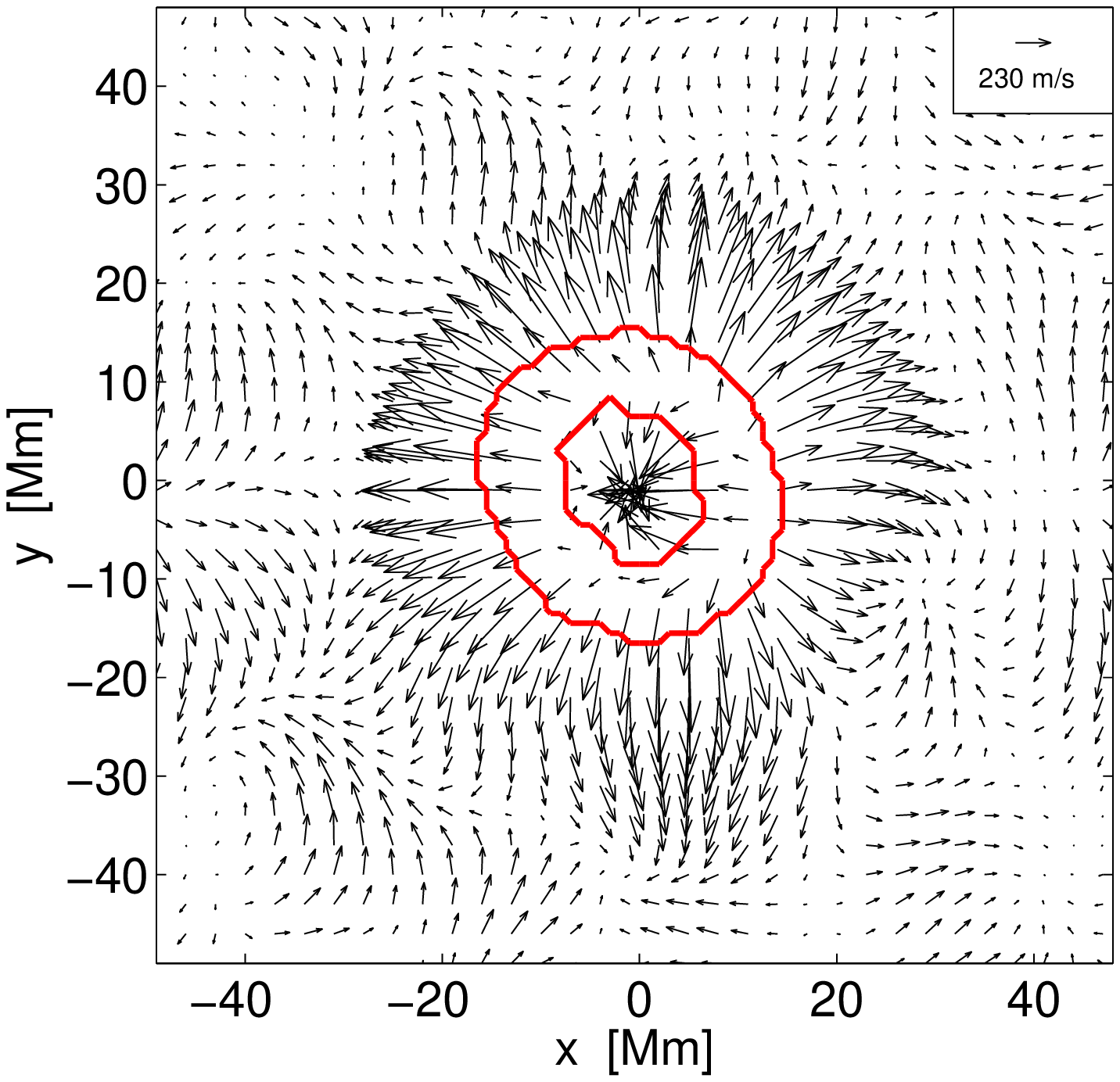} &
\includegraphics[width=0.2\linewidth,clip=]{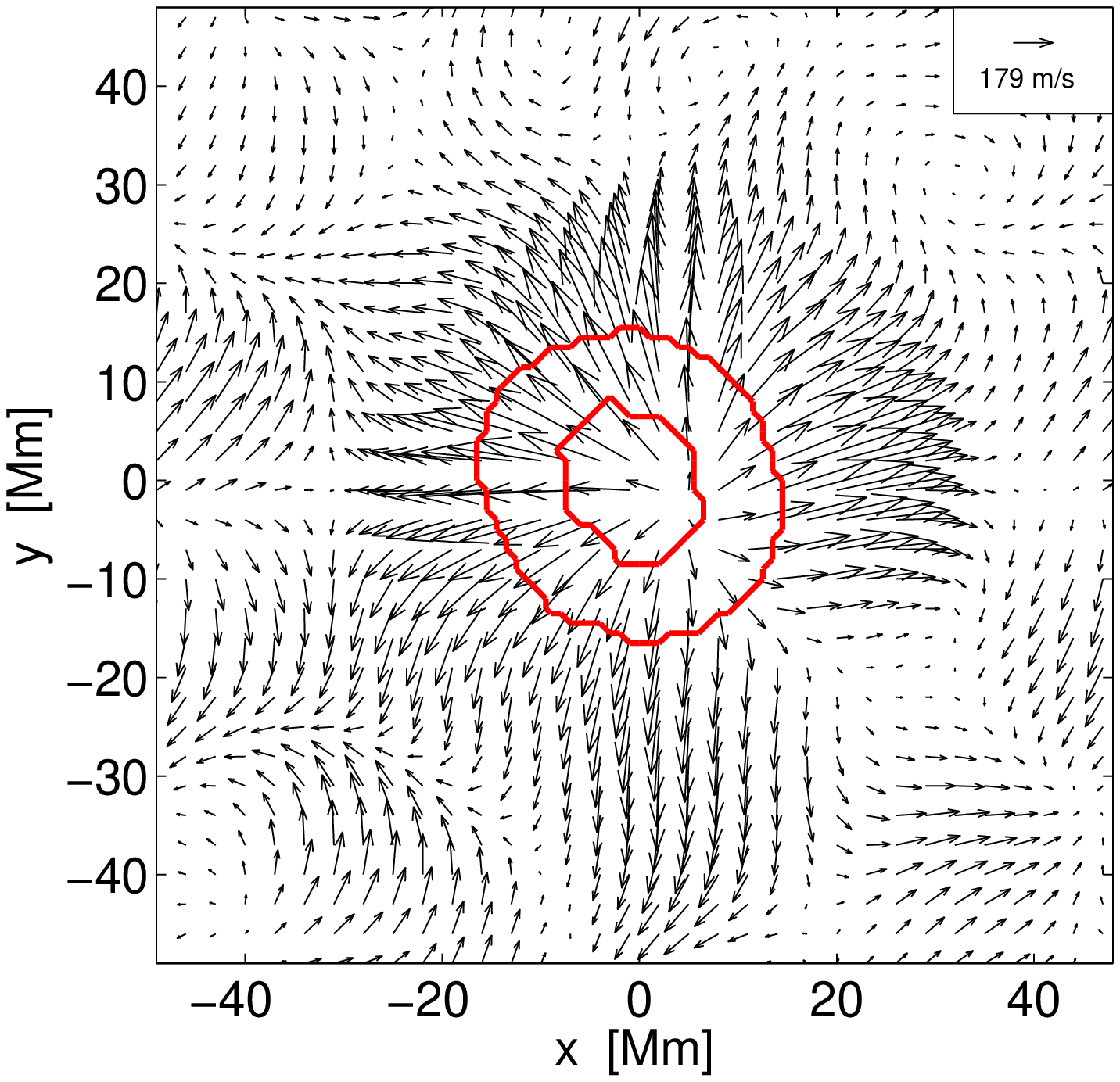} \\
\includegraphics[width=0.2\linewidth,clip=]{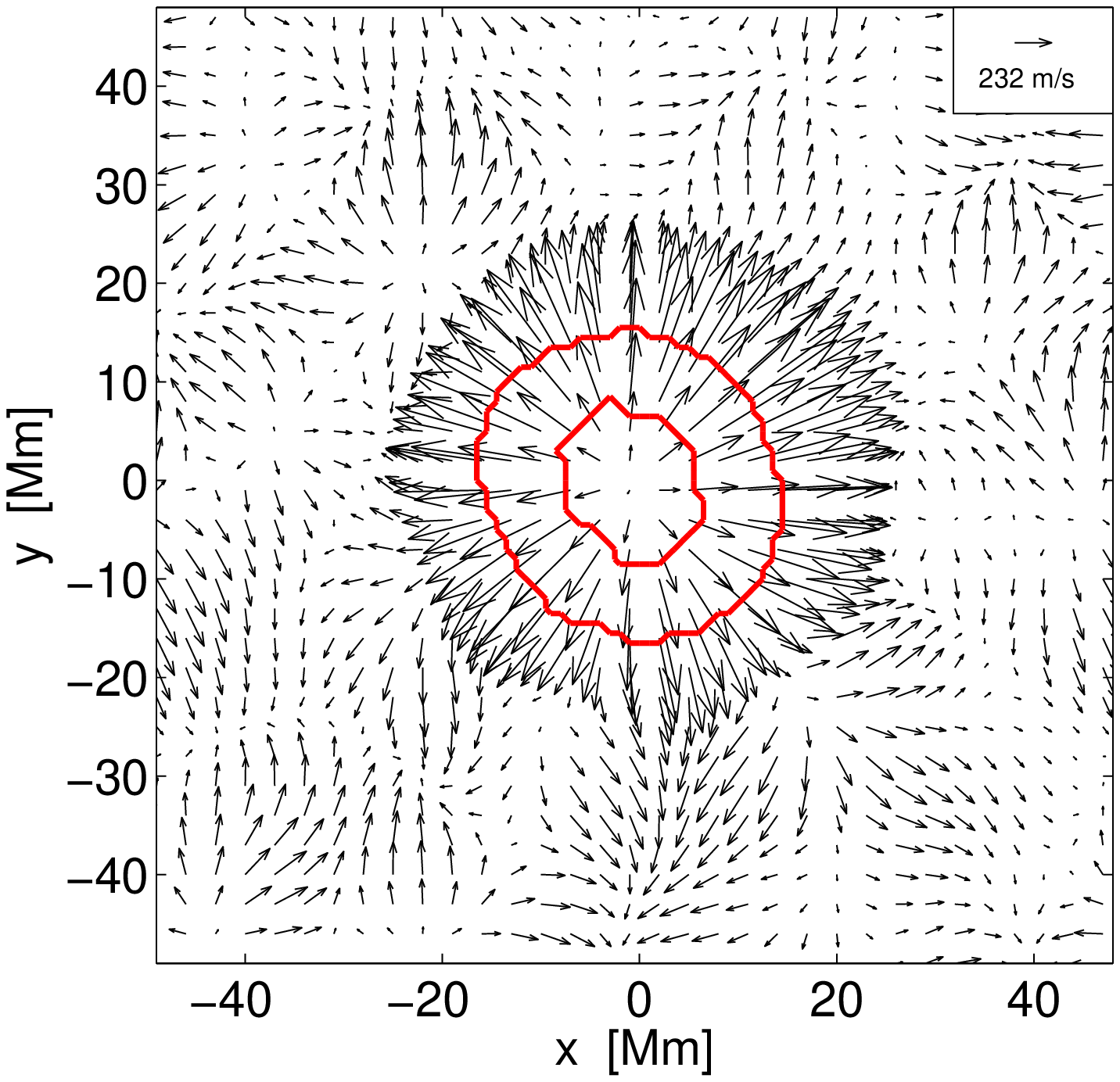} &
\includegraphics[width=0.2\linewidth,clip=]{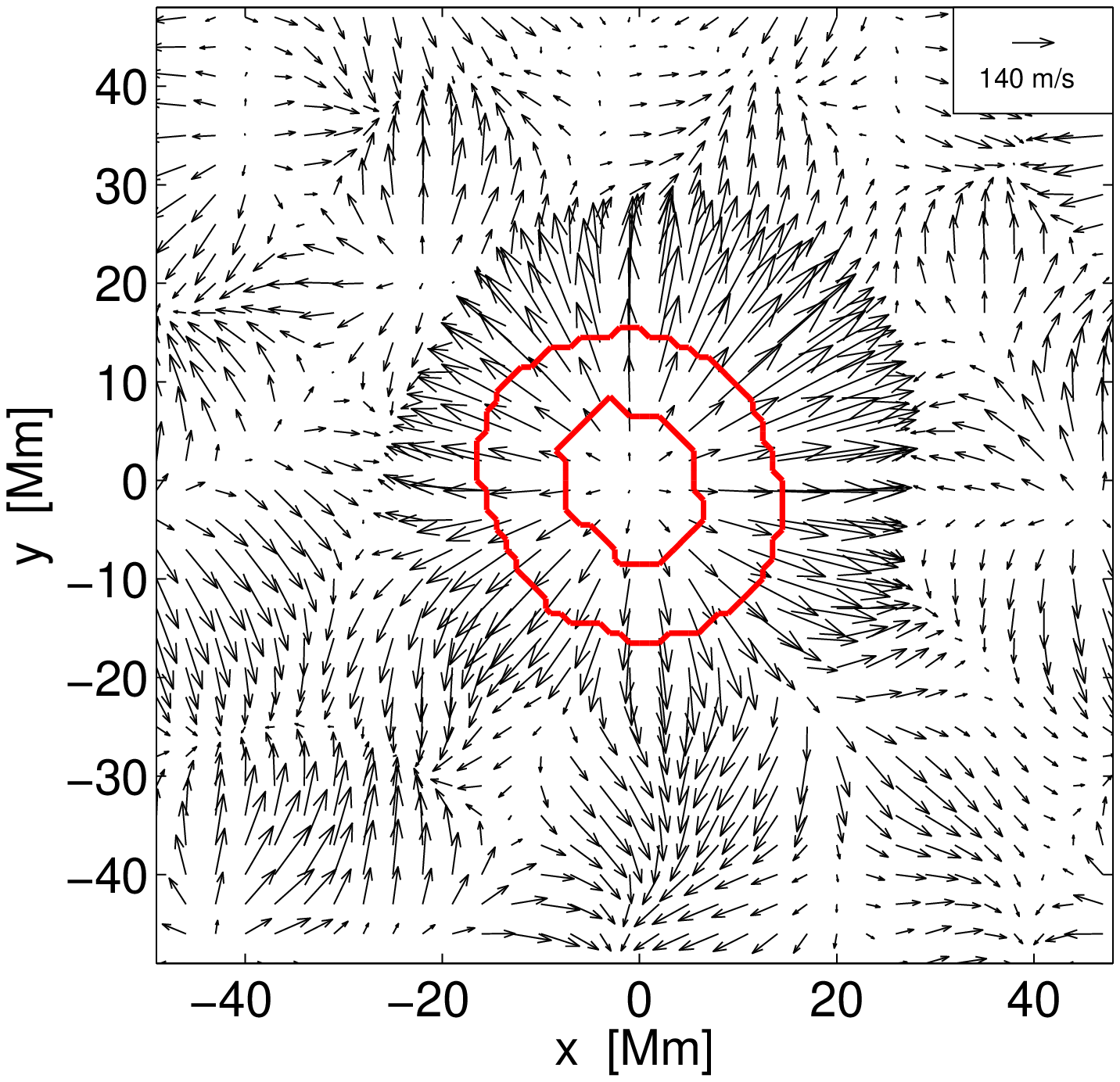} &
\includegraphics[width=0.2\linewidth,clip=]{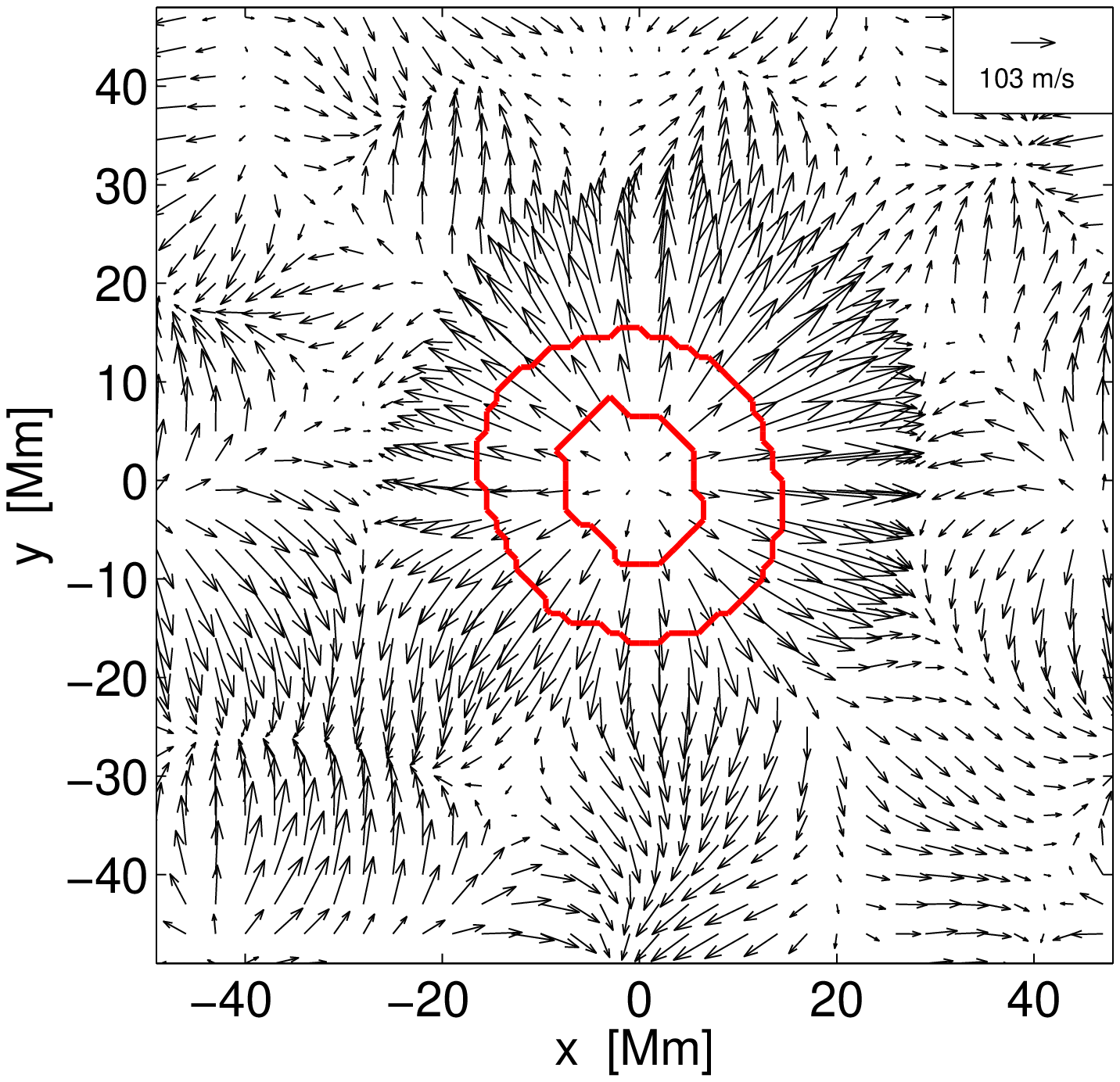}
\end{array}$
\end{center}
\caption{HRes GB02 horizontal $(v_x, v_y)$ inversion flow maps for the ridge (first row), phase-speed (second row), and ridge+phase-speed (third row) travel-time differences for depths (left to right) 1, 3 and 5~Mm. The smoothed simulation flow maps (i.e. $v_{x,y}^{\rm tgt}$) at these depths are shown in the bottom row. The noise for each inversion is $\sim35~\rm{ms^{-1}}$ and the reference arrows represent the RMS velocity corresponding to each flow map. The 2D target function at each depth is shown in the upper lefthand corner of the first row figures. The width of the box corresponds to the horizontal FWHM of each target function and represents the approximate spatial resolution of each flow map. All maps in the same column have identical horizontal resolution. The contours marking the boundaries of the spot umbra and penumbra have been overplotted.}
\label{fig:vxy02SP}
\end{figure}

\begin{figure}
\begin{center}$
\begin{array}{ccc}
\includegraphics[width=0.2\linewidth,clip=]{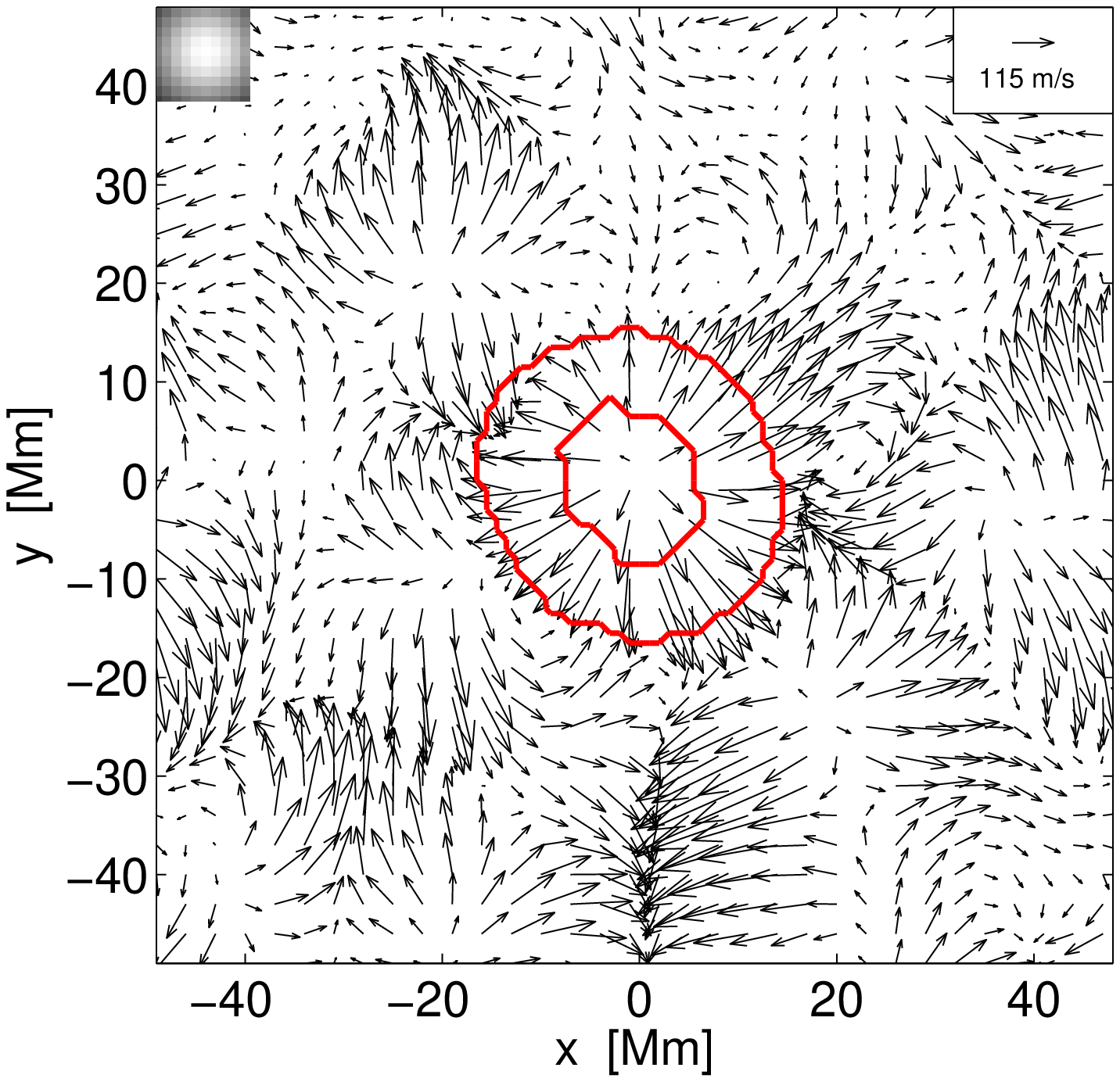} &
\includegraphics[width=0.2\linewidth,clip=]{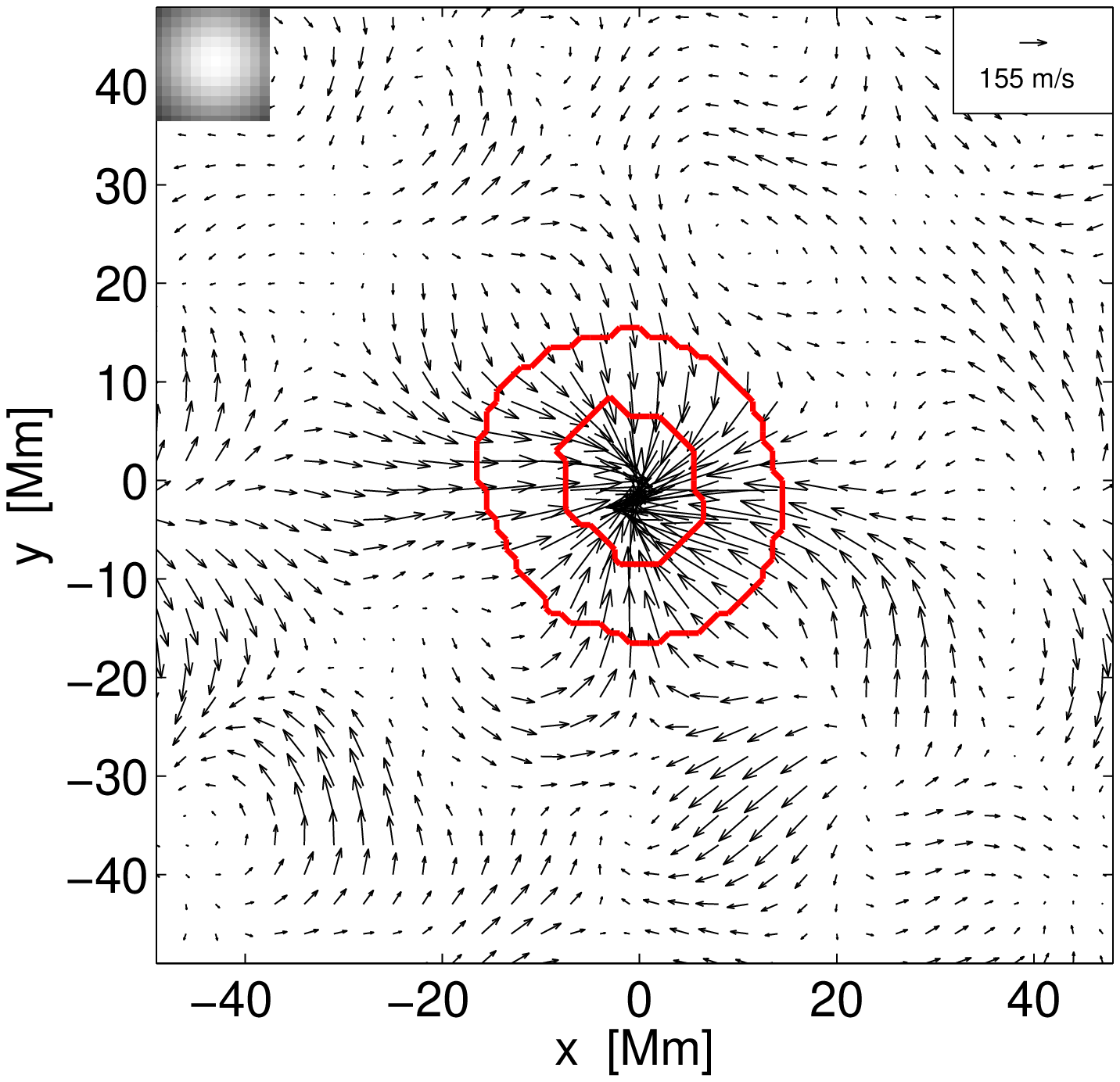} &
\includegraphics[width=0.2\linewidth,clip=]{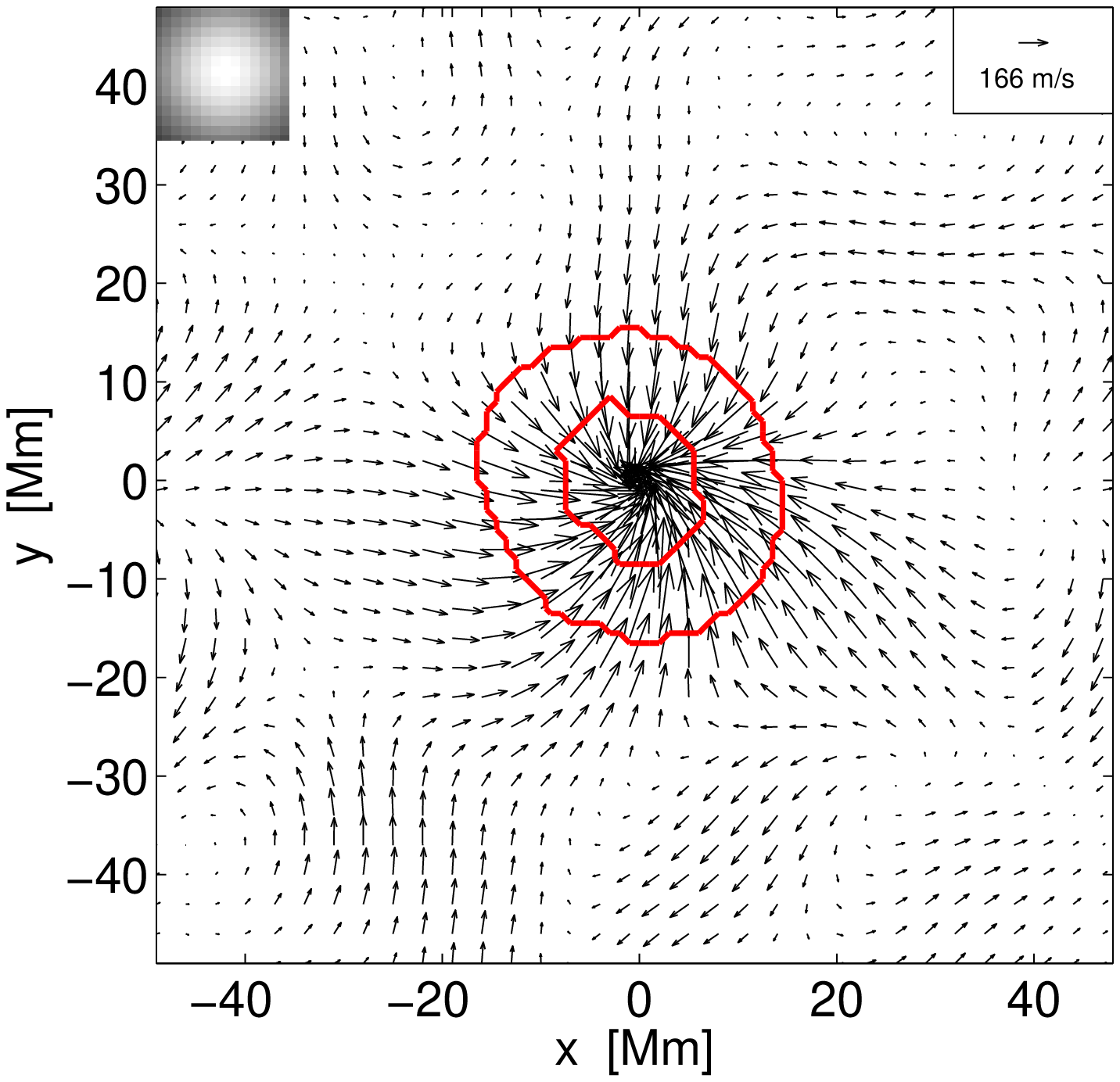} \\
\includegraphics[width=0.2\linewidth,clip=]{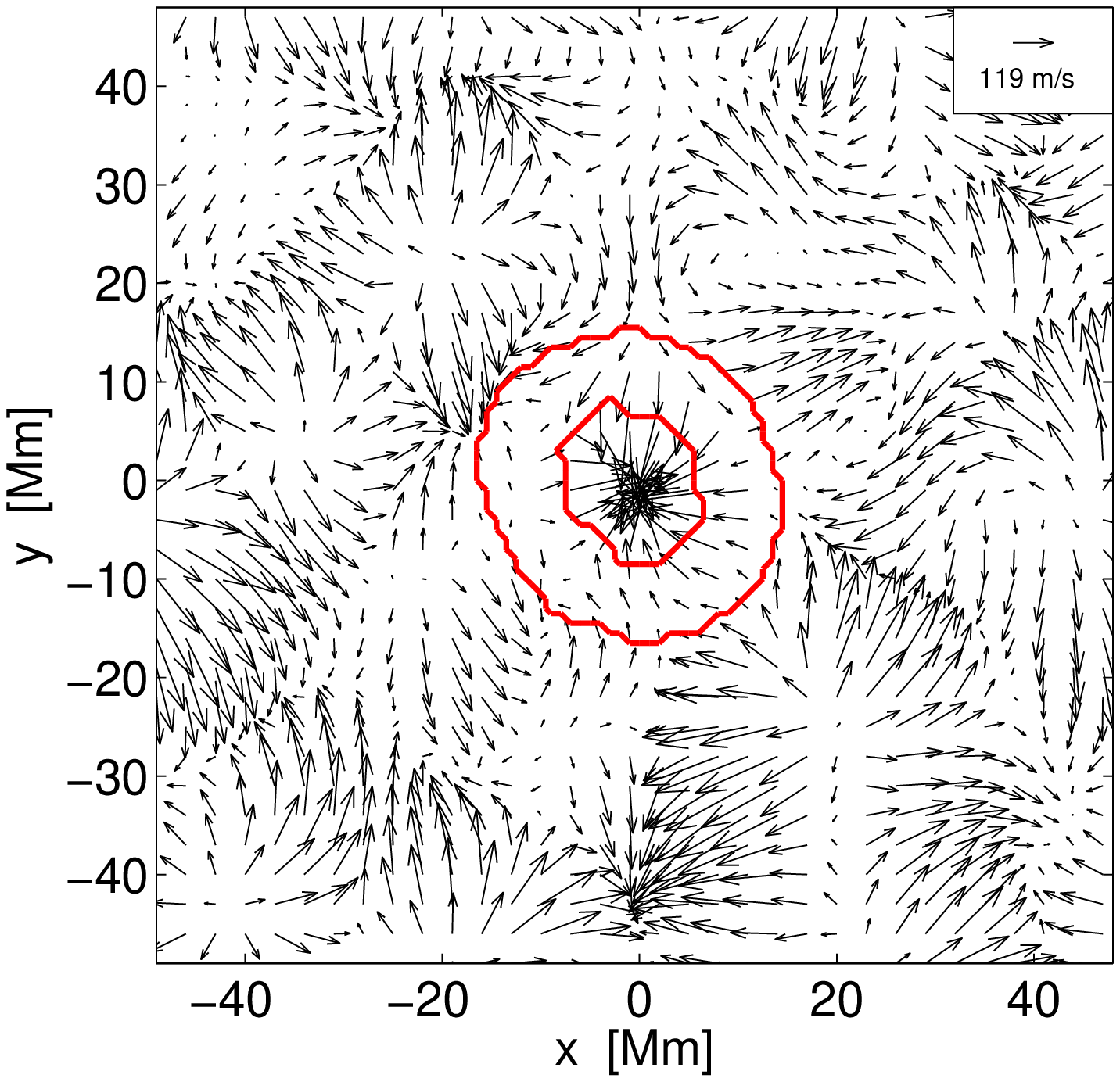} &
\includegraphics[width=0.2\linewidth,clip=]{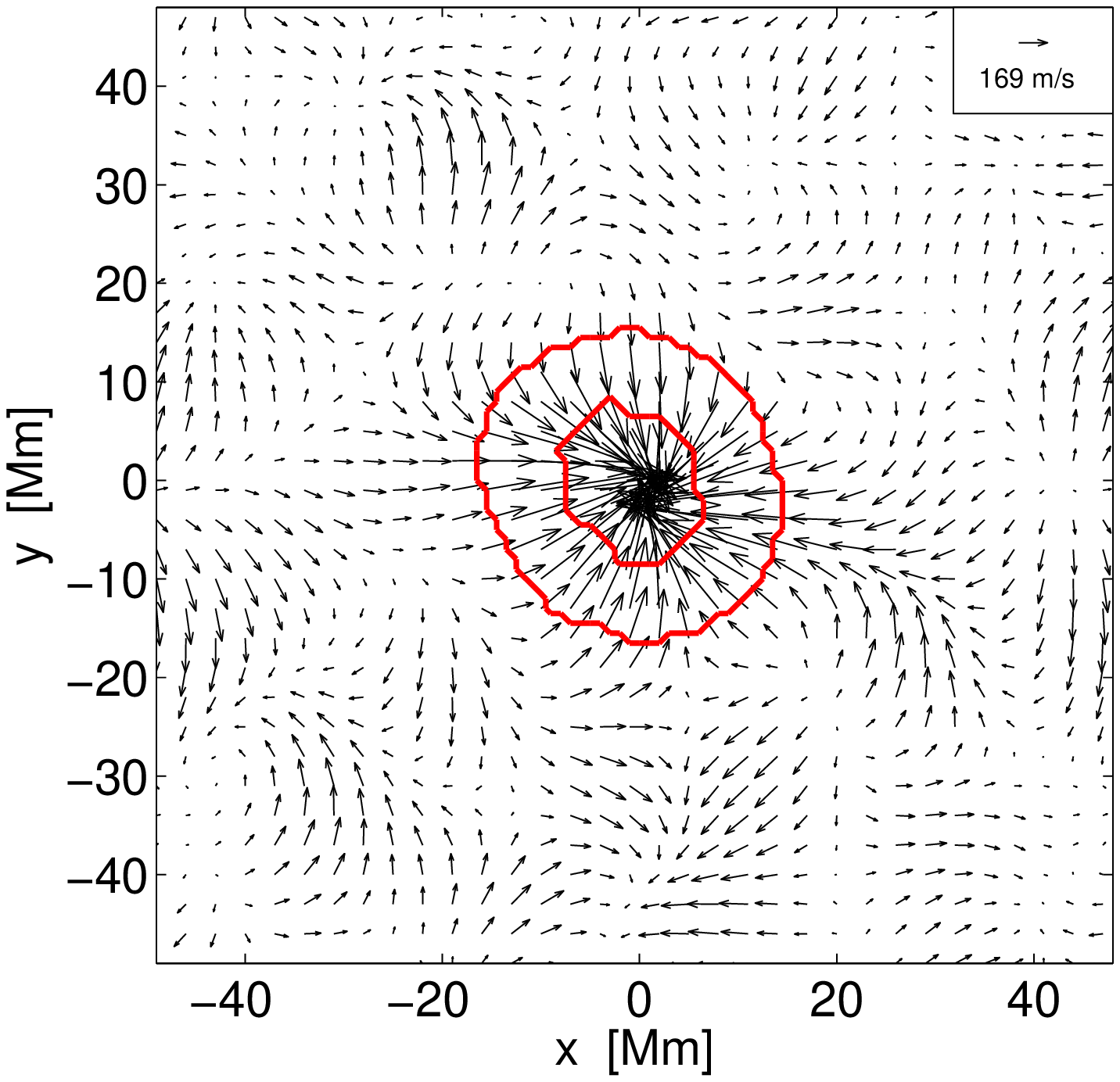} &
\includegraphics[width=0.2\linewidth,clip=]{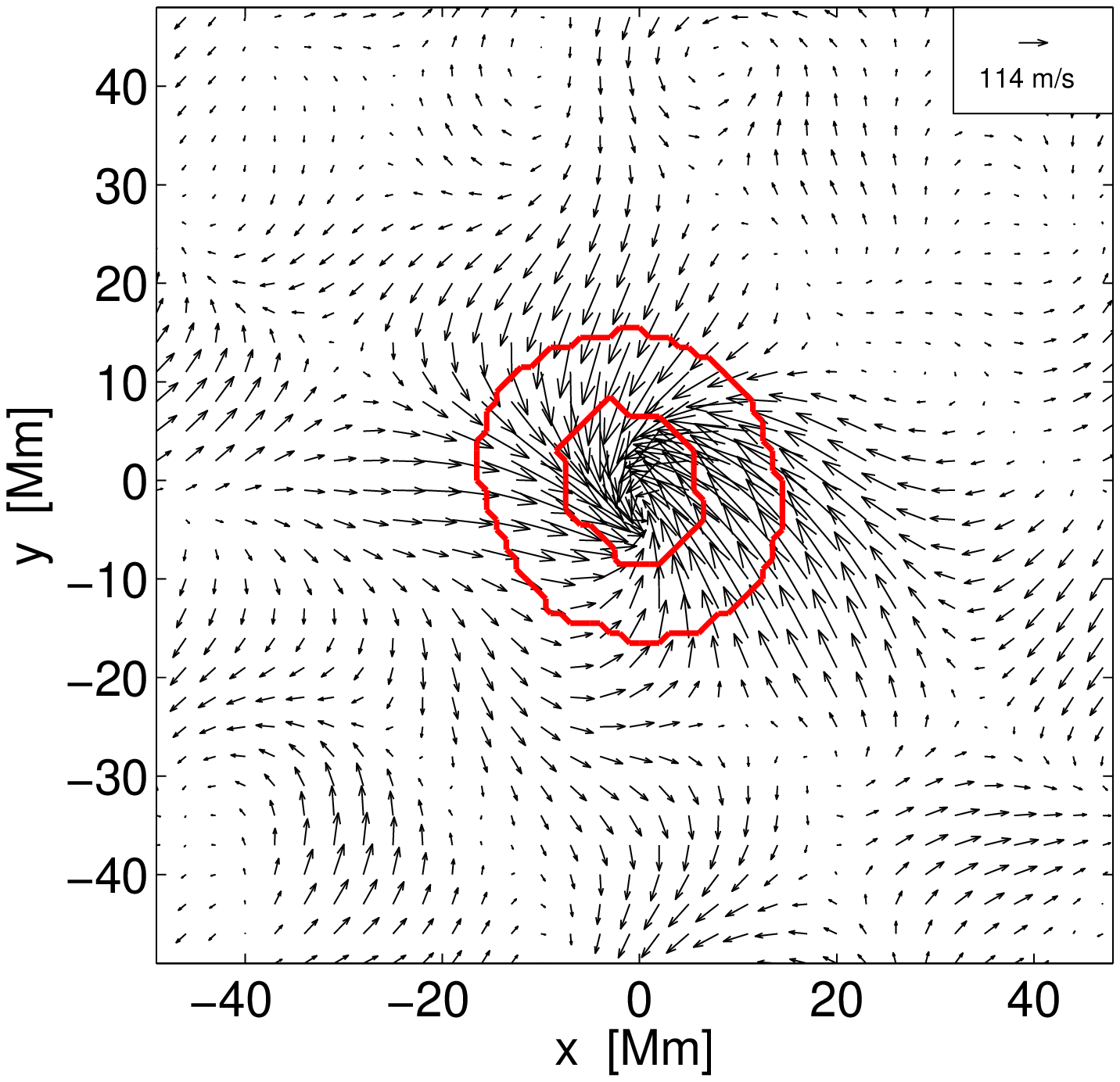} \\
\includegraphics[width=0.2\linewidth,clip=]{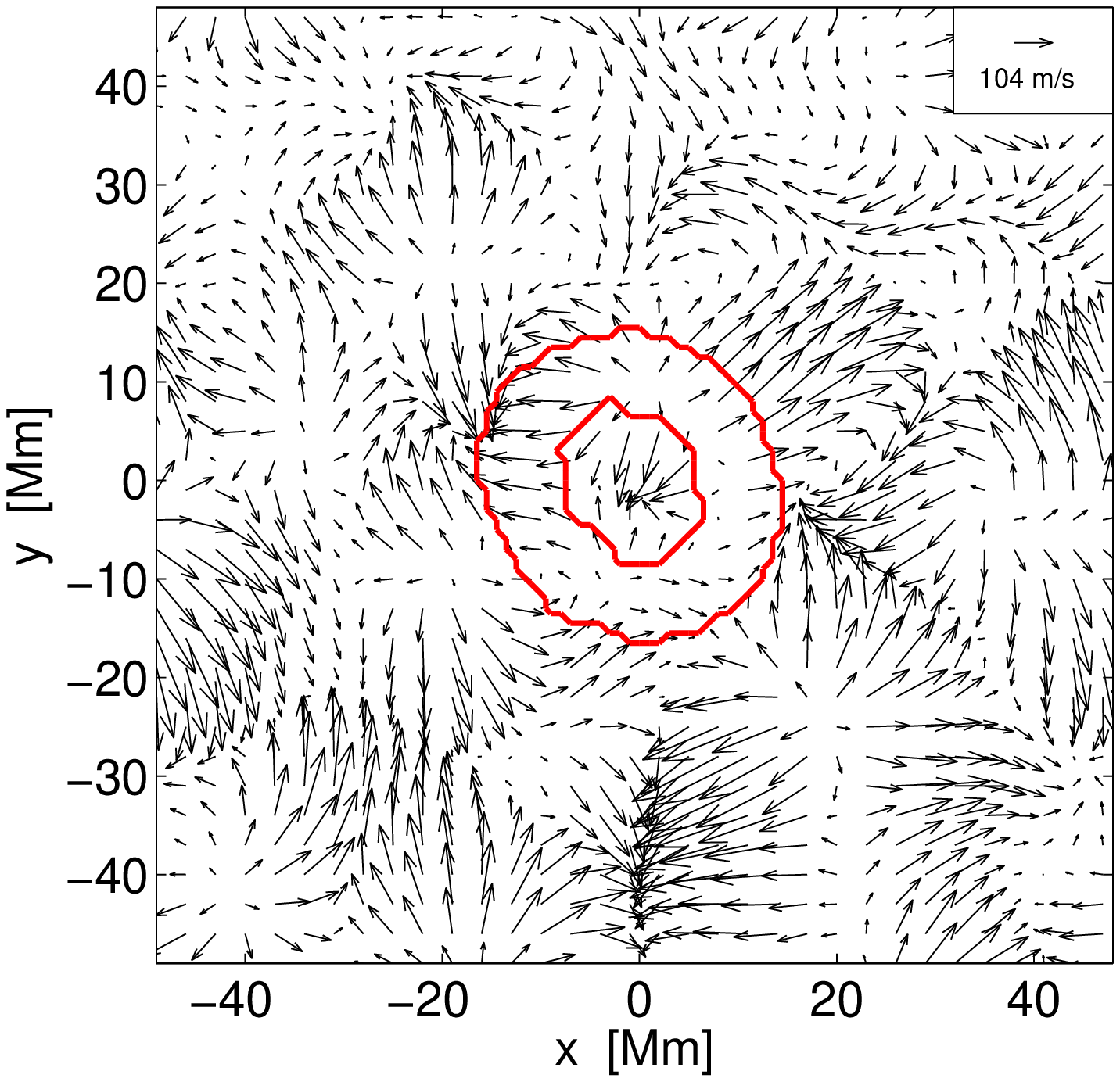} &
\includegraphics[width=0.2\linewidth,clip=]{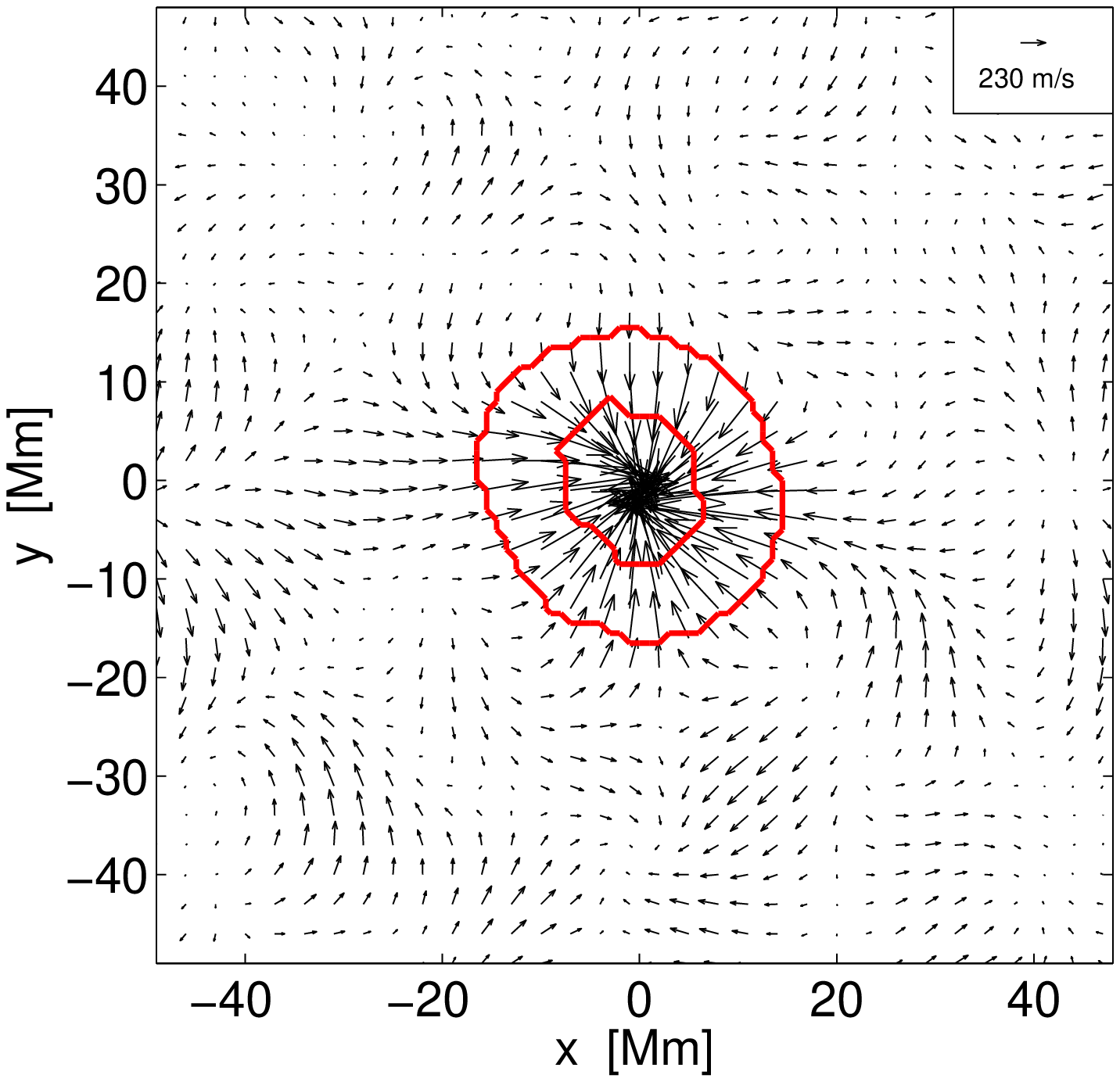} &
\includegraphics[width=0.2\linewidth,clip=]{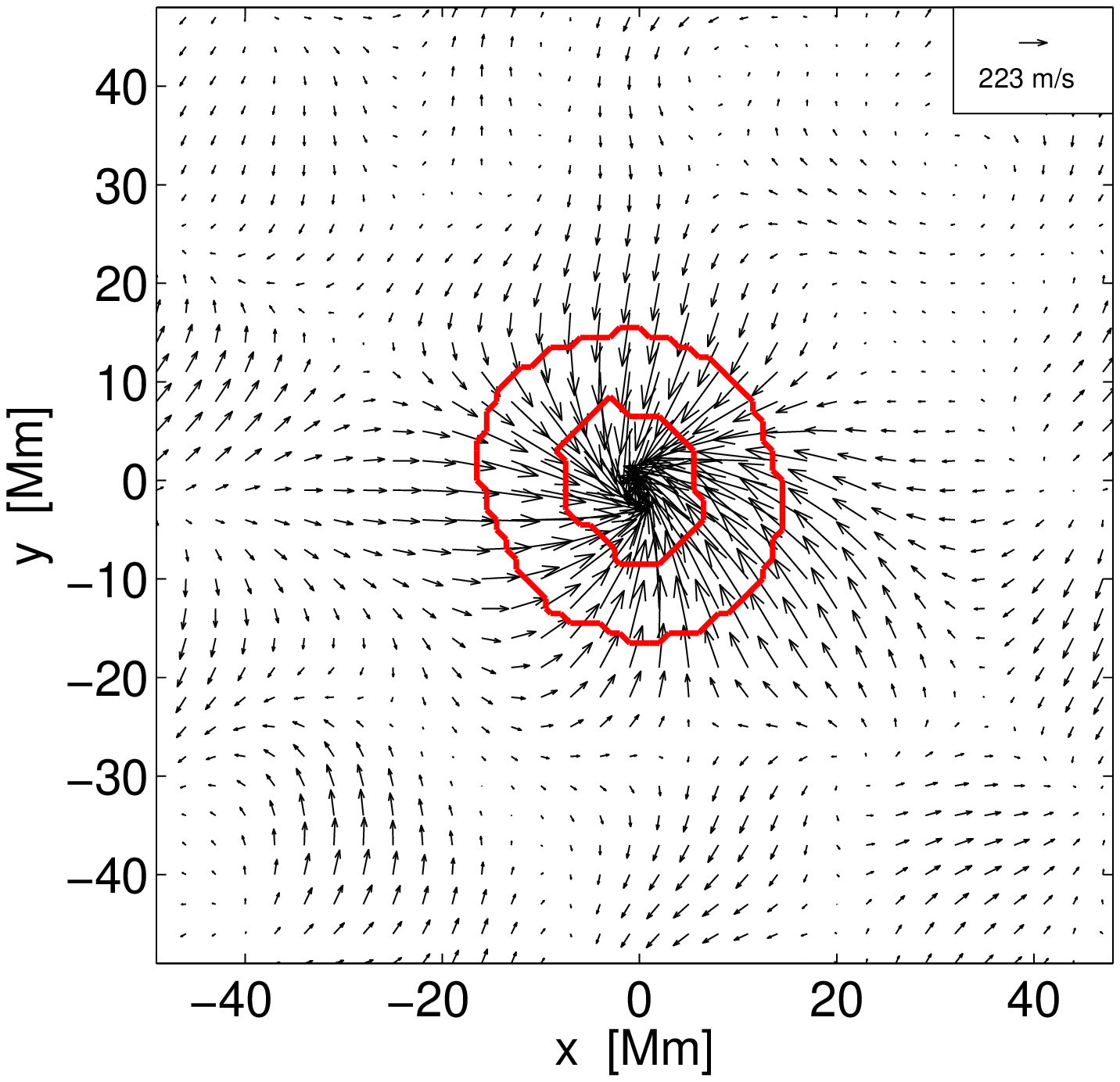} \\
\includegraphics[width=0.2\linewidth,clip=]{f10j.eps} &
\includegraphics[width=0.2\linewidth,clip=]{f10k.eps} &
\includegraphics[width=0.2\linewidth,clip=]{f10l.eps}
\end{array}$
\end{center}
\caption{HRes GB04 horizontal $(v_x, v_y)$ inversion flow maps for the ridge (first row), phase-speed (second row), and ridge+phase-speed (third row) travel-time differences for depths (left to right) 1, 3 and 5~Mm. The smoothed simulation flow maps (i.e. $v_{x,y}^{\rm tgt}$) at these depths are shown in the bottom row. The noise for each inversion is $\sim35~\rm{ms^{-1}}$ and the reference arrows represent the RMS velocity corresponding to each flow map. The 2D target function at each depth is shown in the upper lefthand corner of the first row figures. The width of the box corresponds to the horizontal FWHM of each target function and represents the approximate spatial resolution of each flow map. All maps in the same column have identical horizontal resolution. The contours marking the boundaries of the spot umbra and penumbra have been overplotted.}
\label{fig:vxy04SP}
\end{figure}

Figures~\ref{fig:vxy02SP} and \ref{fig:vxy04SP} show the equivalent HRes $v_{x,y}^{\rm inv}$ horizontal flow maps at each depth using the GB02 and GB04 travel-time definitions respectively. Again, each row represents the separate ridge and phase-speed filter inversions (rows 1 and 2 respectively), and the combined ridge+phase-speed filter inversions (row 3). The noise for each is $\sim35~\rm{m\,s^{-1}}$, and the inversion parameters are identical to the LRes ones presented above. The boundaries of the spot umbra and penumbra are marked by the two contour lines. Though it is difficult to tell by eye, at a depth of 1~Mm we generally find good agreement between GB02 and GB04 inversion flow maps away from the spot, which correlate reasonably well with the simulation $v_{x,y}^{\rm tgt}$ flow maps. The inversions are also able to recover some of the surrounding structure at depths of 3 and 5~Mm. As seen for the case of LRes, HRes GB04 maps tend to show increased inflow within the spot in these deeper layers, consistent with the issues regarding the travel-time measurements discussed in Sec.~\ref{sec:ttcomp}. At these depths, however, GB02 generally exhibits inflow only within the umbra, with outflow occurring through parts of the penumbra and surrounding area. We see that the GB02 phase-speed and ridge+phase-speed inversions are able to capture the divergent moat flow of the simulation $v_{x,y}^{\rm tgt}$ maps to some degree at every depth.

\begin{figure}
\begin{center}$
\begin{array}{ccc}
\includegraphics[width=0.3\linewidth,clip=]{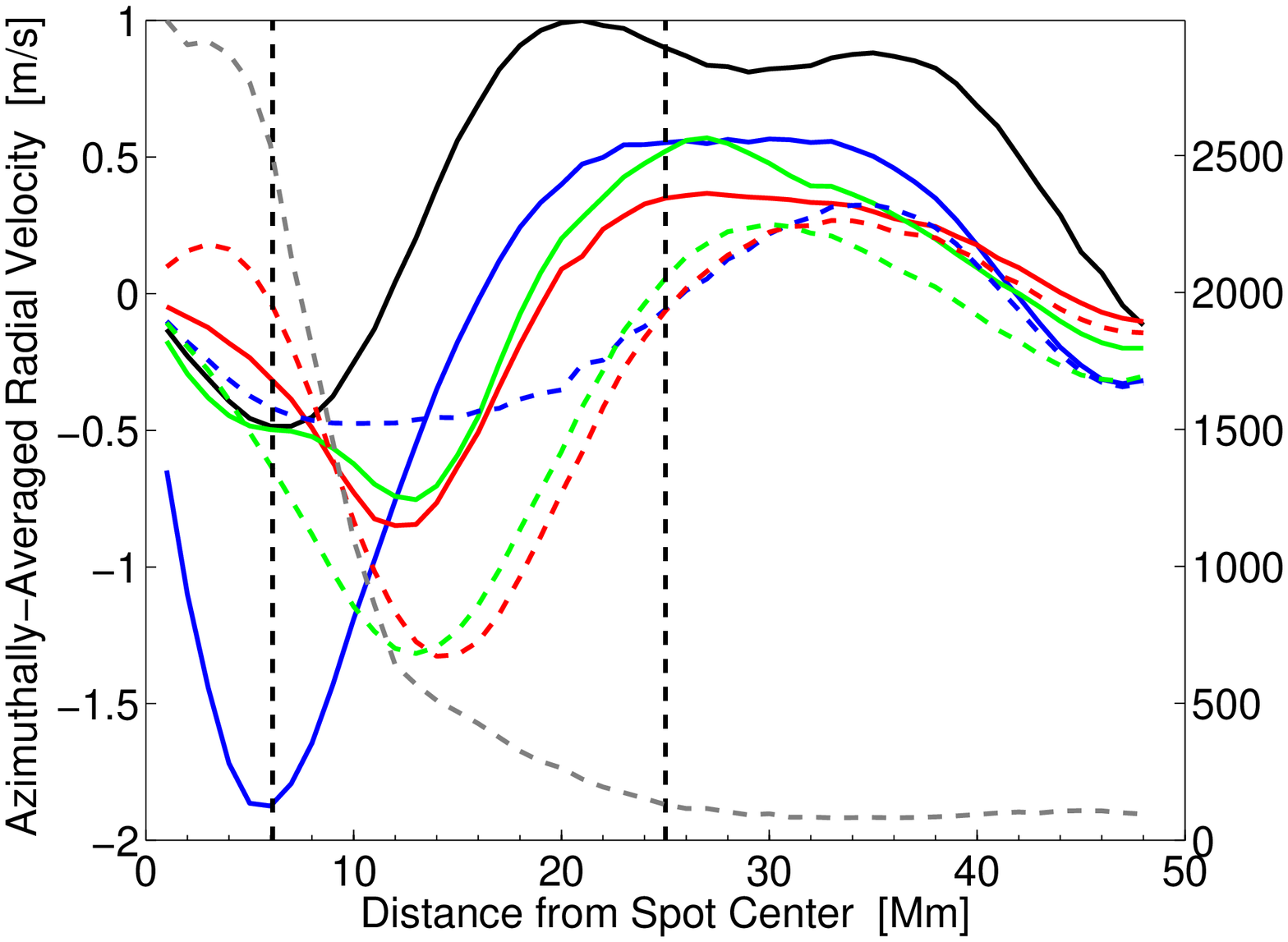} &
\includegraphics[width=0.3\linewidth,clip=]{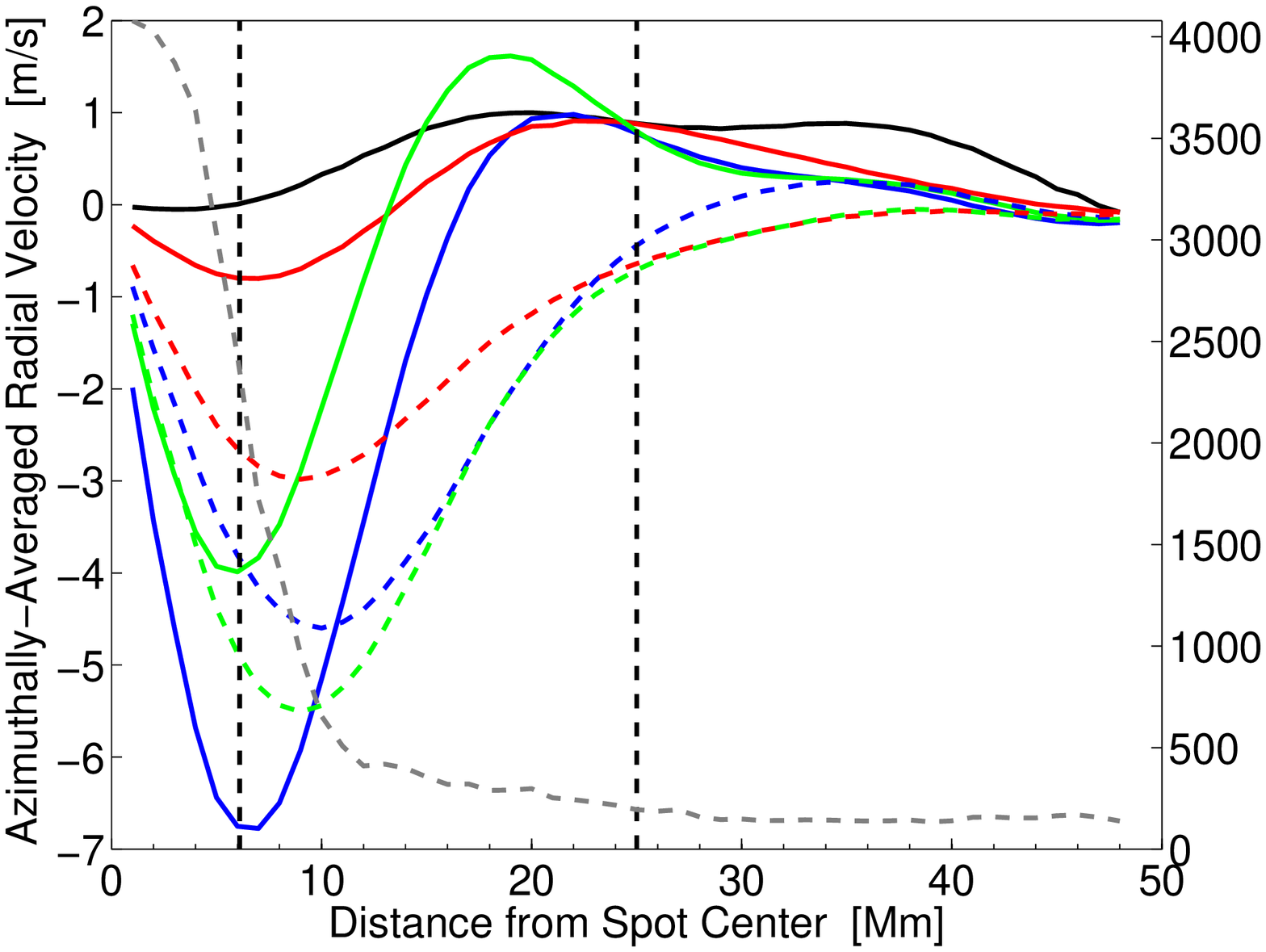} &
\includegraphics[width=0.3\linewidth,clip=]{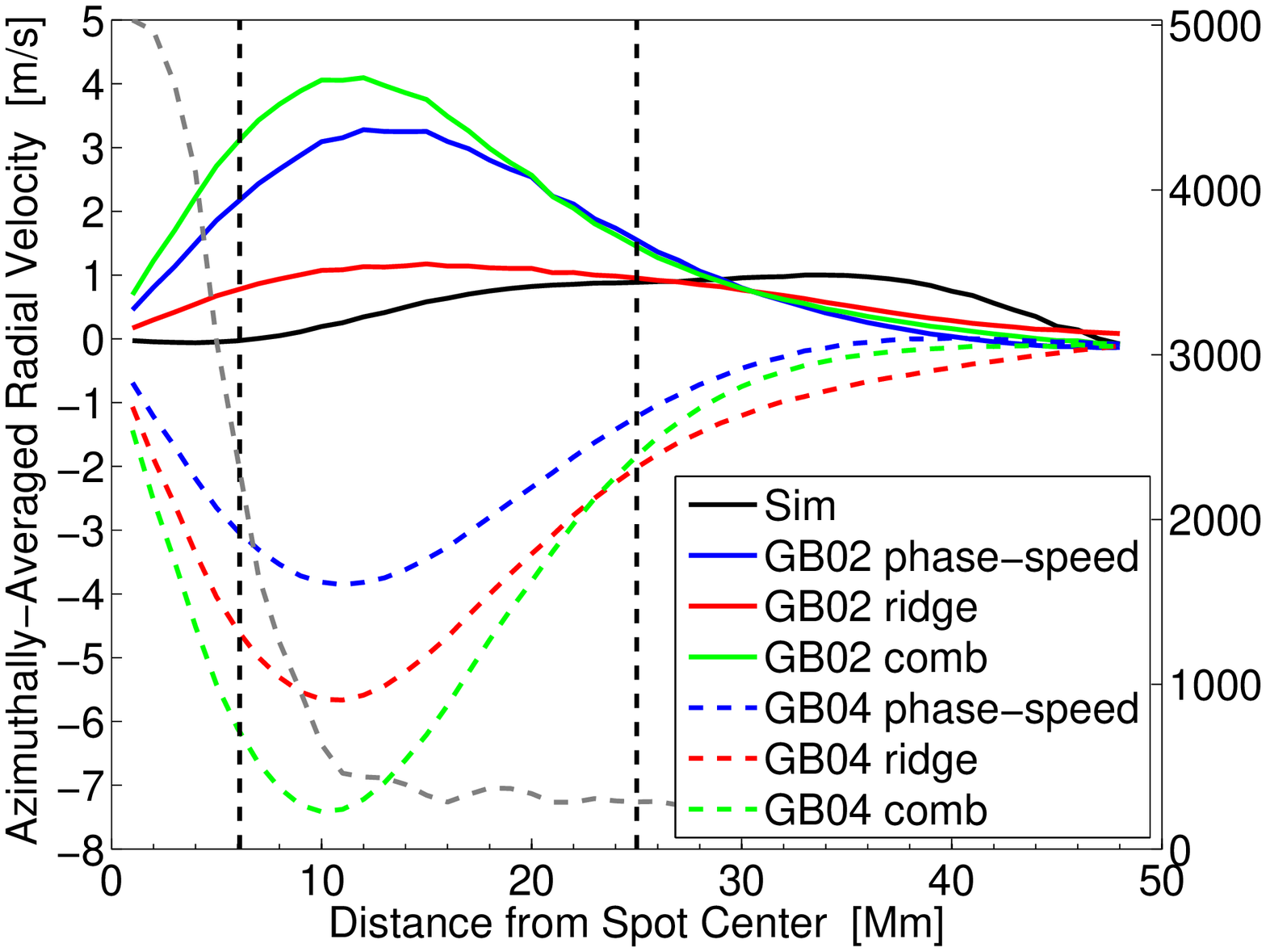} \\
\includegraphics[width=0.3\linewidth,clip=]{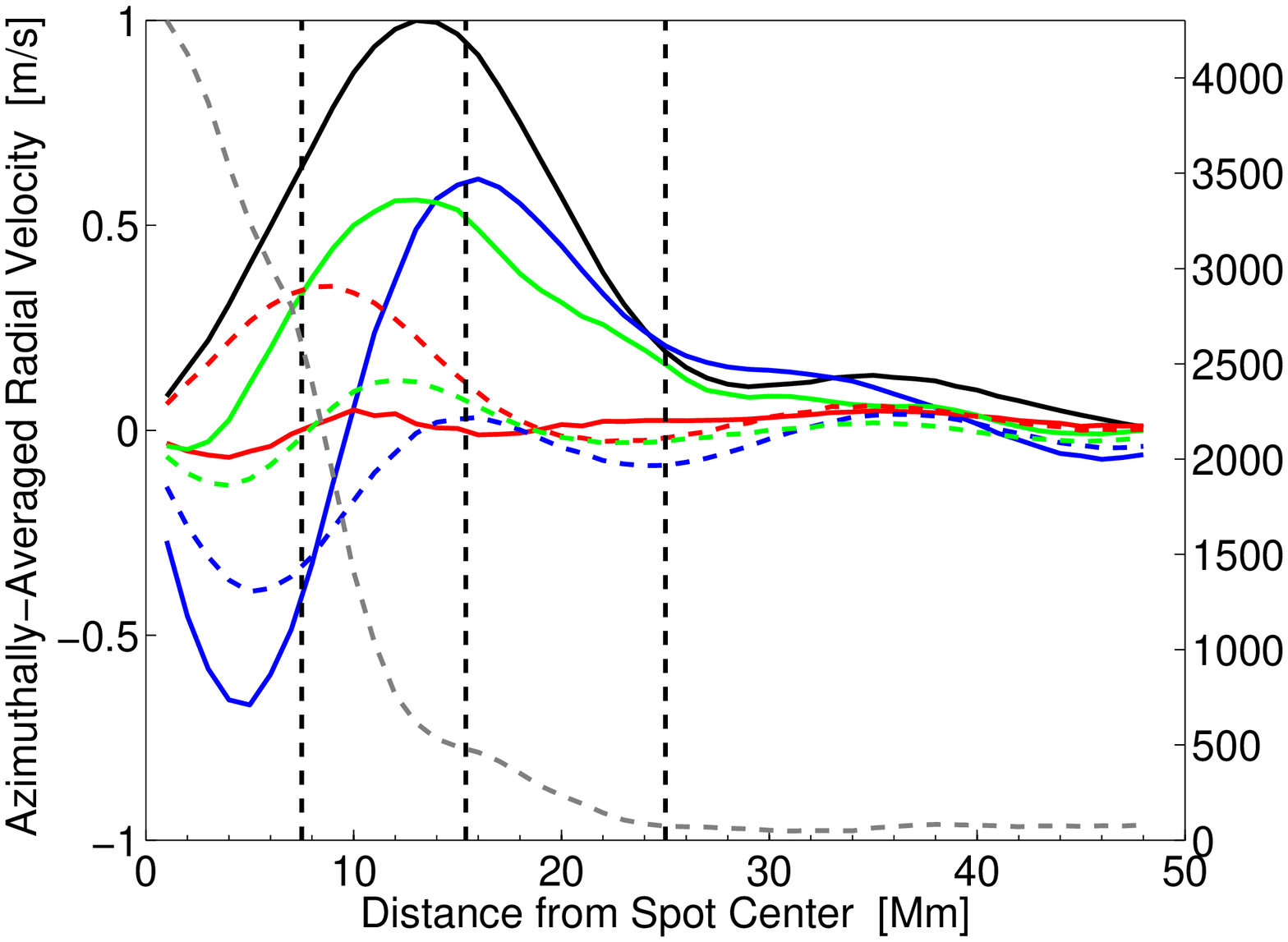} &
\includegraphics[width=0.3\linewidth,clip=]{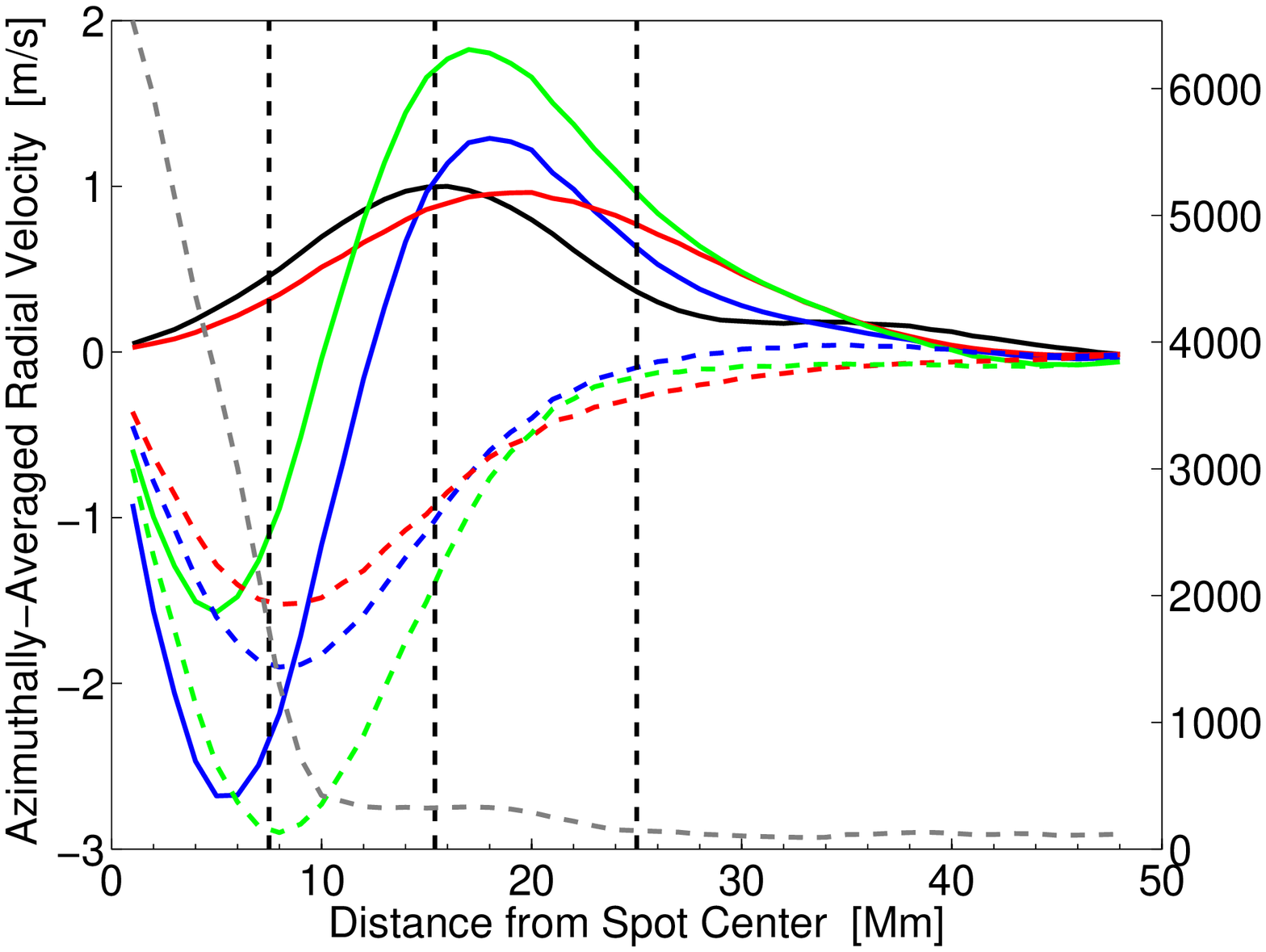} &
\includegraphics[width=0.3\linewidth,clip=]{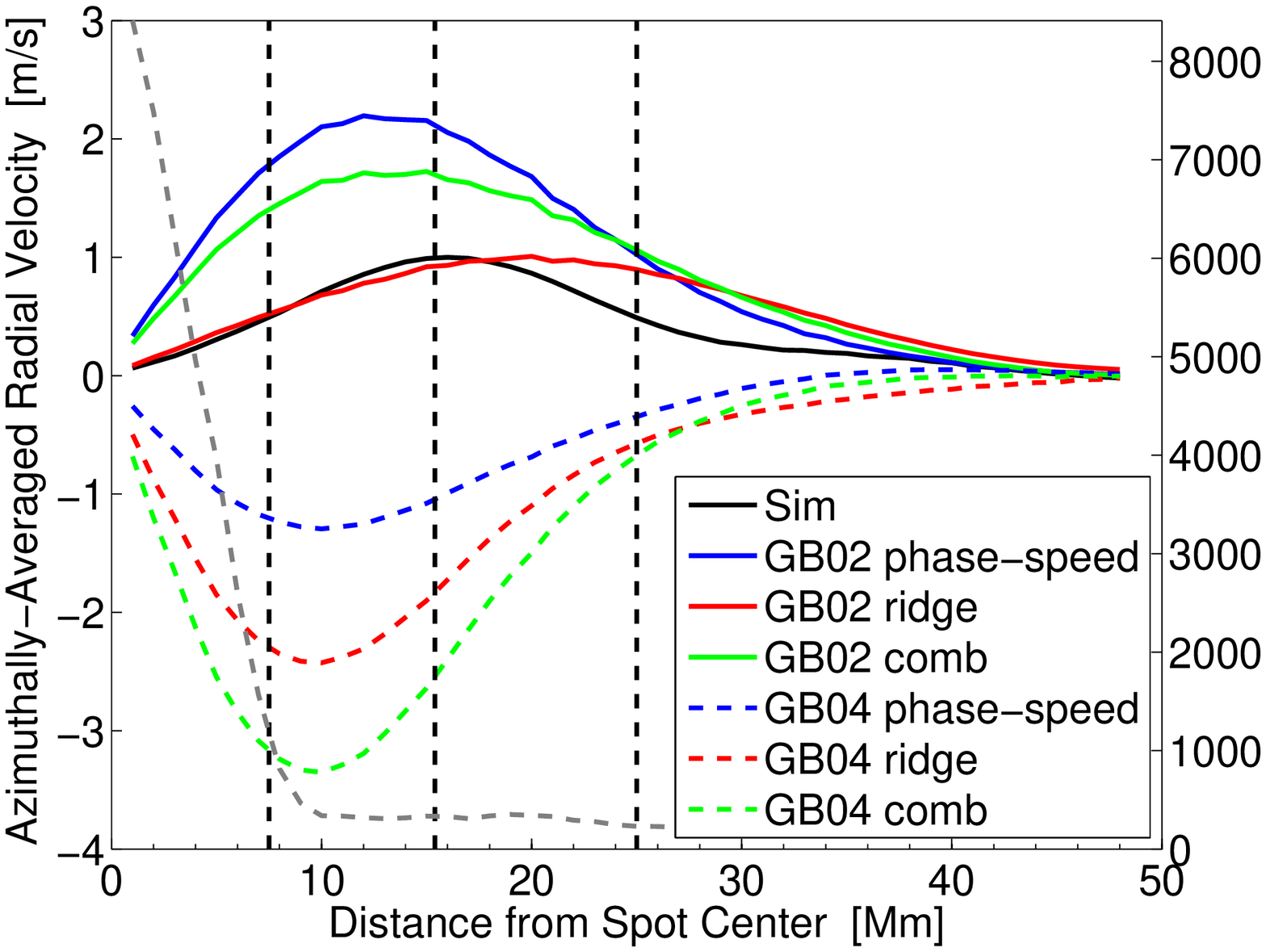}
\end{array}$
\end{center}
\caption{Azimuthally-averaged radial velocity profiles computed from the $v_{x,y}^{\rm inv}$ flow maps for LRes (top row) and HRes (bottom row) at depths (left to right) 1, 3 and 5~Mm. Results obtained using both GB02 and GB04 travel-time definitions are shown together in the same figures. The solid black lines correspond to the azimuthally-averaged radial velocity computed from the smoothed simulation flow maps (i.e. $v_{x,y}^{\rm tgt}$) at each depth. All profiles have been normalized by the largest absolute velocity value of the simulation profile in each figure for easier comparison. The gray dashed lines represent the magnetic field profile (i.e. $|B|=\sqrt{B_x^2+B_y^2+B_z^2}$) at each depth as a function of distance from spot center corresponding to the right-most y-axis in units of Gauss. The vertical dashed lines represent the boundaries of the simulation umbra, penumbra, and the circular mask.}
\label{fig:vazim}
\end{figure}

The radial velocities from the spot center were computed for each horizontal flow map and averaged azimuthally to give the profiles in Figure~\ref{fig:vazim}. These plots are useful for several reasons, namely they give a sense of how well the inversions converge to the expected answer as one moves away from the magnetic structure into the quiet Sun, and how well the amplitude is recovered in the inversions. All profiles have been scaled by the largest absolute velocity value of the simulation profile in each panel. One notices that near the surface, the inverted velocities mostly underestimate the flow amplitude, sometimes up to a factor of two, as was noted in \citet{degrave2014}, and as expected, the inverted GB02 travel times more closely match the model. Deeper down, flow estimates are typically larger than the models due to increased effects of noise, yet the inverted ridge-filtered GB02 measurements are consistent with some of the flow profiles within the influence of the magnetic perturbations, while the phase-speed GB02 results are better beyond it. The GB04 cases almost always give flows with the wrong sign.



\begin{figure}
\begin{center}$
\begin{array}{cc}
\includegraphics[width=0.5\linewidth,clip=]{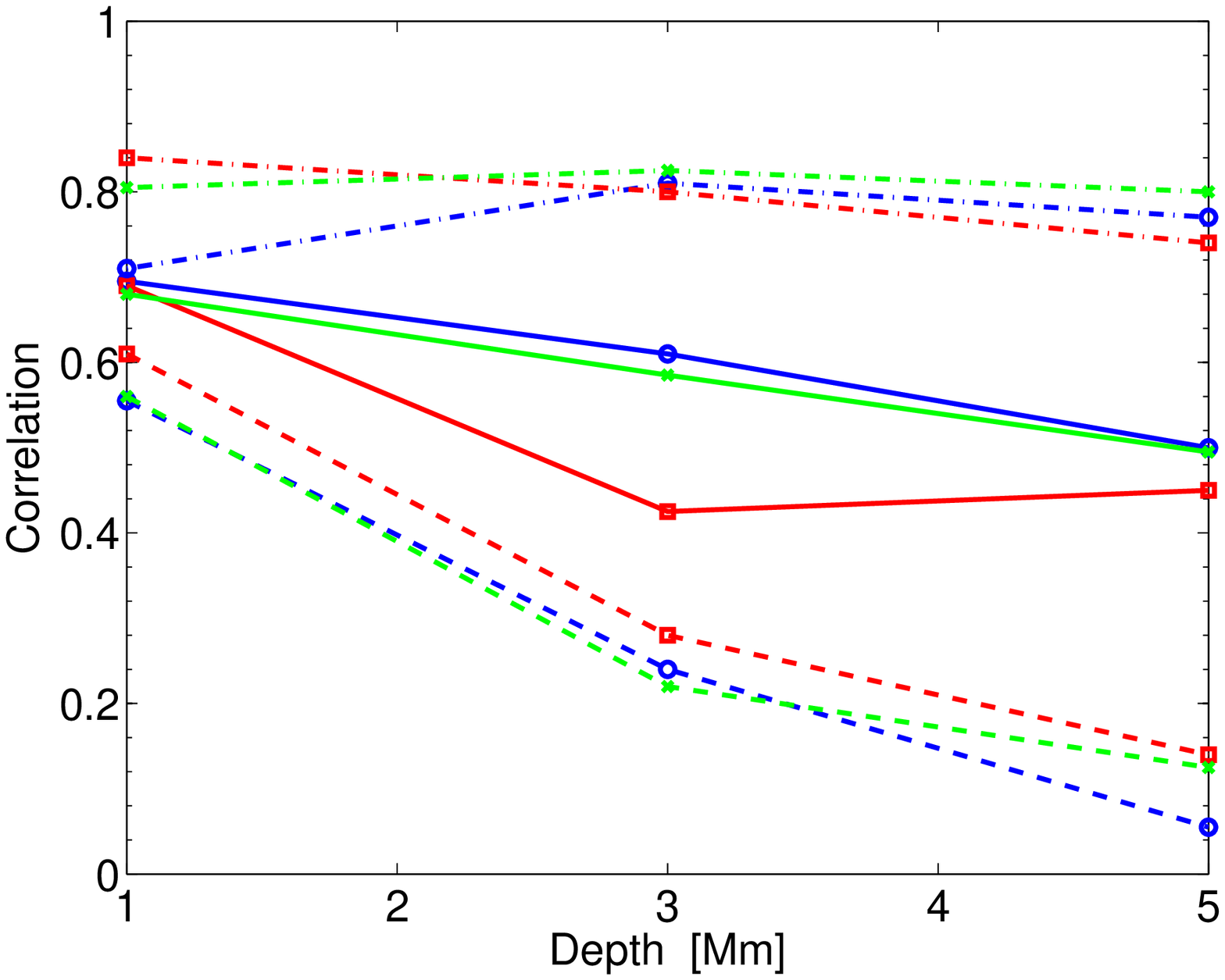} &
\includegraphics[width=0.5\linewidth,clip=]{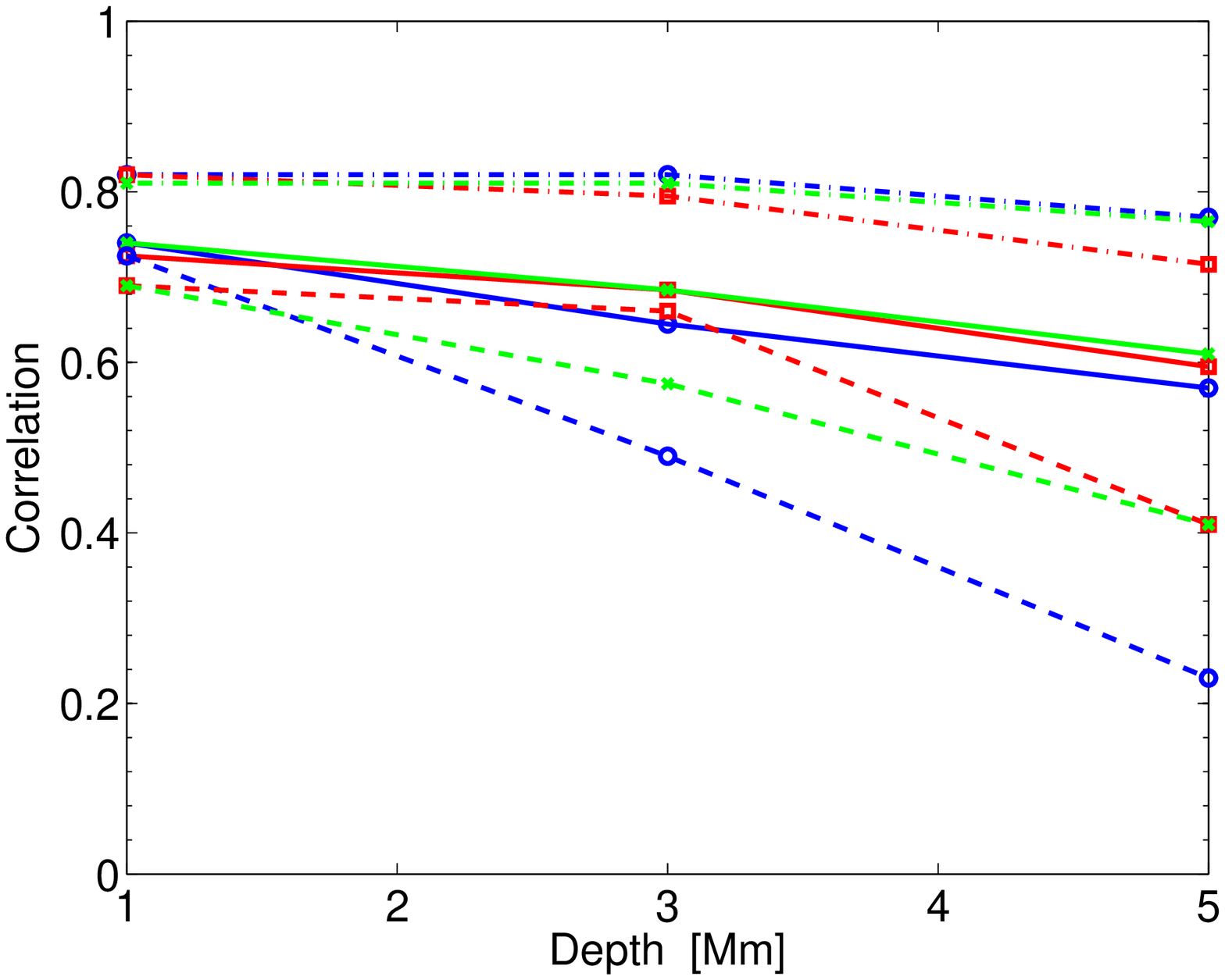} \\
\includegraphics[width=0.5\linewidth,clip=]{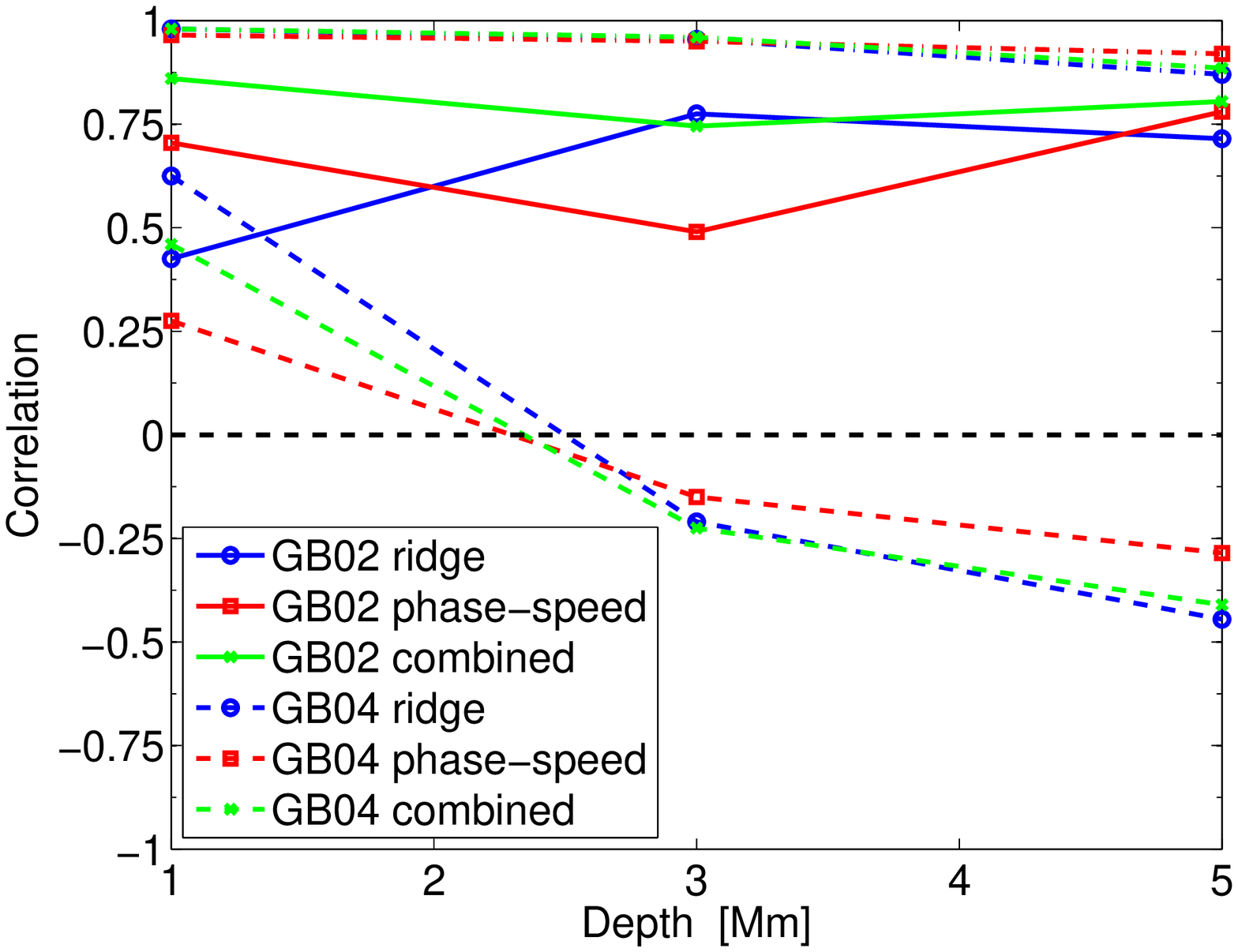} &
\includegraphics[width=0.5\linewidth,clip=]{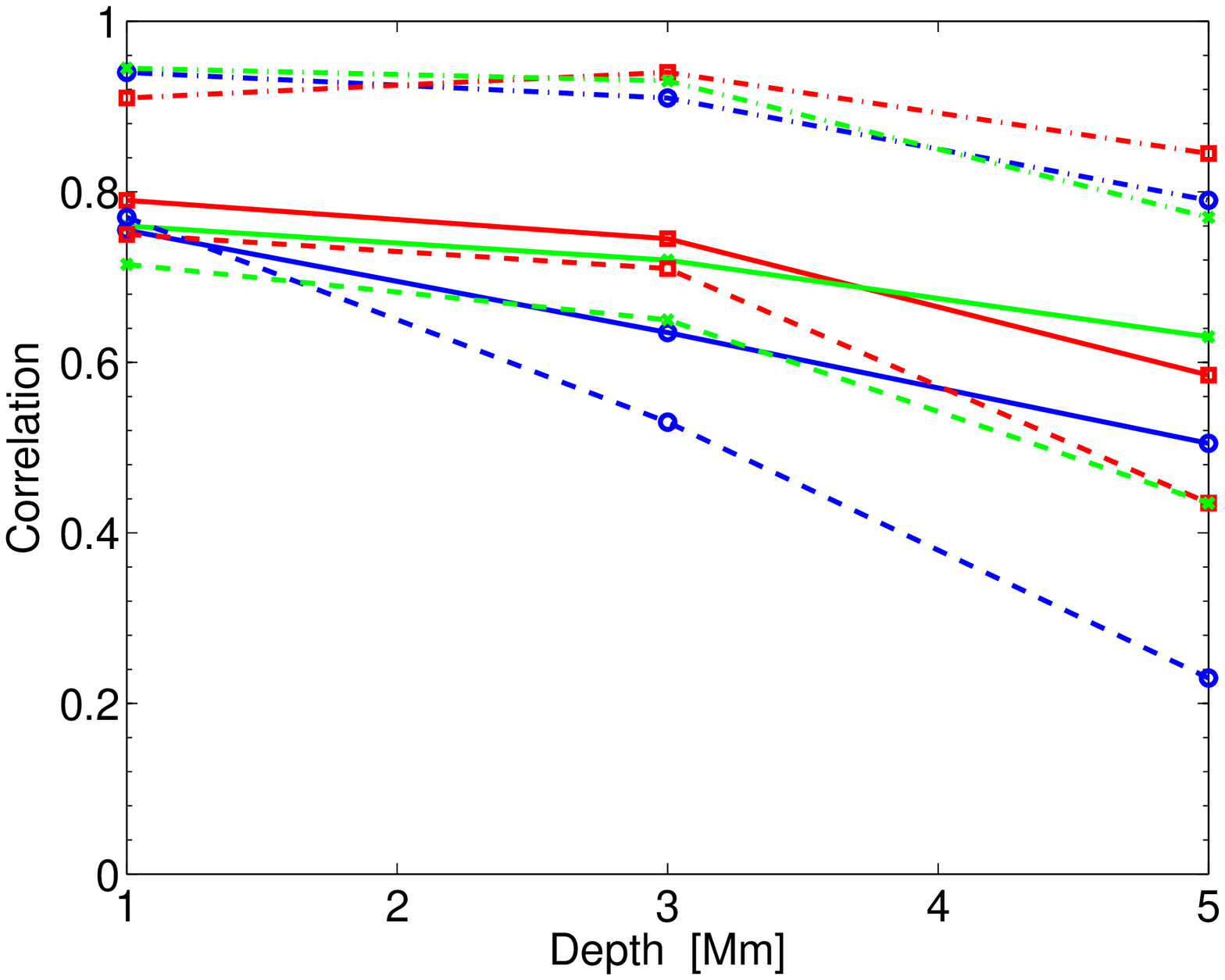}
\end{array}$
\end{center}
\caption{The 2D correlation between horizontal inversion flow maps, $v_{x,y}^{\rm inv}$, and simulation target flow maps, $v_{x,y}^{\rm tgt}$ for LRes (top row) and HRes (bottom row) before (left column) and after (right column) applying the circular mask to each map to eliminate the spot. All values are computed from the mean of the individual $v_{x}^{\rm inv}$ and $v_{y}^{\rm inv}$ correlations, as these flow components differ substantially in several cases. The solid lines represent the GB02 inversions, while the dashed lines represent the GB04 inversions. Also plotted using the dash-dot (- .) lines are the correlations found using the inverted forward-modeled travel times. Line color represents the particular filtering scheme used in the inversions as defined in the legend.}
\label{fig:vxinvsimcorr}
\end{figure}

Spatial correlations were also computed between inversion and model flow maps. Figure~\ref{fig:vxinvsimcorr} shows these correlation values for both simulations as a function of depth using both travel-time definitions and every filtering scheme. For comparison, this was done before and after applying the circular mask to eliminate the spot from every map. This is reminiscent of the ``cookie cutter" tests carried out by \citet{zhao2003} and \citet{korzennik2006}. Here, however, the masking is done to the flow maps themselves, rather than to the measurements before the flows are computed. Before masking, it's clear that the GB04 correlations were overall substantially worse than those of GB02 for both simulations. In fact, the HRes GB04 inversions are quite anticorrelated at depth, reflecting the inability to capture the flow divergence in the spot. After masking, many correlations at depth were significantly improved, most notably those of GB04. Interestingly, the masked correlations show very similar trends between the two simulations. We find that the masked correlations are now reasonably good ($\sim0.5$--0.8) in the upper 3~Mm, and are approaching the values found by \citet{degrave2014} using quiet-Sun simulation data. A slight improvement is seen when using the phase-speed filters in both travel-time definitions, as was observed by \citet{degrave2014}.

\begin{figure}
\begin{center}$
\begin{array}{cc}
\includegraphics[width=0.5\linewidth,clip=]{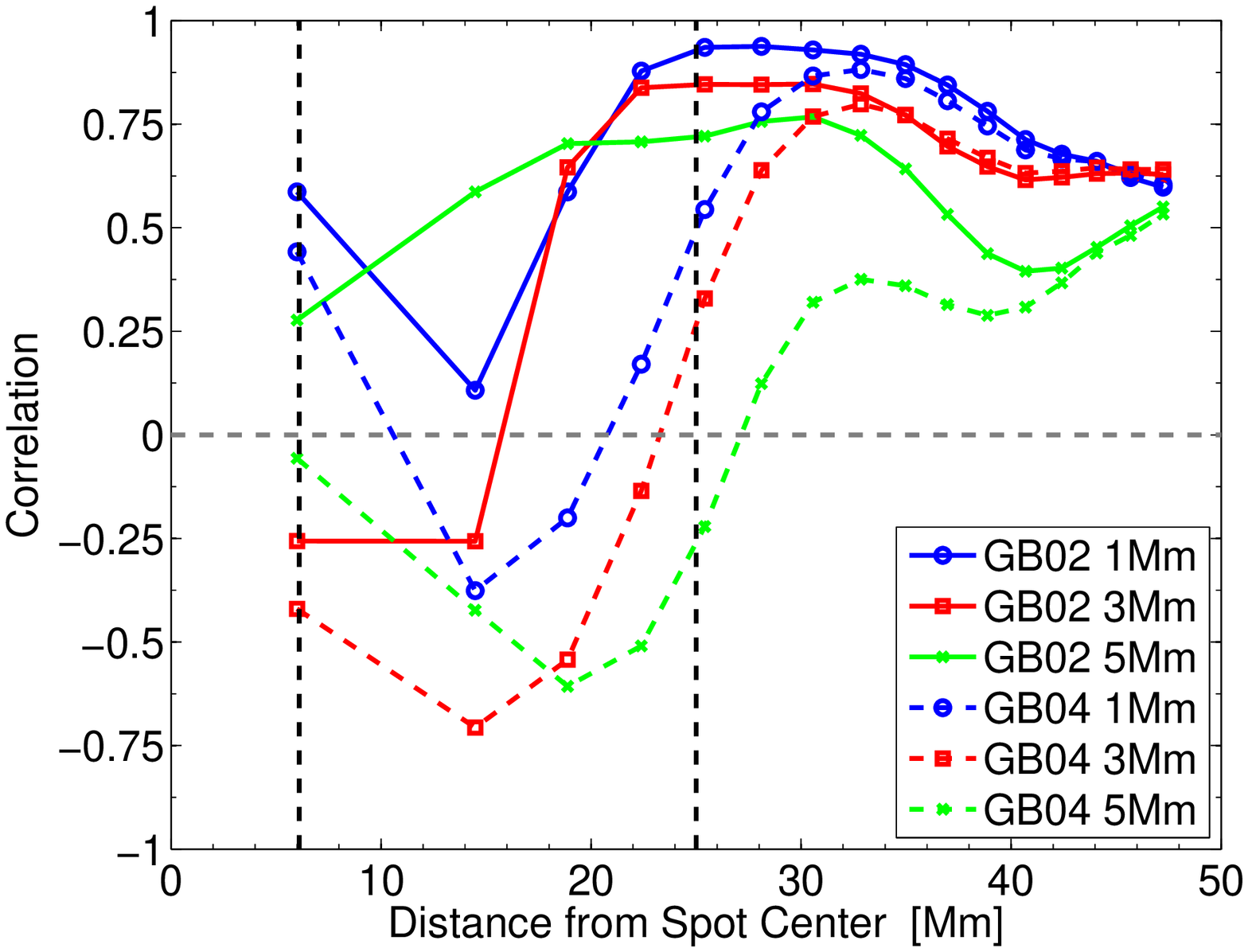} &
\includegraphics[width=0.5\linewidth,clip=]{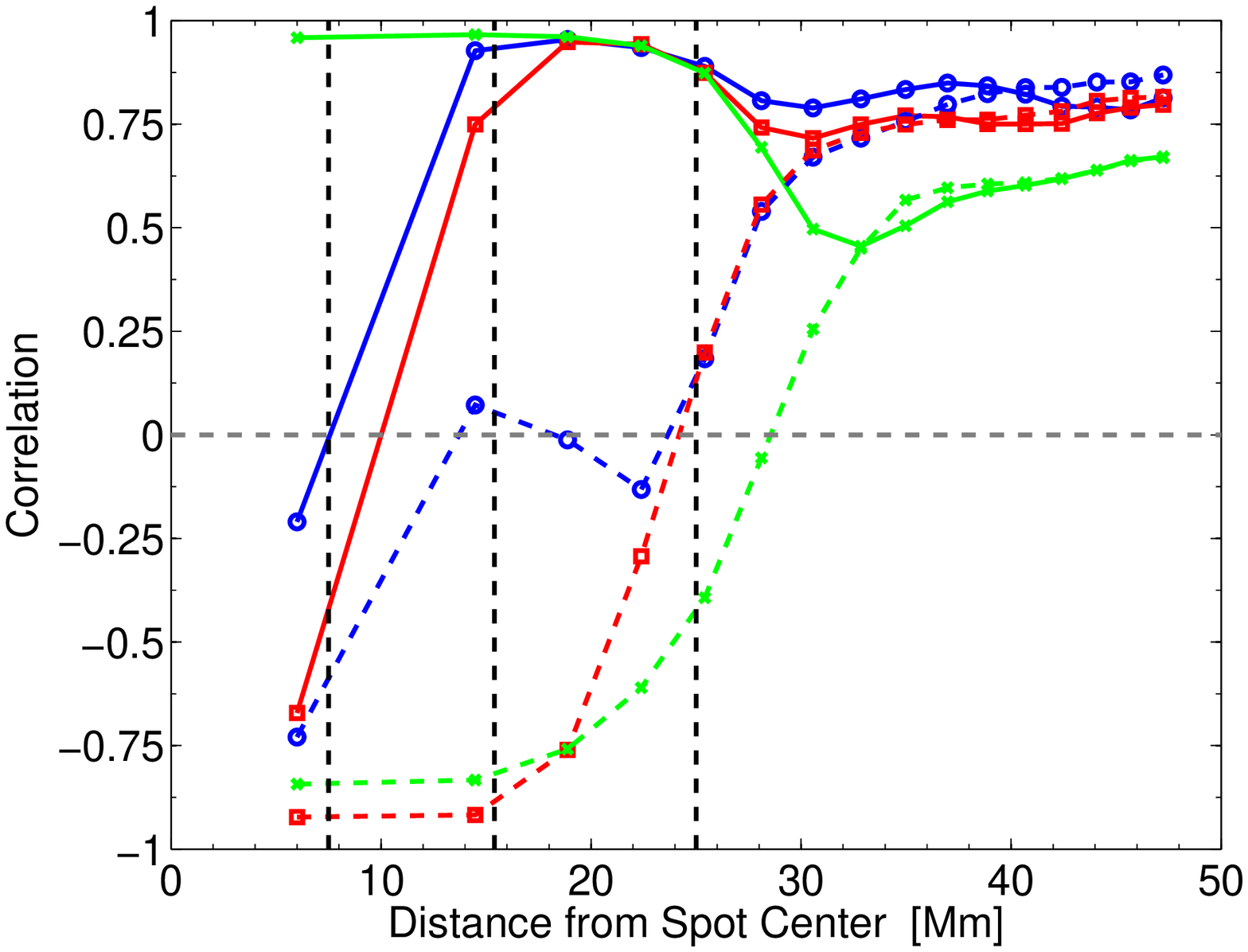}
\end{array}$
\end{center}
\caption{The 2D correlation between LRes (left) and HRes (right) $v_{x,y}^{\rm tgt}$ and phase-speed $v_{x,y}^{\rm inv}$ flow maps at each depth as a function of distance from spot center. These correlations were computed over annuli of increasing inner and outer radii of roughly equal area centered about each spot. Values shown here are the mean of the individual $v_x$ and $v_y$ flow component correlations. The distance from spot center is taken to be the average of the inner and outer radii for a particular annulus. The vertical dashed lines represent the boundaries of the simulation umbra, penumbra, and the circular mask.}
\label{fig:anncorr}
\end{figure}

We also investigated how horizontal flow correlation changed as one moves away from spot center. Figure~\ref{fig:anncorr} shows the correlation values between $v_{x,y}^{\rm tgt}$ and phase-speed $v_{x,y}^{\rm inv}$ flow maps using both GB02 and GB04 travel-time definitions computed over annuli of increasing inner and outer radii centered about each spot. All annuli were constructed to have roughly the same area. We find that GB02 and GB04 flow correlations typically begin to agree at a radius of $\sim30$~Mm, especially in the upper 3~Mm of the simulations.

Additionally, forward-modeled travel times were also inverted to compare with the results from the measured ones. We see from the high correlation values of Figure~\ref{fig:vxinvsimcorr} (dash-dot lines) that the inversions from the modeled travel times are able to reproduce the simulation target maps exceptionally well even before masking out the spots, with virtually no filter dependence. These values can be directly compared to Figure~10 in \citet{degrave2014} where similarly high correlation values were found for the quiet-Sun simulations. It is apparent from this that the self-consistency of the inversion procedure when using forward-modeled travel times is respected, and that the low correlation values shown elsewhere is due to our inability to accurately measure travel times near active regions. Despite the good agreement in overall flow structure, however, we note that our inversions using forward-modeled travel times always give lower root-mean-square (RMS) velocities than both the model and inversion flow maps using measured travel times at depths larger than 1~Mm.


\subsection{Vertical Flow Inversions}
\begin{figure}
\begin{center}$
\begin{array}{ccc}
\includegraphics[width=0.3\linewidth,clip=]{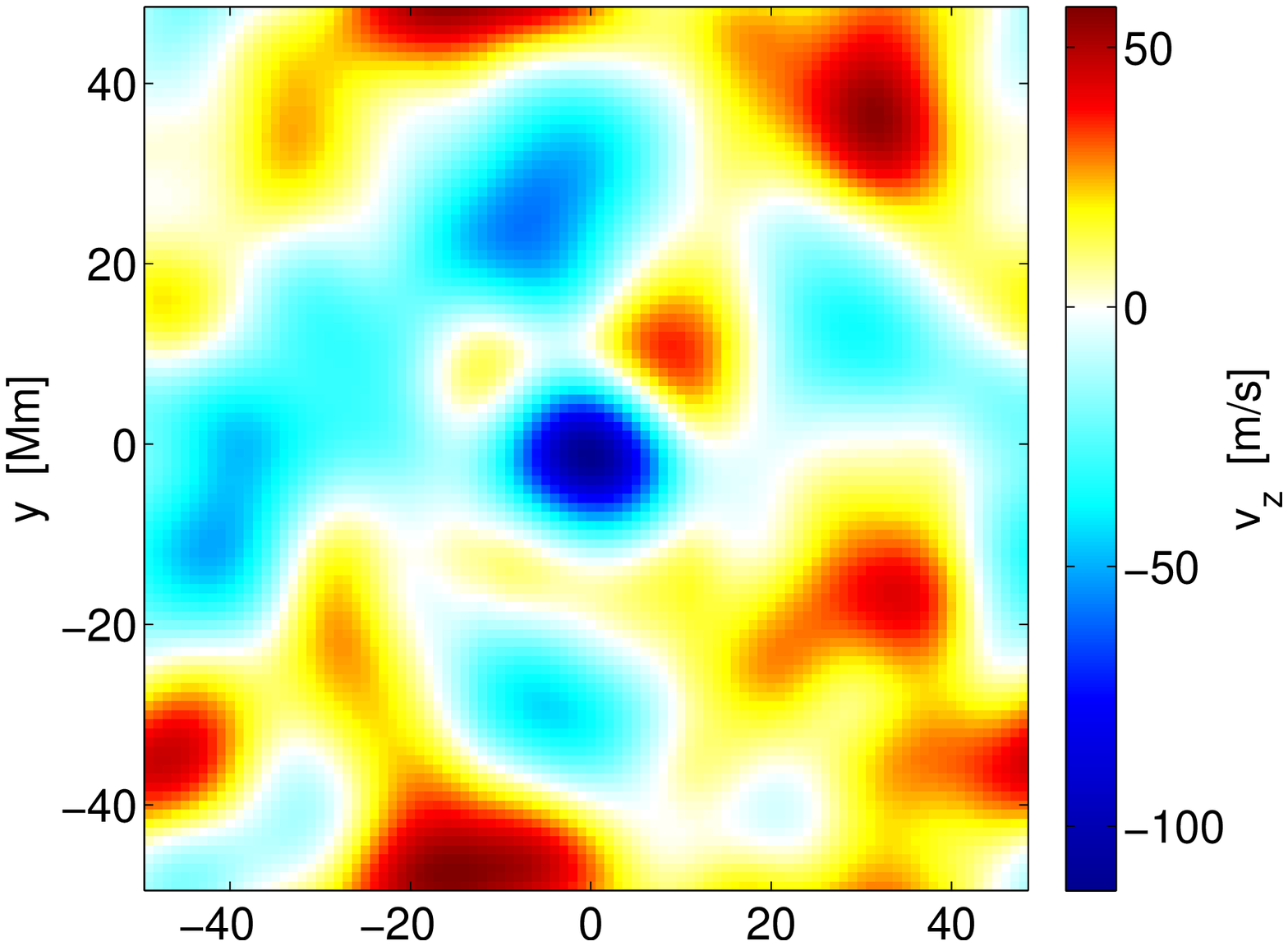} &
\includegraphics[width=0.3\linewidth,clip=]{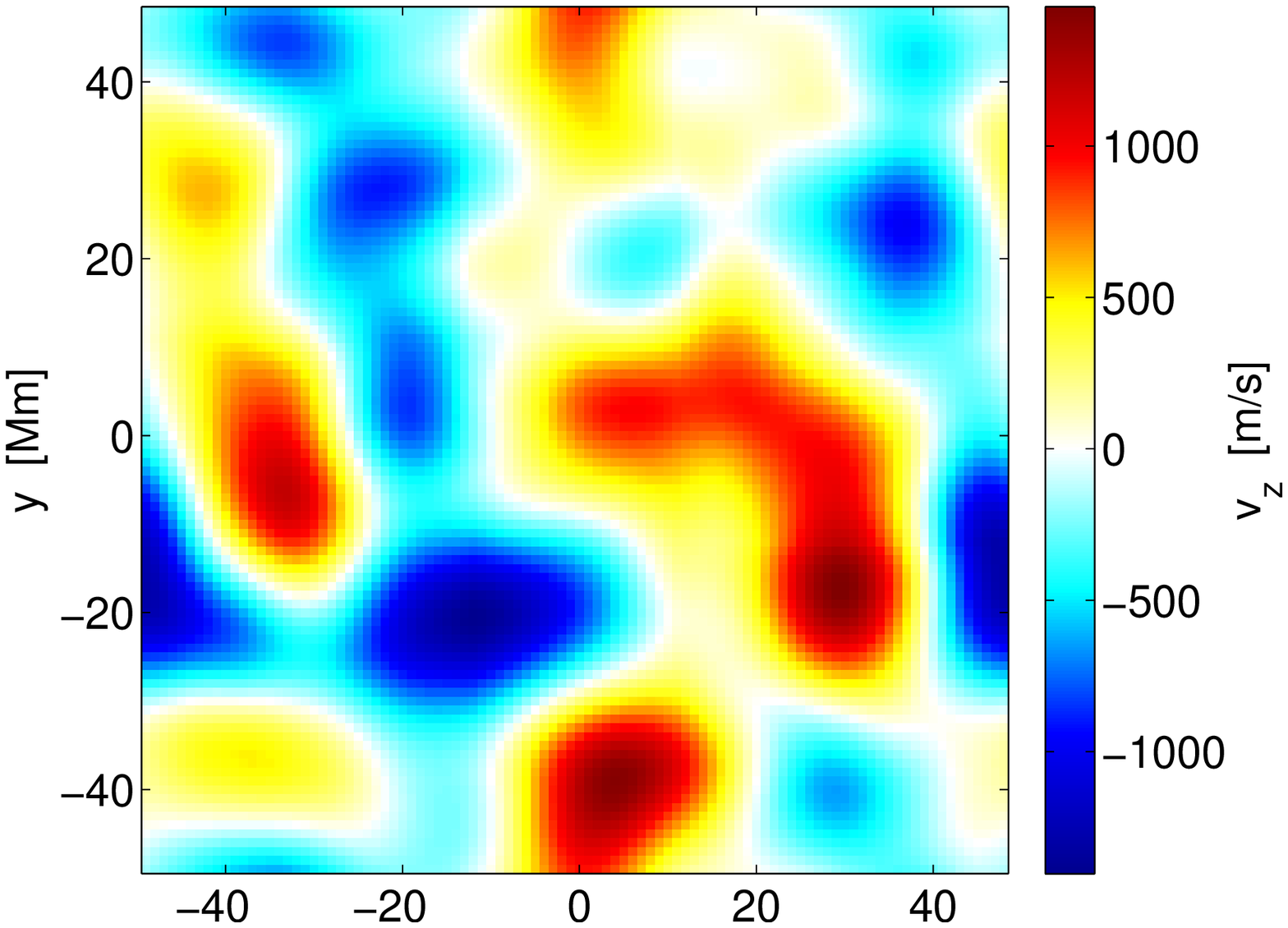} &
\includegraphics[width=0.3\linewidth,clip=]{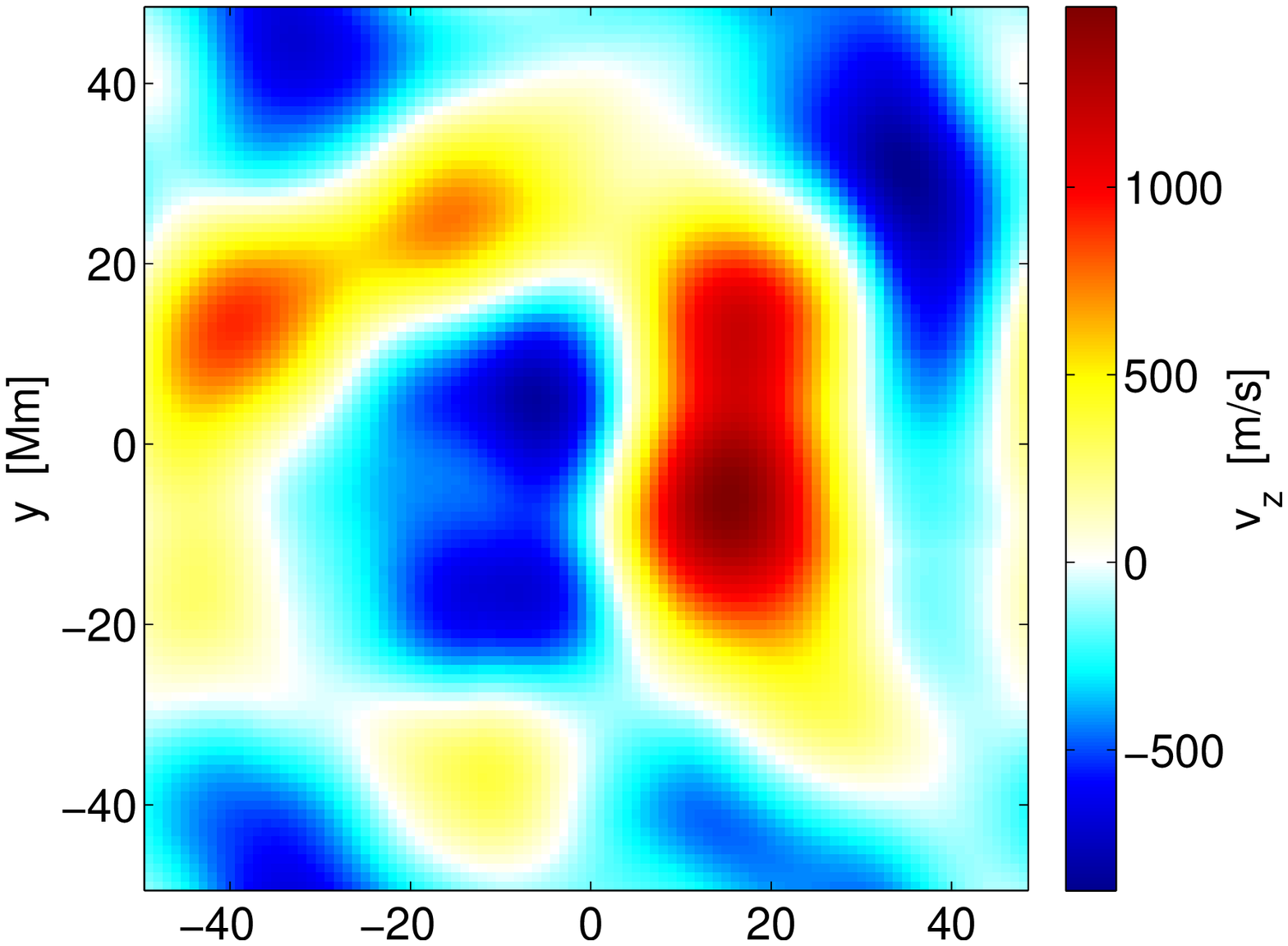} \\
\includegraphics[width=0.3\linewidth,clip=]{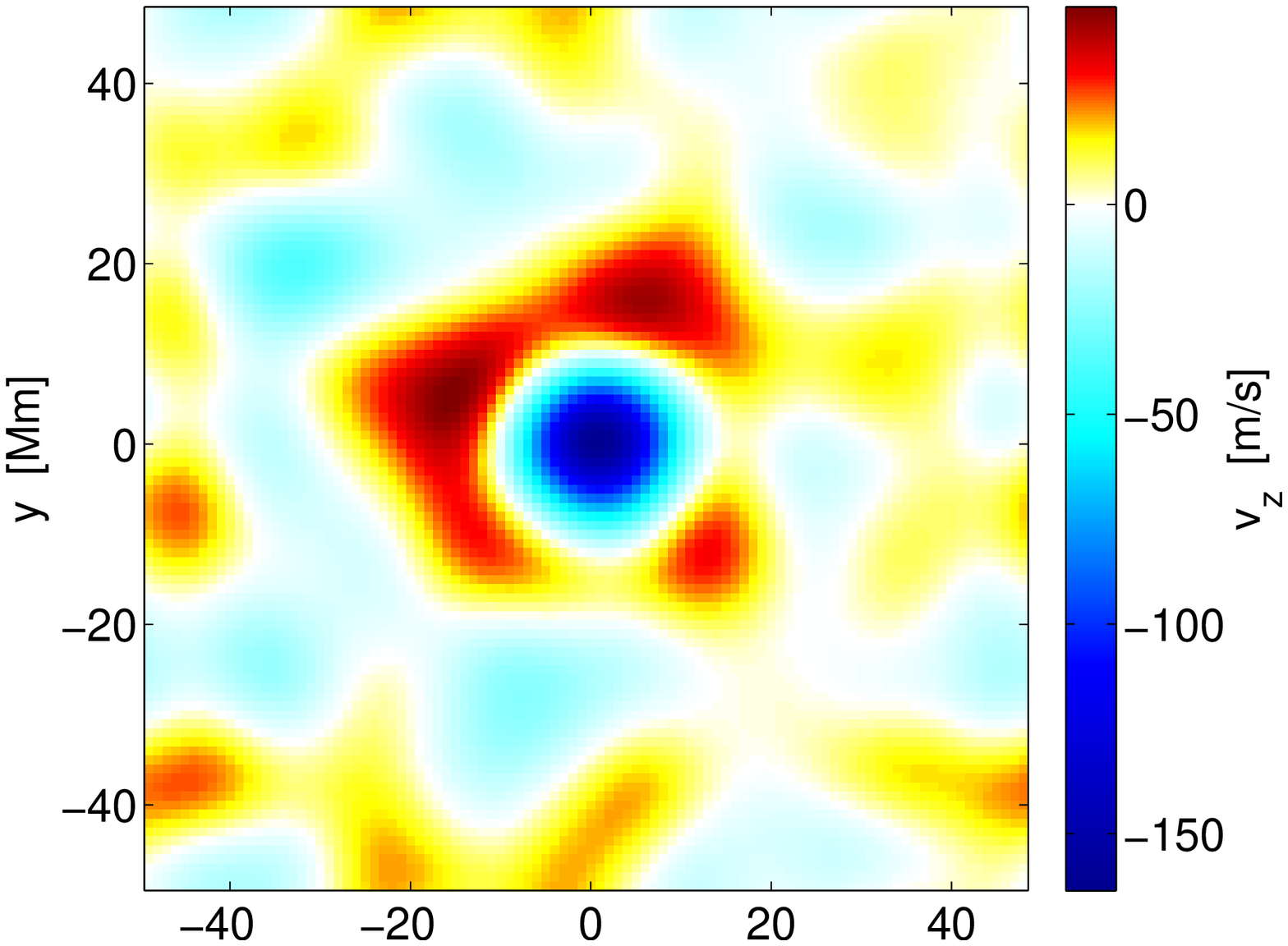} &
\includegraphics[width=0.3\linewidth,clip=]{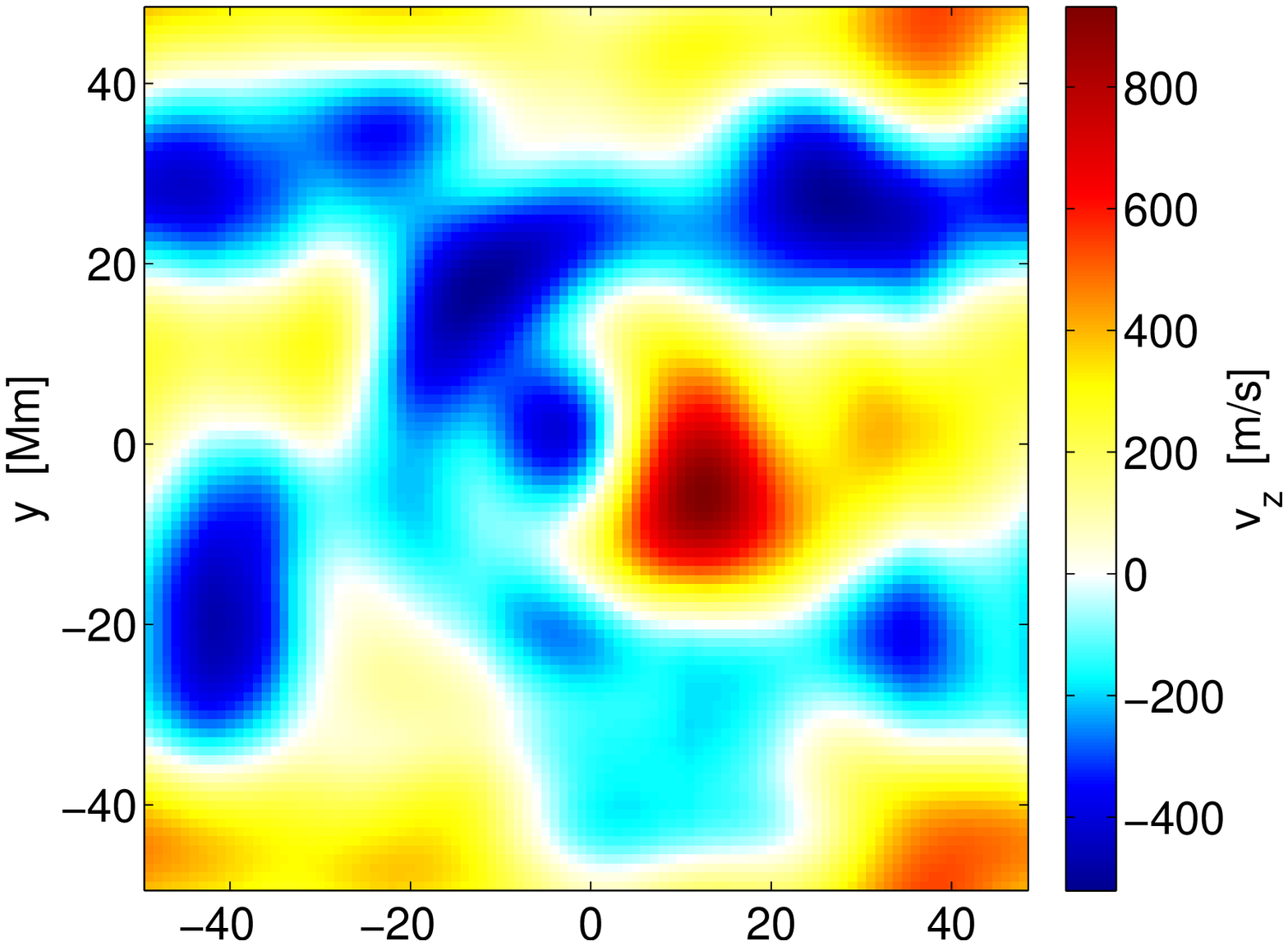} &
\includegraphics[width=0.3\linewidth,clip=]{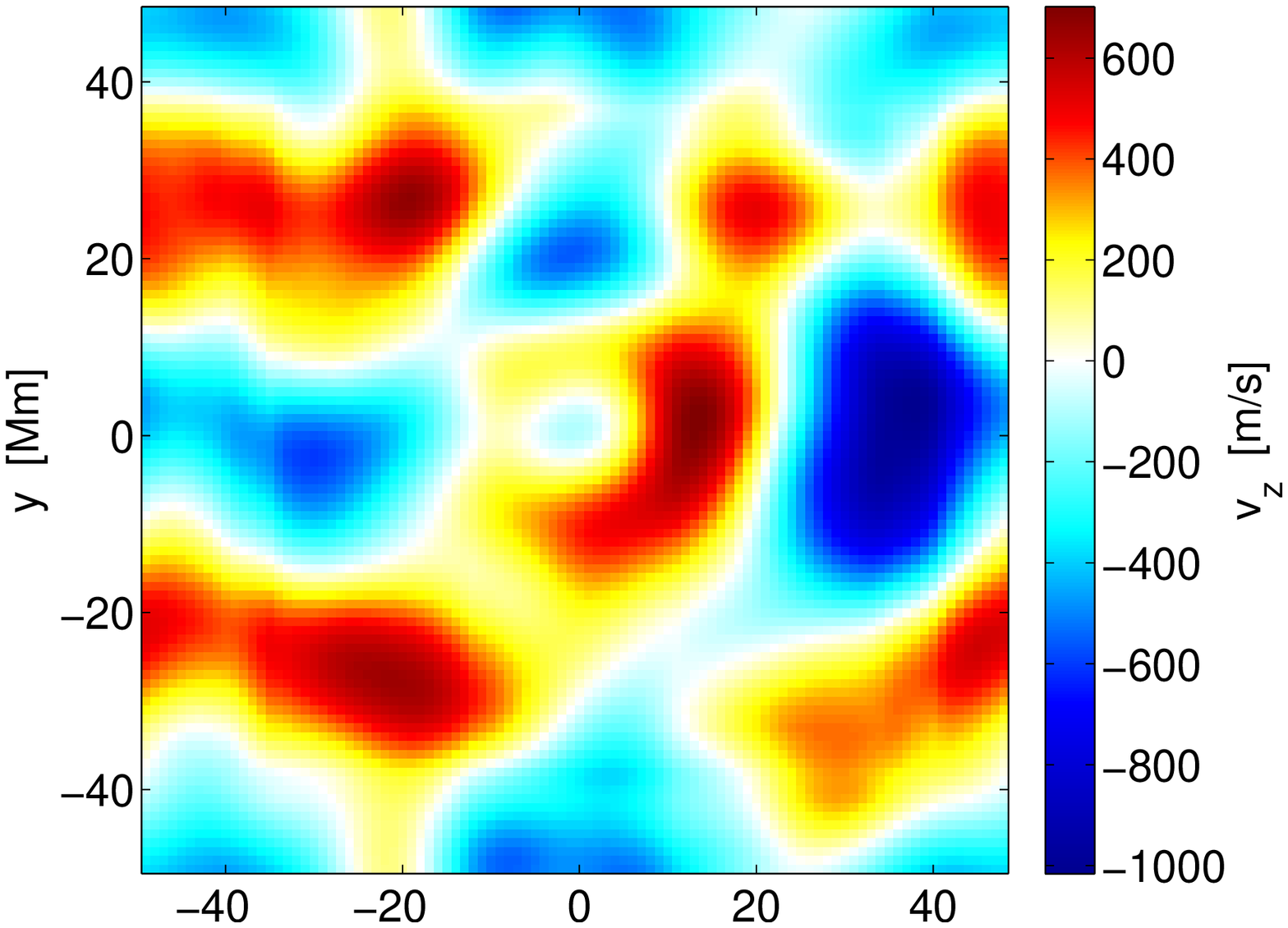} \\
\includegraphics[width=0.3\linewidth,clip=]{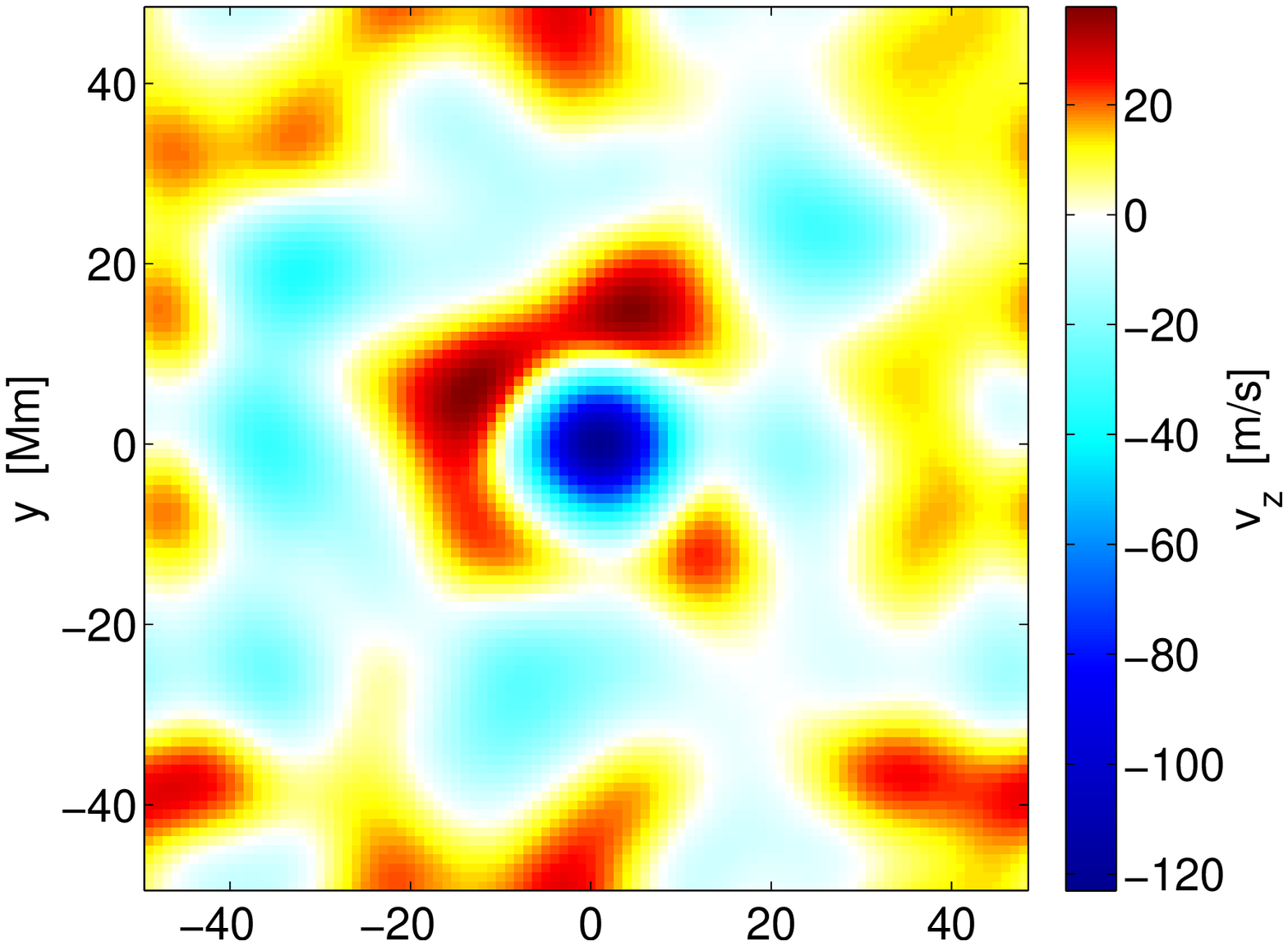} &
\includegraphics[width=0.3\linewidth,clip=]{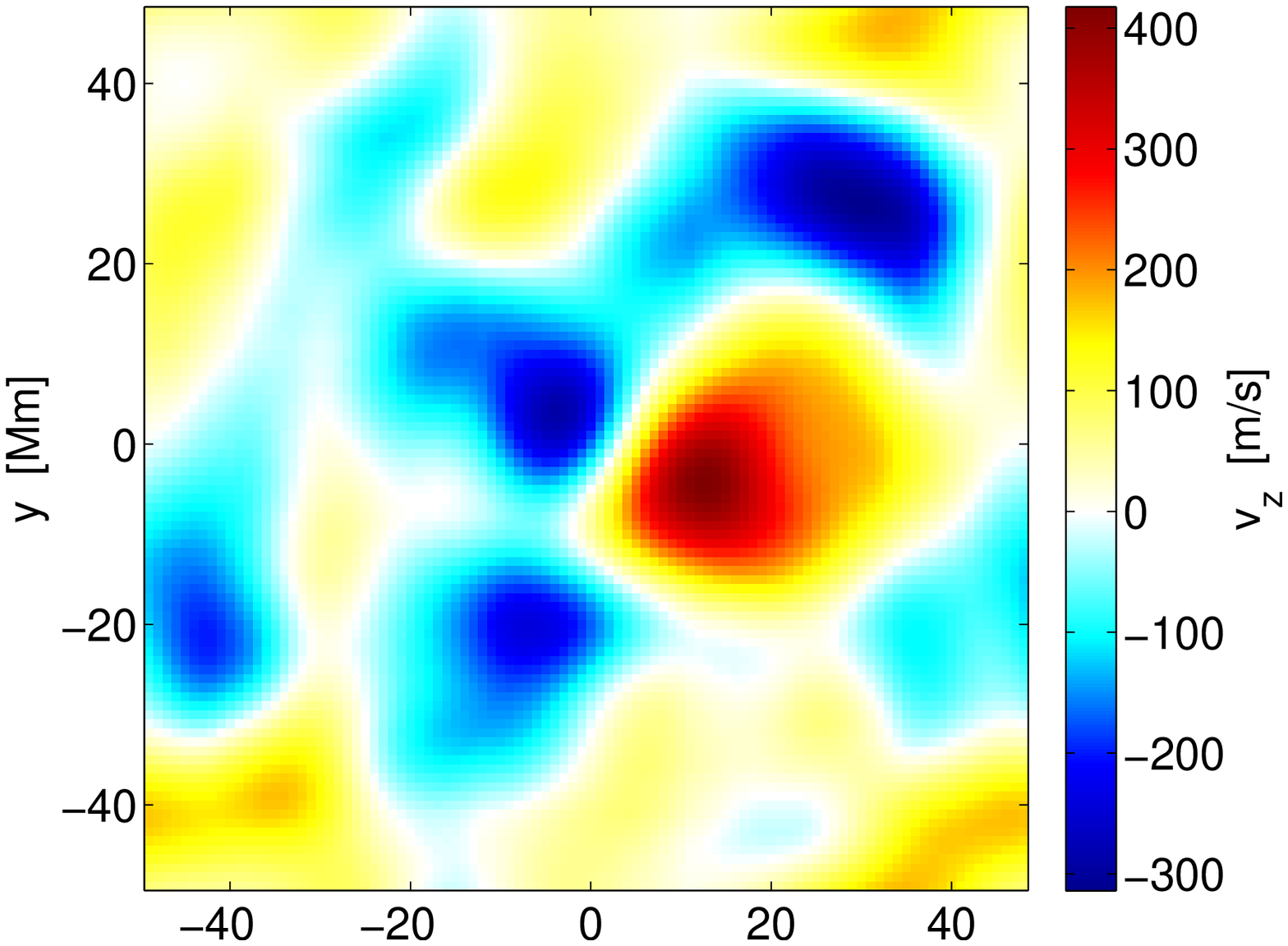} &
\includegraphics[width=0.3\linewidth,clip=]{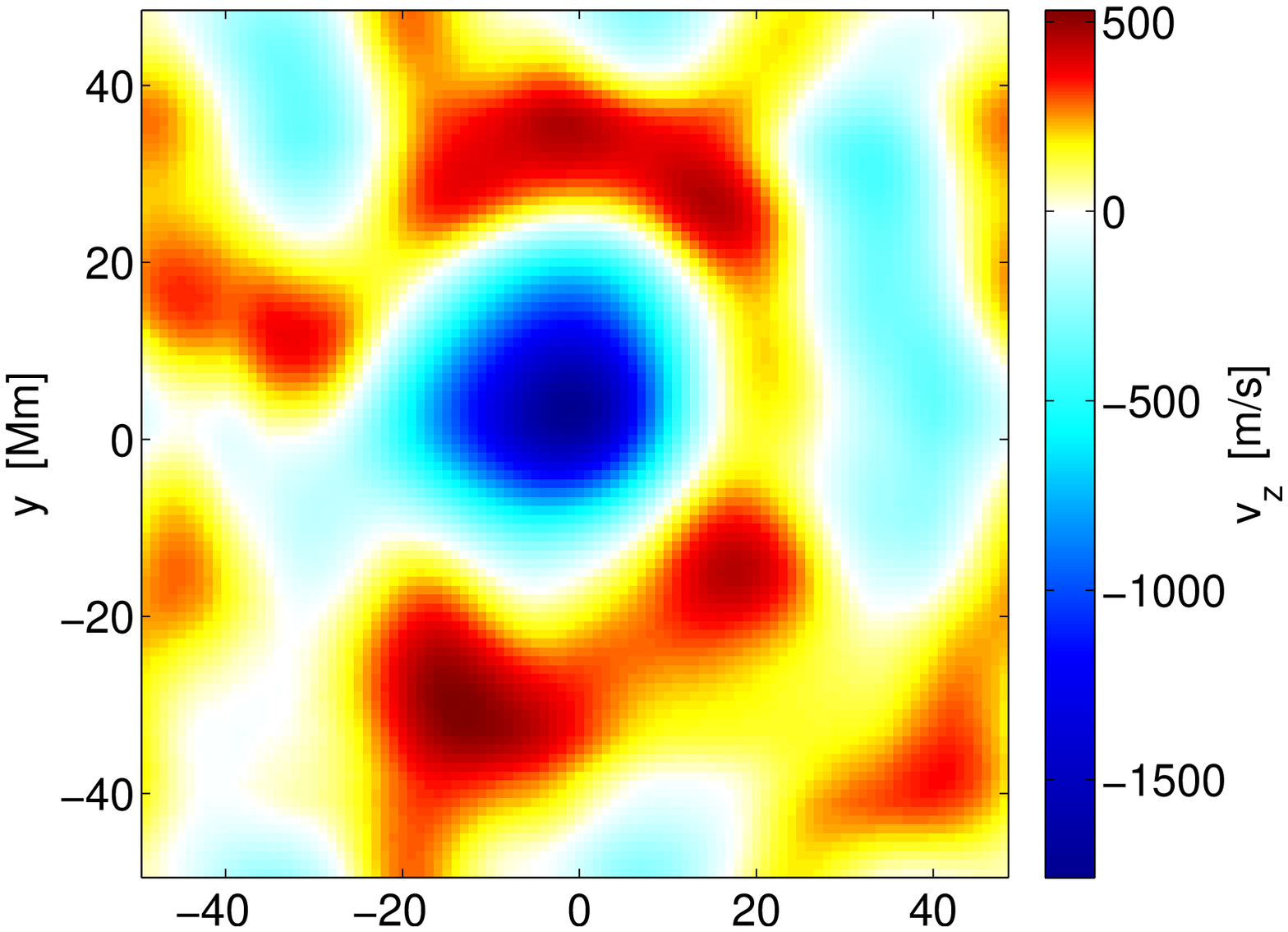} \\
\includegraphics[width=0.3\linewidth,clip=]{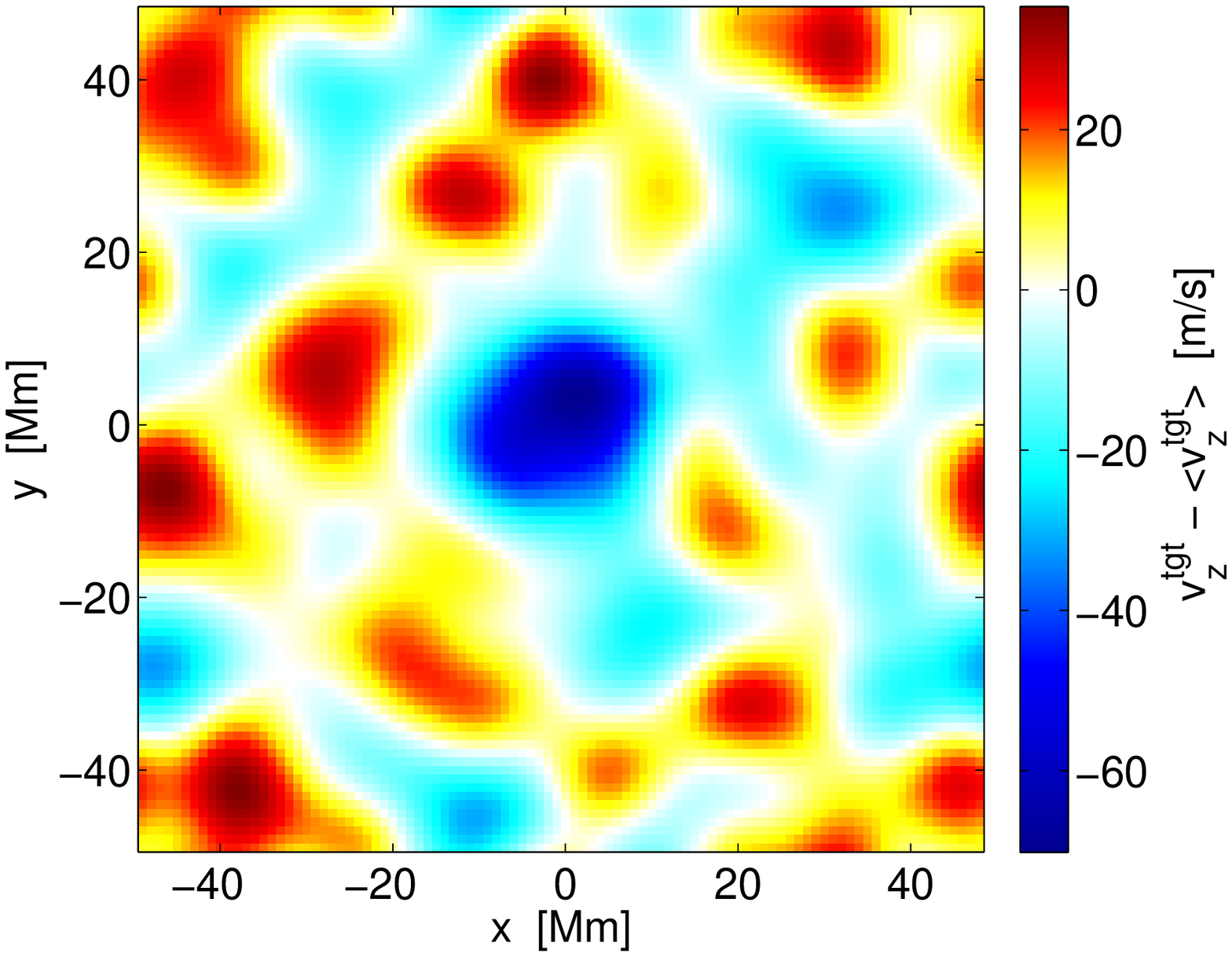} &
\includegraphics[width=0.3\linewidth,clip=]{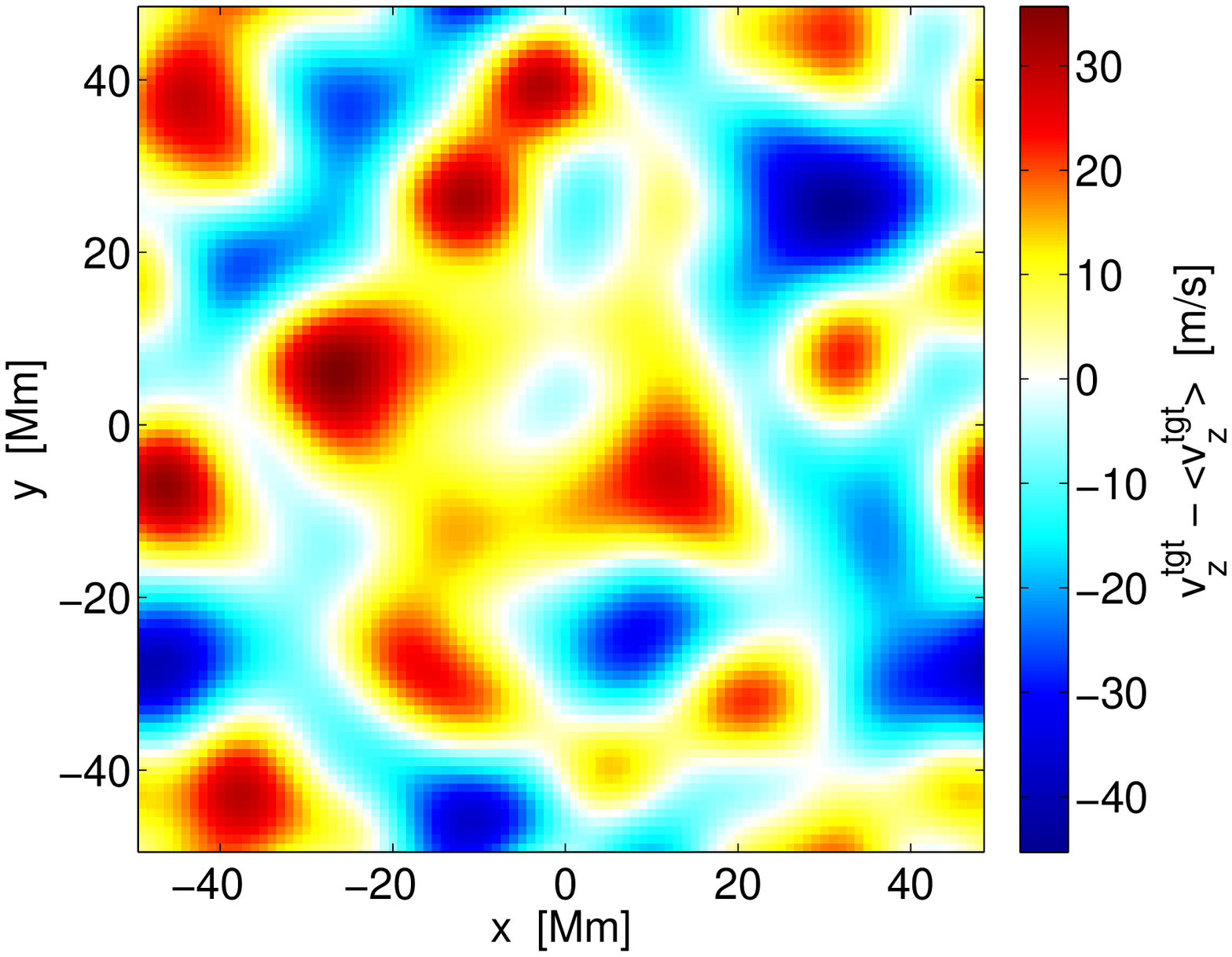} &
\includegraphics[width=0.3\linewidth,clip=]{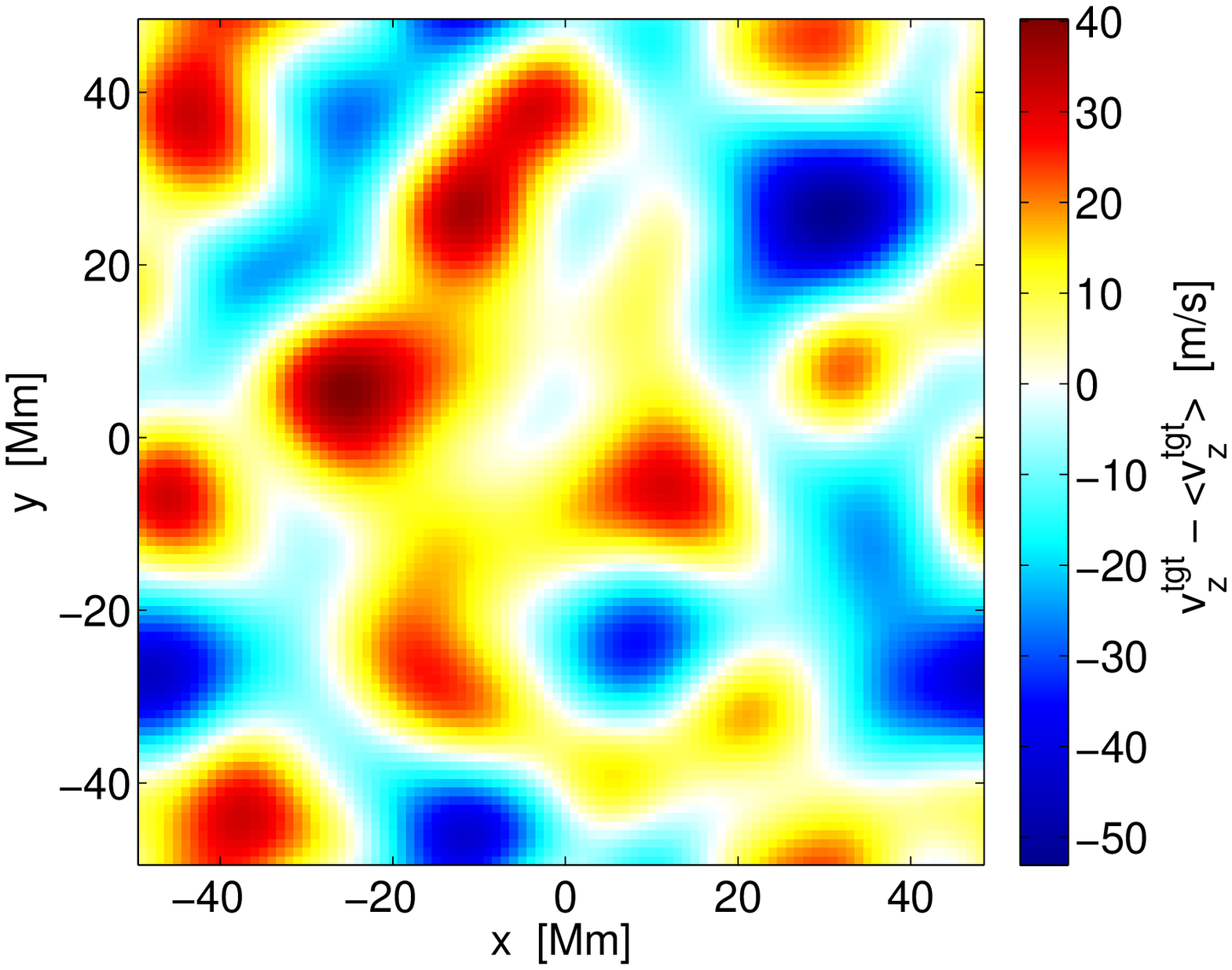}
\end{array}$
\end{center}
\caption{LRes GB02 vertical velocity inversion maps for the ridge (first row), phase-speed (second row), and ridge+phase-speed (third row) travel-time differences for depths (left to right) 1, 3 and 5~Mm. The smoothed simulation flow maps (i.e. $v_{z}^{\rm tgt}$) at these depths are shown in the bottom row.}
\label{fig:vz02SC}
\end{figure}

\begin{figure}
\begin{center}$
\begin{array}{ccc}
\includegraphics[width=0.3\linewidth,clip=]{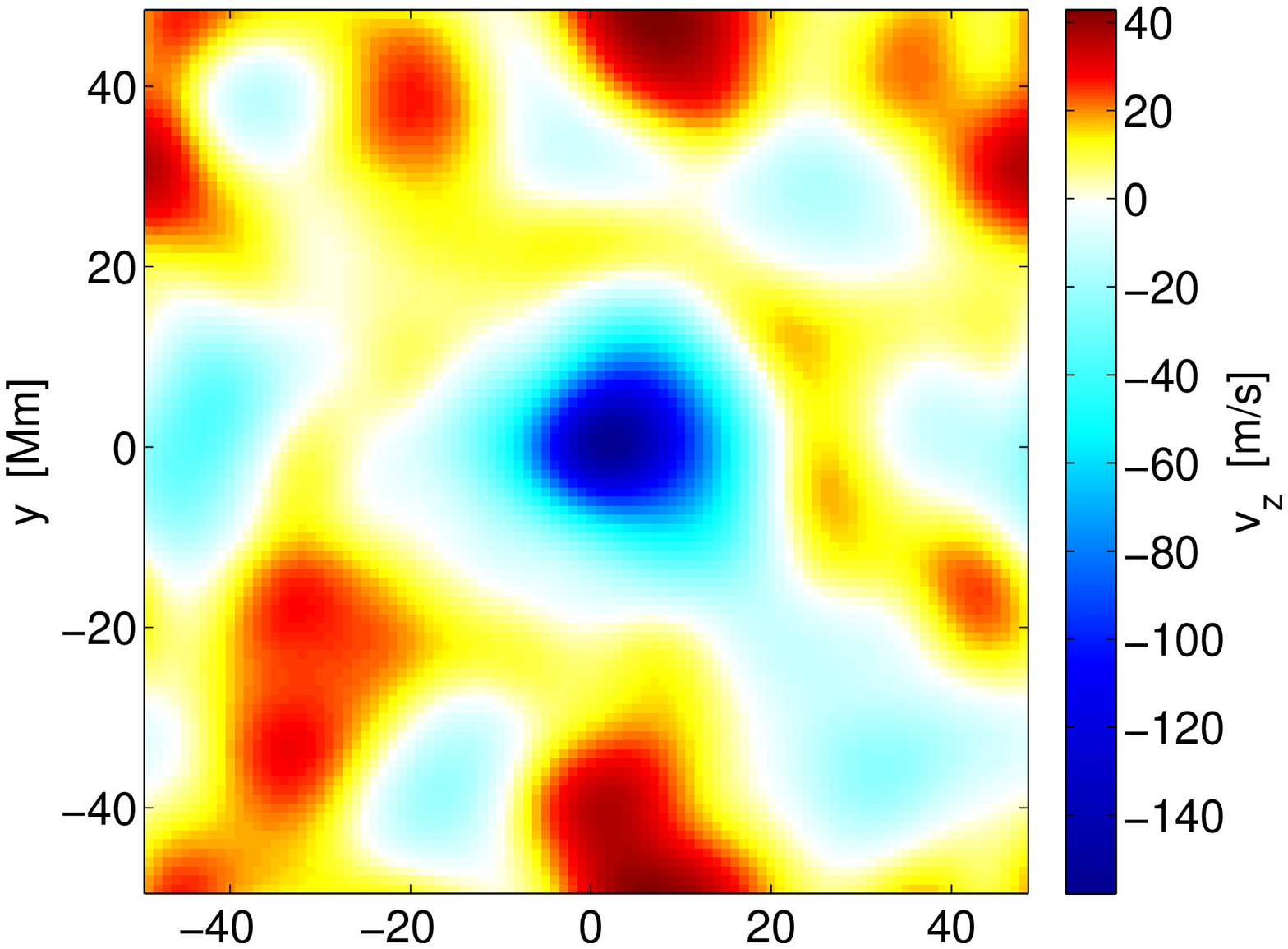} &
\includegraphics[width=0.3\linewidth,clip=]{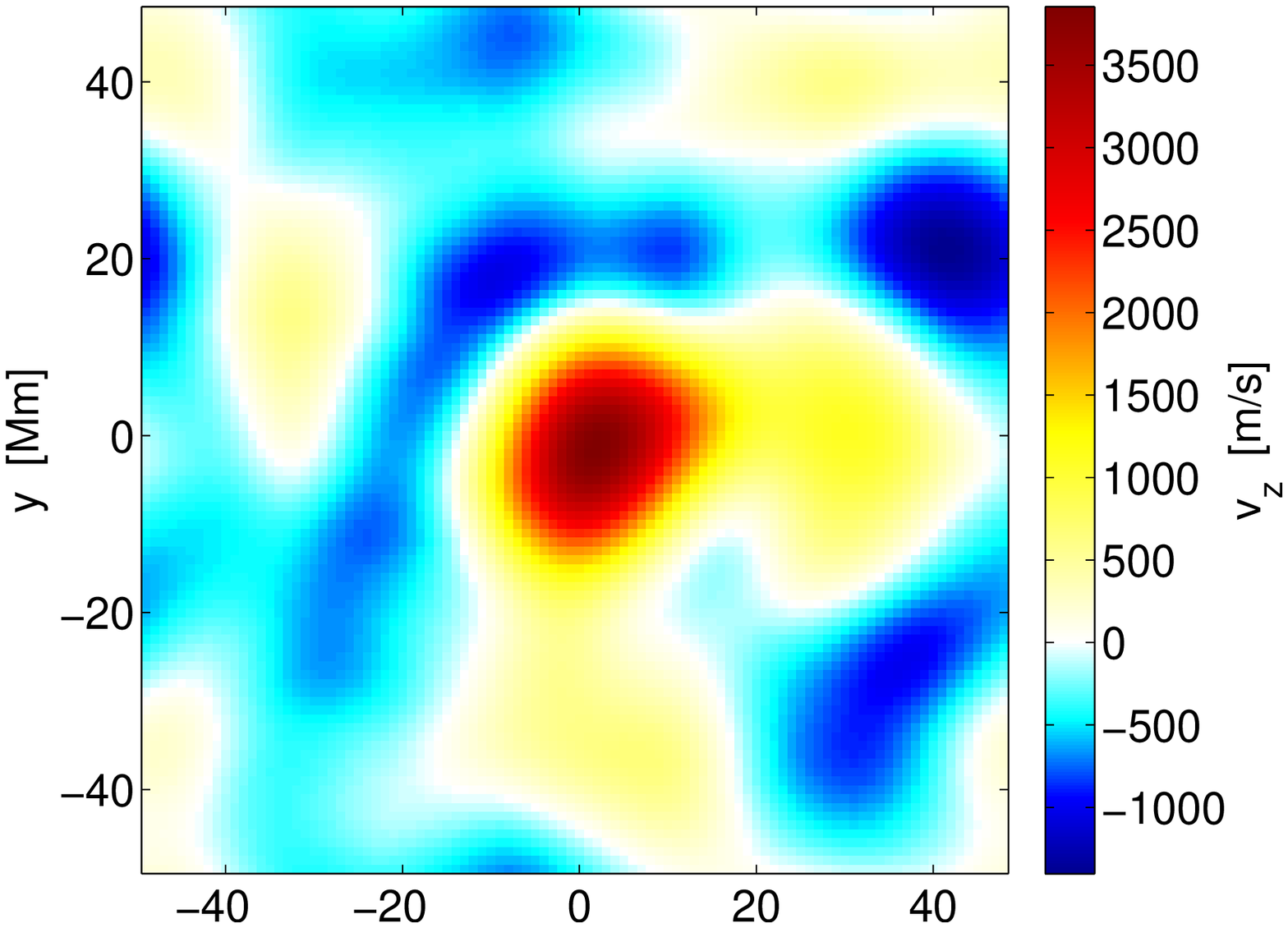} &
\includegraphics[width=0.3\linewidth,clip=]{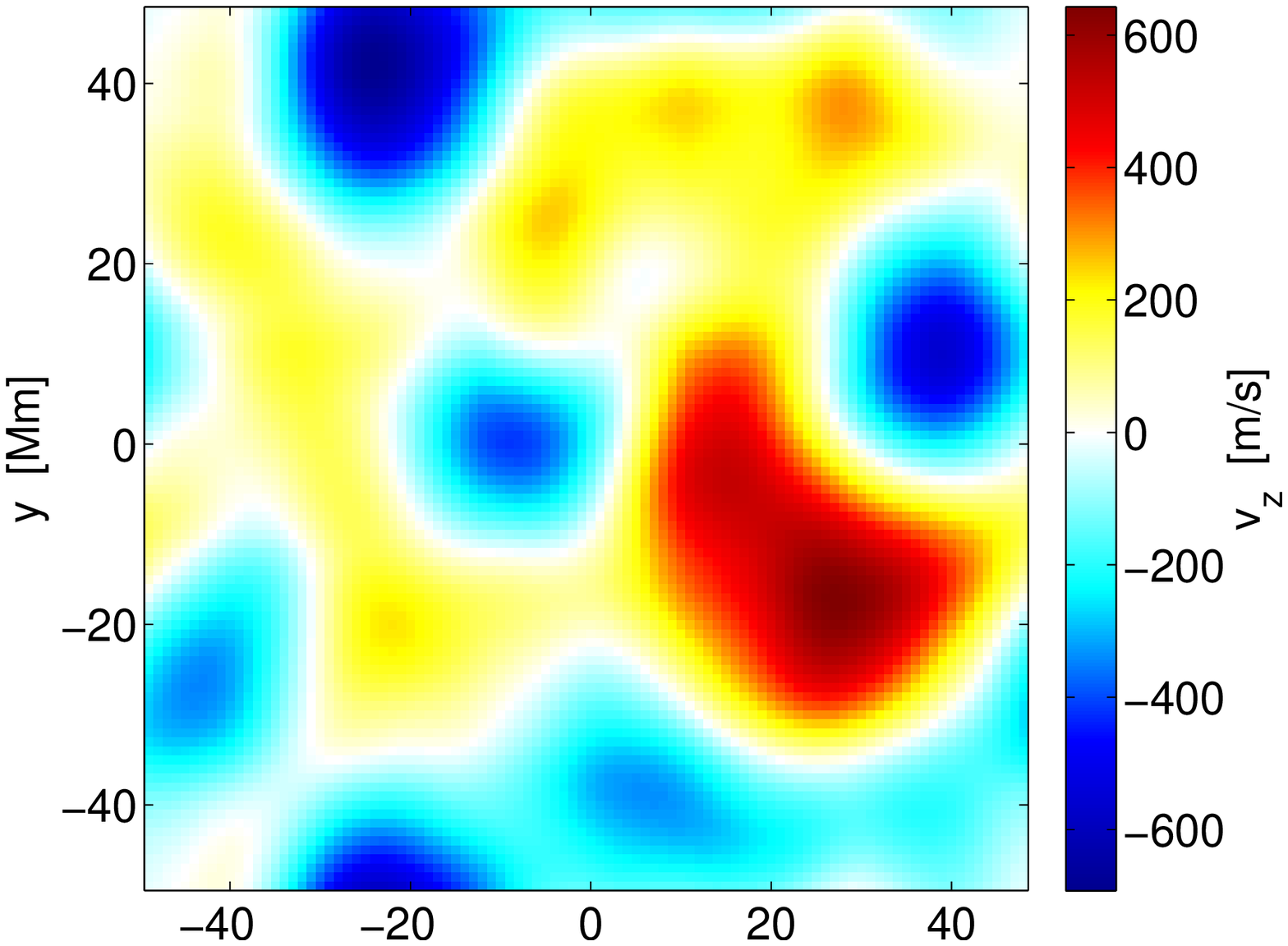} \\
\includegraphics[width=0.3\linewidth,clip=]{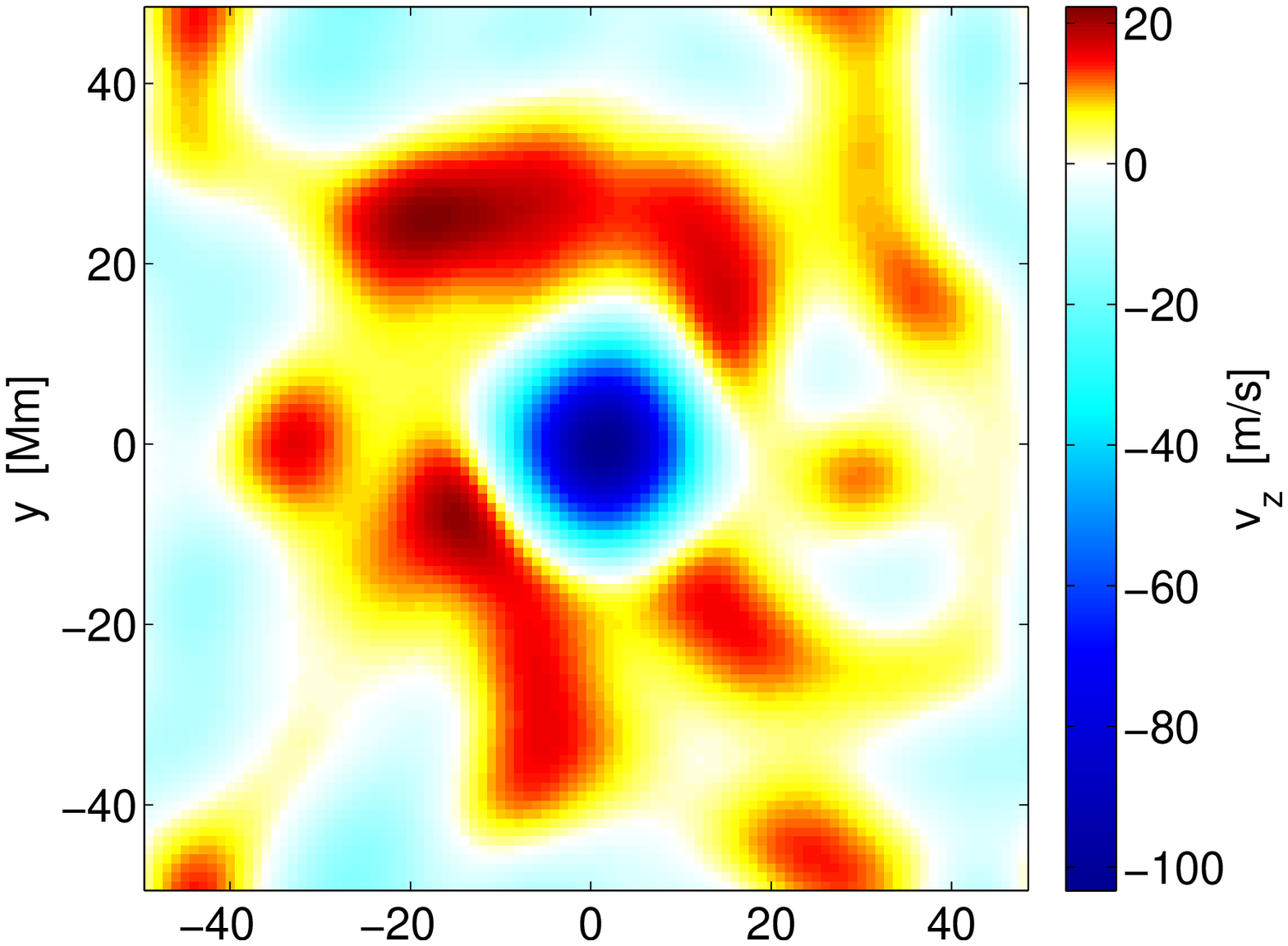} &
\includegraphics[width=0.3\linewidth,clip=]{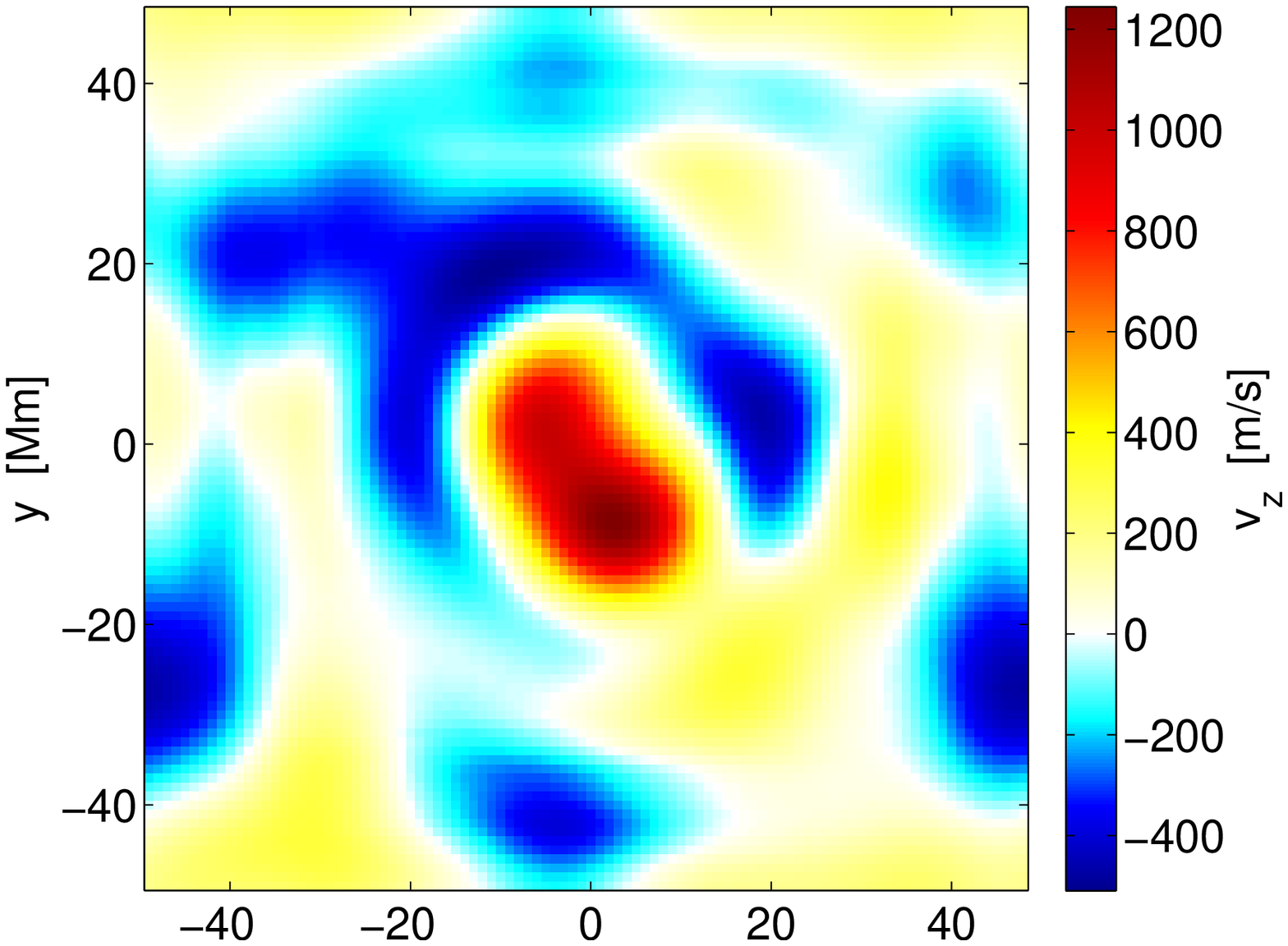} &
\includegraphics[width=0.3\linewidth,clip=]{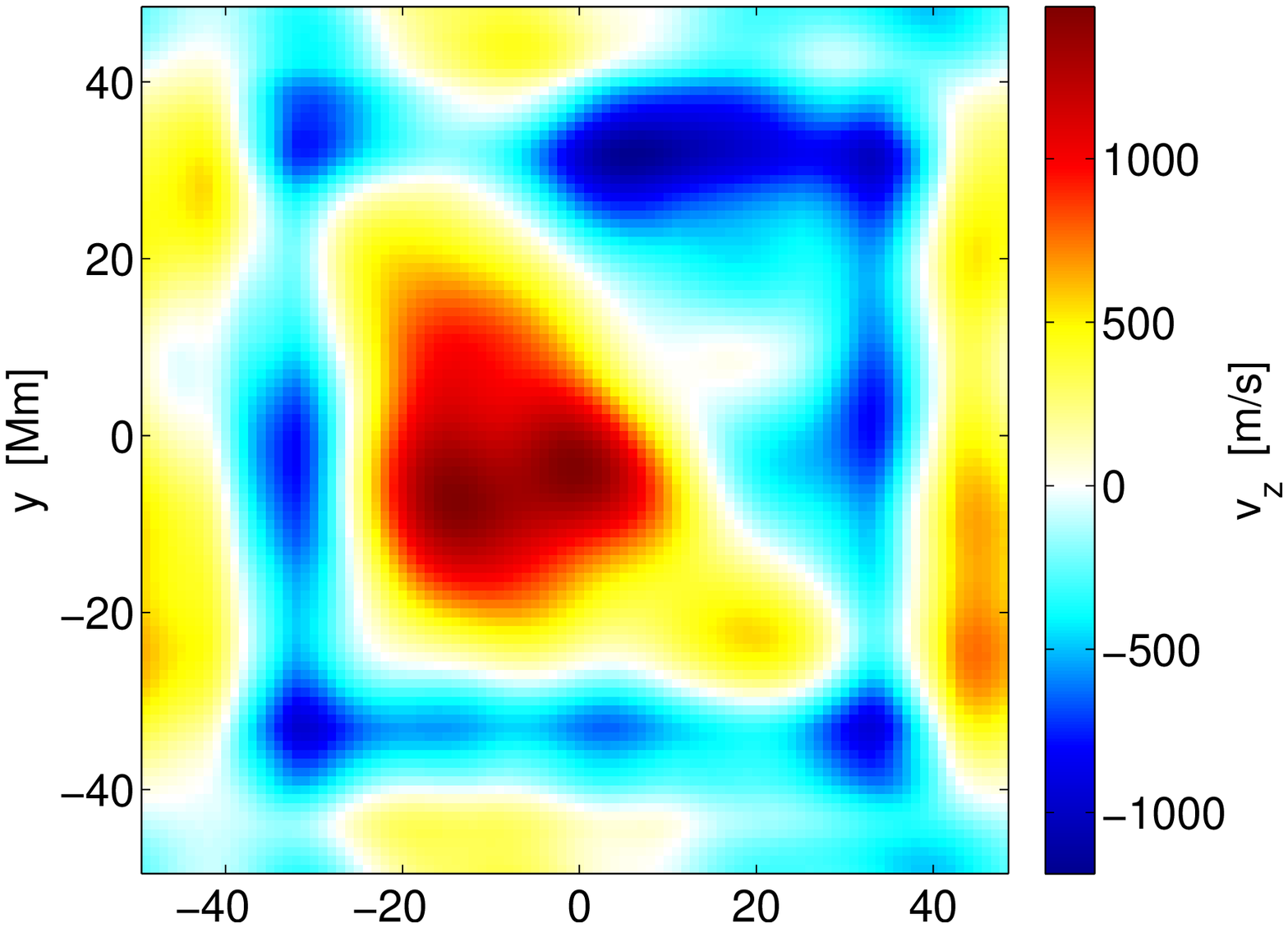} \\
\includegraphics[width=0.3\linewidth,clip=]{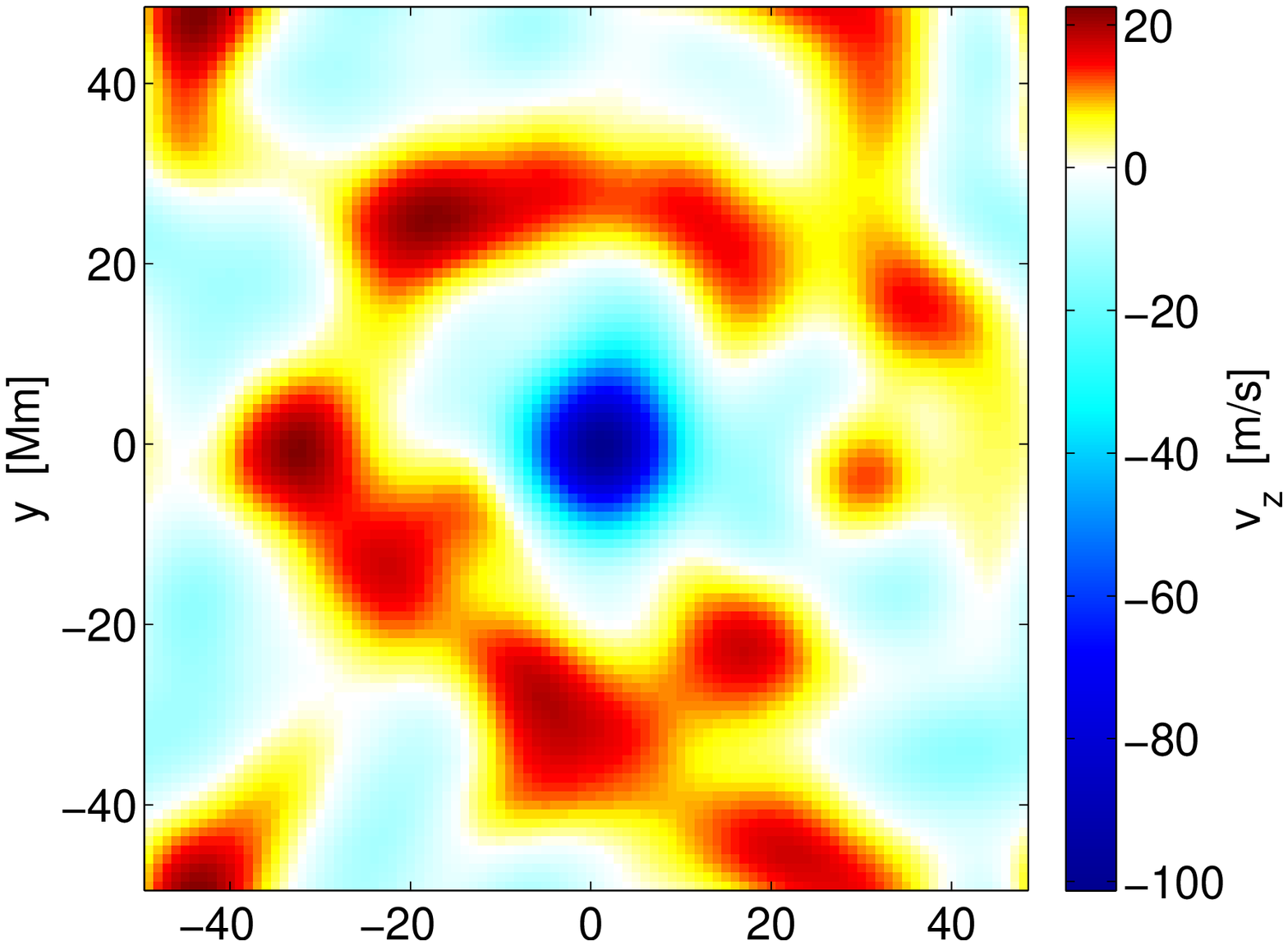} &
\includegraphics[width=0.3\linewidth,clip=]{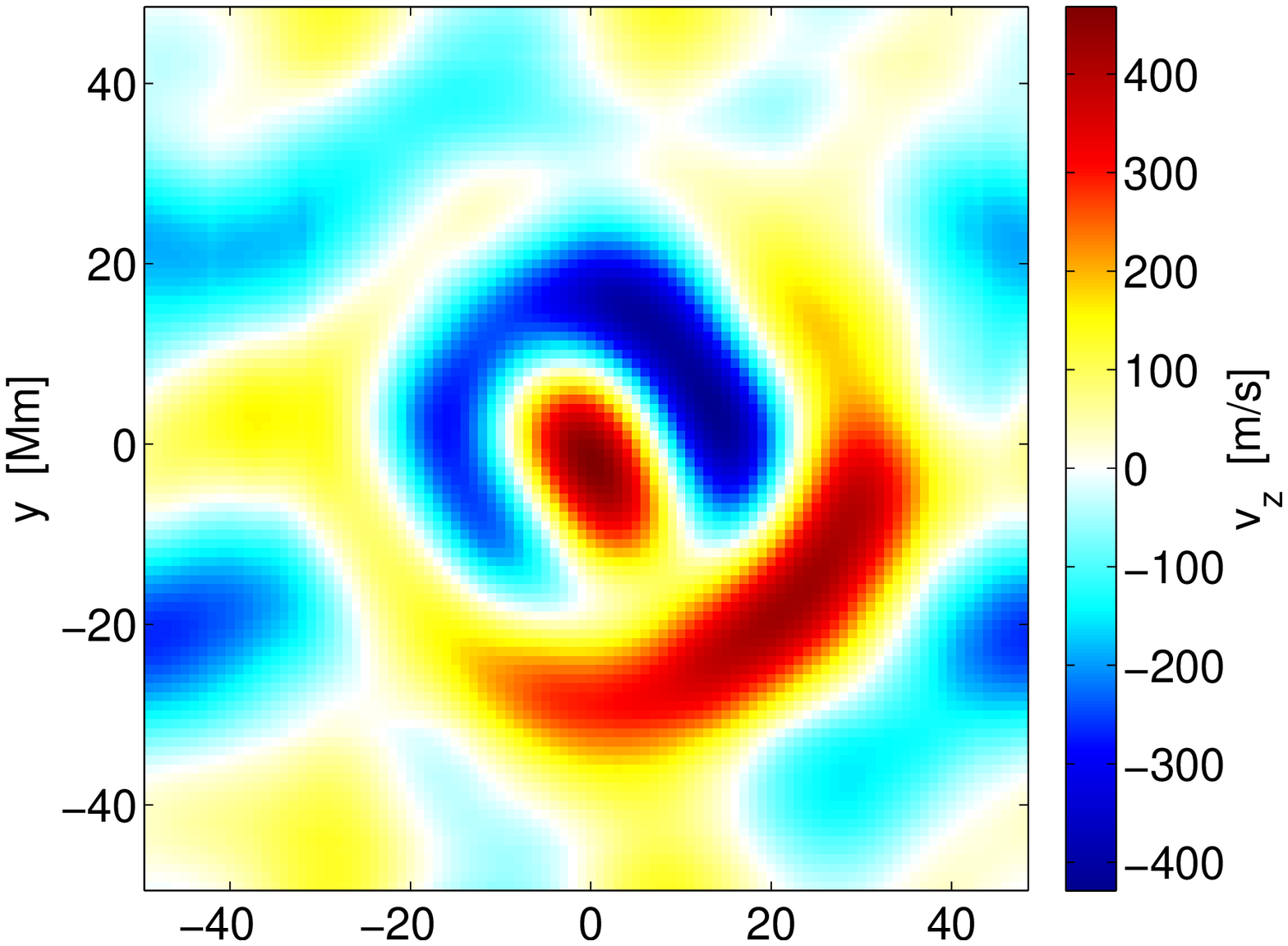} &
\includegraphics[width=0.3\linewidth,clip=]{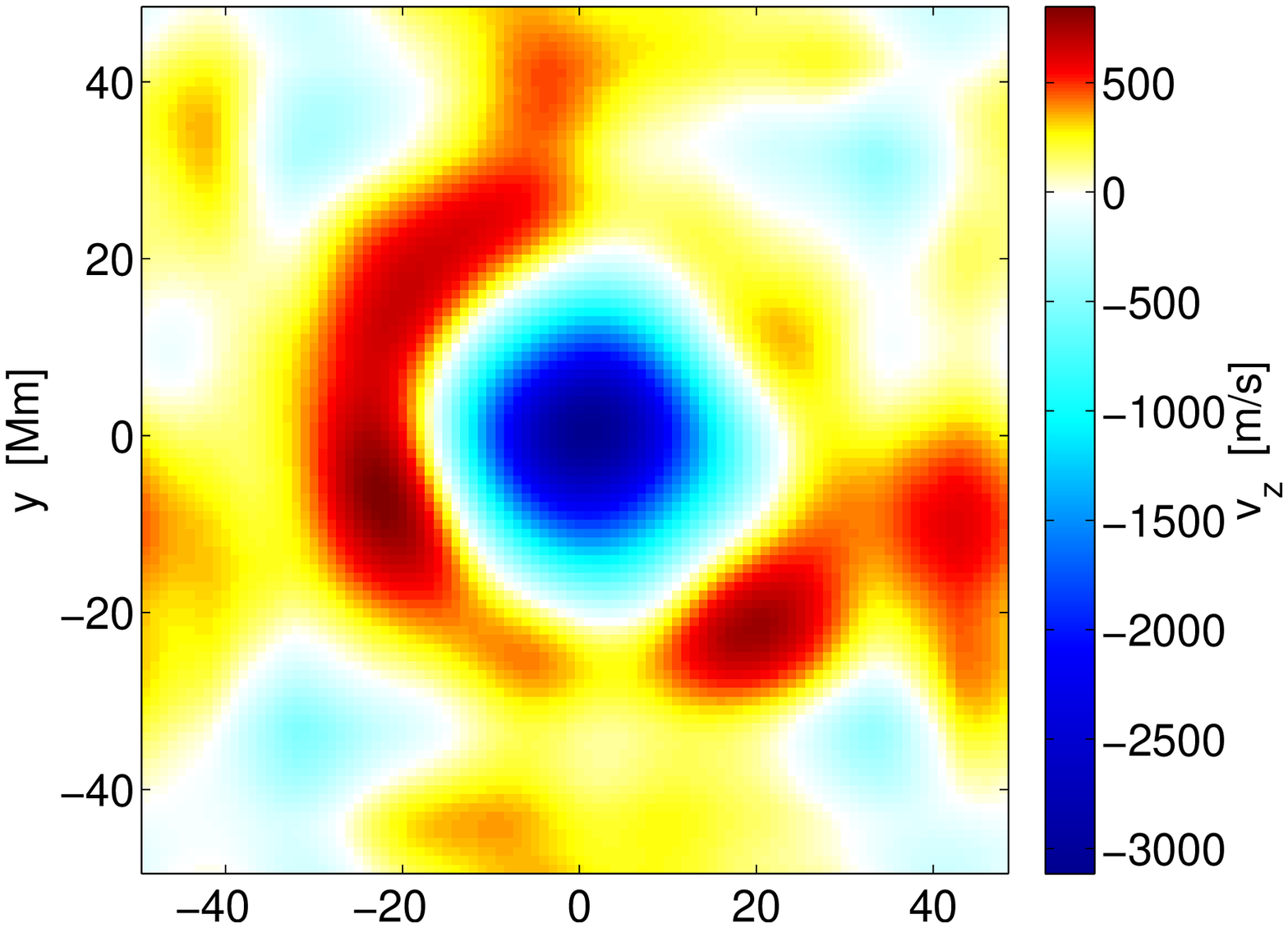} \\
\includegraphics[width=0.3\linewidth,clip=]{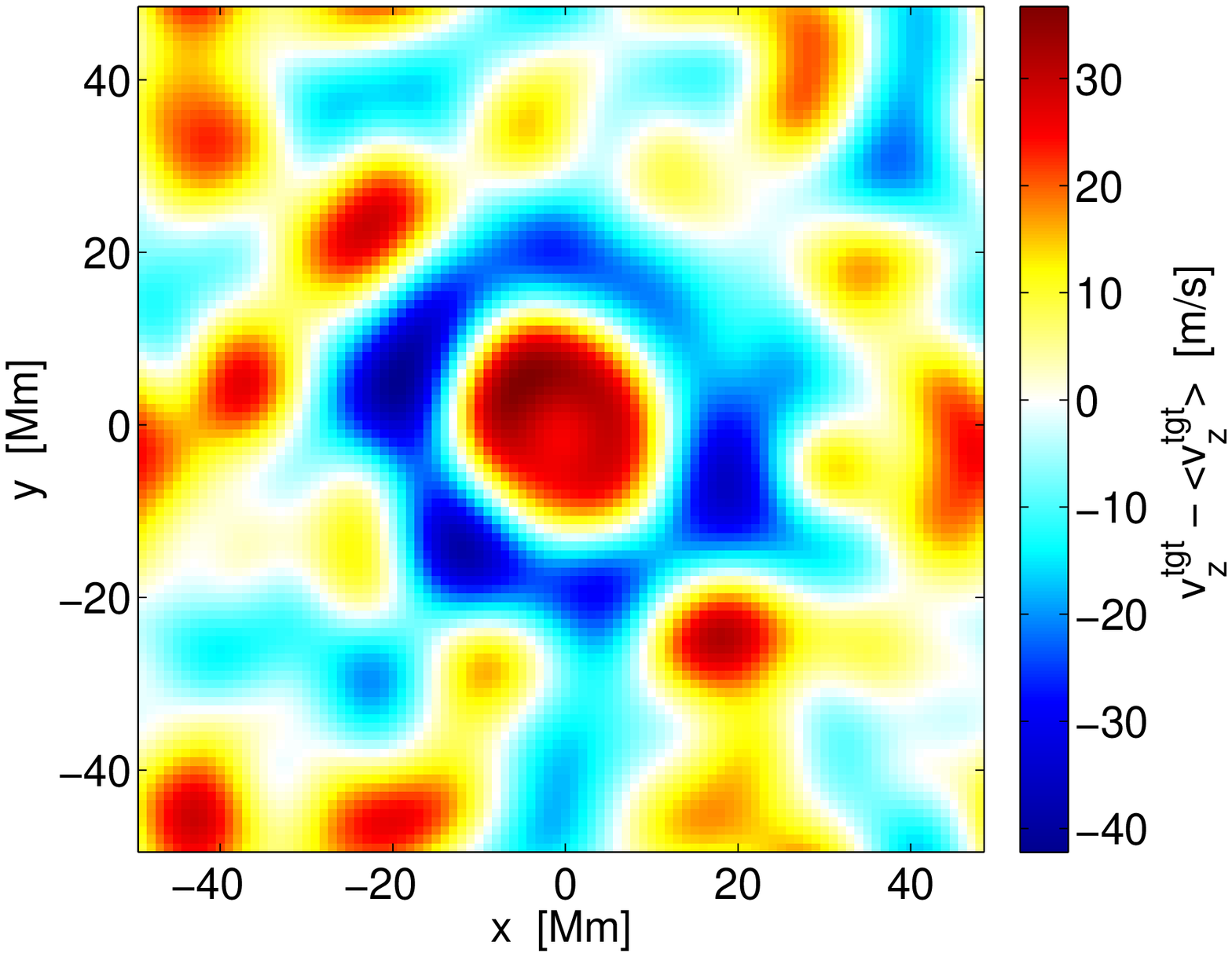} &
\includegraphics[width=0.3\linewidth,clip=]{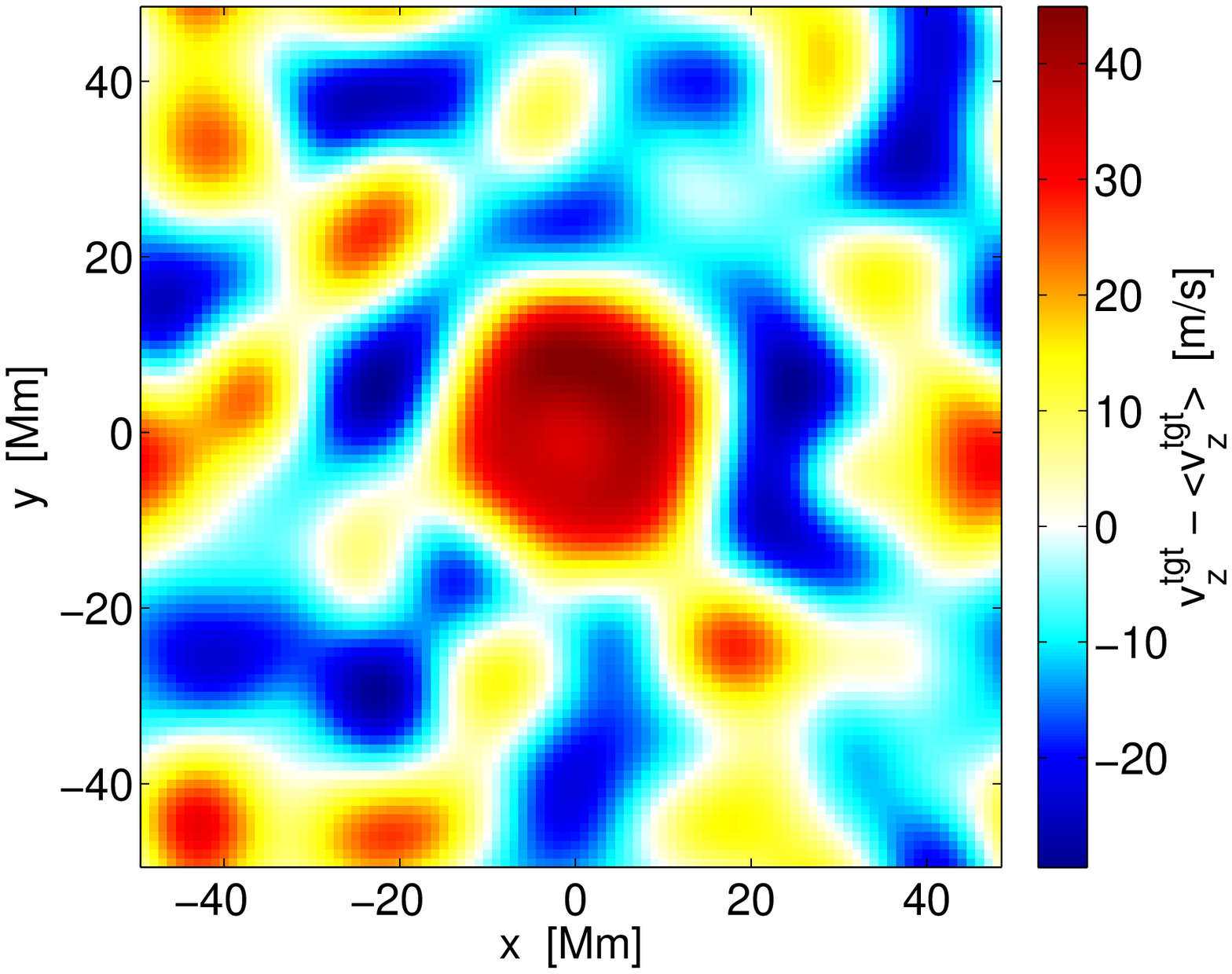} &
\includegraphics[width=0.3\linewidth,clip=]{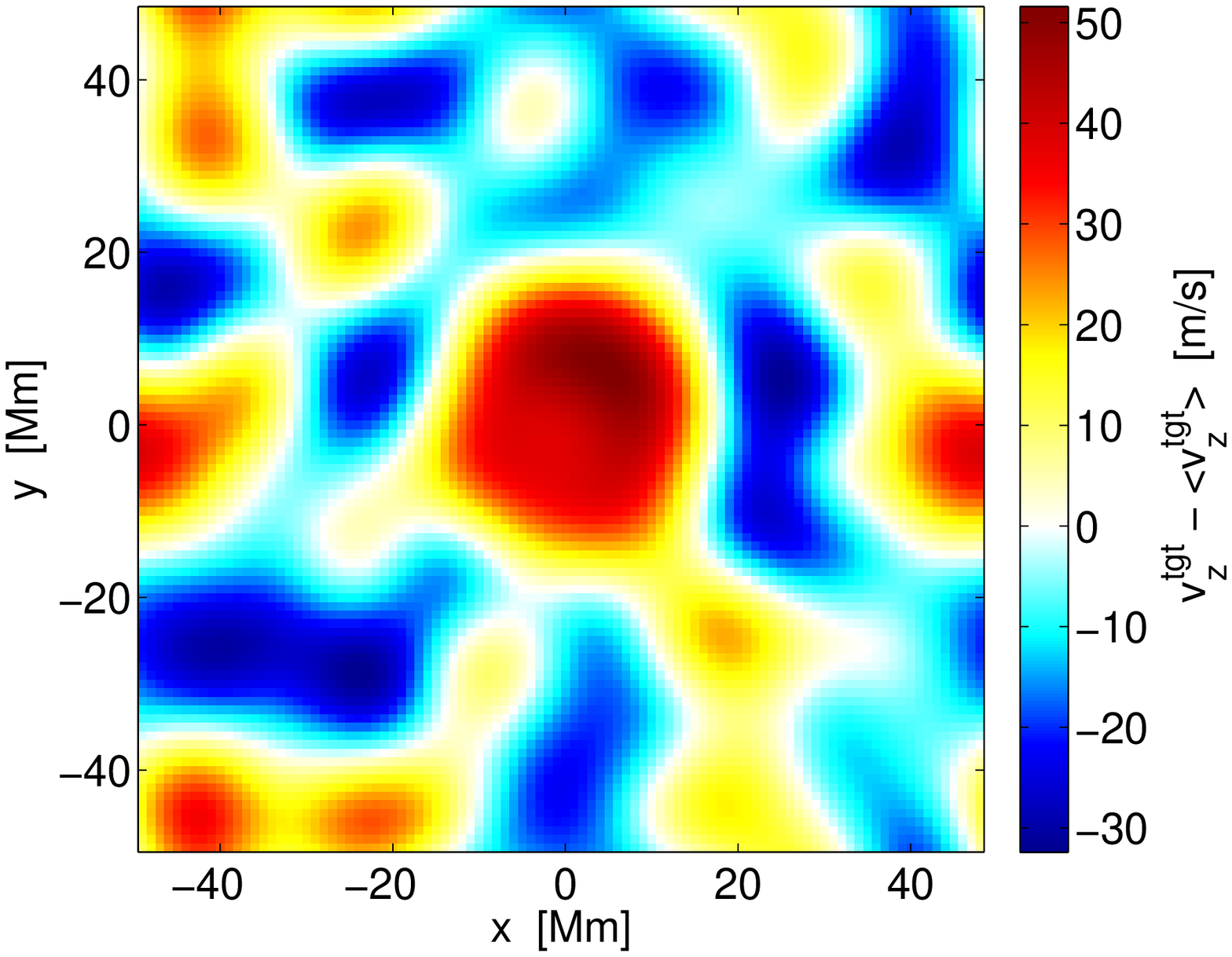}
\end{array}$
\end{center}
\caption{HRes GB02 vertical velocity inversion maps for the ridge (first row), phase-speed (second row), and ridge+phase-speed (third row) travel-time differences for depths (left to right) 1, 3 and 5~Mm. The smoothed simulation flow maps (i.e. $v_{z}^{\rm tgt}$) at these depths are shown in the bottom row.}
\label{fig:vz02SP}
\end{figure}

Inversions for the vertical flow component were also carried out for both simulation domains, the parameters of which are given in Table~1, inversion set~3 of \citet{degrave2014}. Figures \ref{fig:vz02SC} and \ref{fig:vz02SP} show these results along with the simulation target flow maps for LRes and HRes respectively, using only the GB02 travel times.

The LRes model shows a relatively strong downflow in the center of the spot at a depth of approximately 1~Mm, but concentrated upflows in most other areas. The inversion maps, while displaying their own downflow center (perhaps serendipitously), fail to recover the quiet-Sun structure. At deeper layers, the inversions are dominated by noise. In the HRes case, an upflow (consistent with a diverging cell) dominates the spot area, yet again, the inversions show a downflow in the shallowest layer. The upflow features in the simulation beyond the penumbra are somewhat recovered from the inversions, particularly using phase-speed measurements. In both cases, filtering changes the answers significantly, but likely only because it is altering the noise properties amid a very weak signal.

\section{Discussion}\label{sec:dis}

We have carried out time-distance helioseismology using artificial data from two simulations containing strong magnetic perturbations: one resembling a pore-like structure and the other closer to a typical sunspot. As may be expected, confidence in the inversions of the horizontal flows is only obtained in the parts of the domain beyond the influence of the magnetic features. Combined with an identical analysis from two quiet-Sun diffuse magnetic simulations in \citet{degrave2014}, the goal has been to validate current seismic inversions for flows and determine some of the conditions under which standard analysis fails using these realistic 
simulations.

Based on these studies, a few general conclusions can be made:

\begin{enumerate}

\item Horizontal flow structure away from strong magnetic features can be determined in the upper 3~Mm or so in the convection zone using 25~hr of data. Here, the spatial correlation coefficients between the inferred and model velocities are typically high. Poor correlations are found at 5~Mm depths.

\item Velocity amplitudes are usually underestimated by up to 50\%. Since the structure of the recovered flows is usually consistent (as given by the high correlations), this conclusion may result from non-linearities caused by strong flows, as suggested in \citet{degrave2014}.

\item There is no unambiguously ``best'' way to filter the data. We have tried three approaches, consistent among each analysis, but this is by no means an exhaustive set of filters that can be used.

\item The GB02 travel-time measurement procedure is the more robust, in that it agrees with the GB04 results in quiet regions and is more accurate in larger perturbative regimes. This has been seen before. However, it is slower to compute, and is not necessarily consistent with the sensitivity kernel definition, which linearly relates travel times to perturbations.

\item Inversions of forward-modeled travel times consistently outperform those of our measured travel times in terms of horizontal flow correlations. This suggests that our inability to recover flow structure near these simulated active regions is largely due to the fact that we are incapable of accurately measuring travel times around strong magnetic features.

\item Vertical velocity inversions are very difficult with the data sets used here. The trade off between noise, spatial resolution, and the cross-talk effects from the stronger horizontal velocities really limits these inferences in the relatively short times series available. Furthermore, acceptable inversions of simulation data with less complex vertical flow structure \citep[e.g.,][]{dombroski2013} have been possible.

\item Caution is needed that these conclusions may be conservative and that the simulations studied here may have stronger flows than the Sun.

\end{enumerate}

Because of the stochastic nature of solar oscillations and the significant noise generated, a substantial amount of work is underway of a statistical nature instead of focusing on individual events or features. For example, the ``average supergranule'' studies \citep{duvall2010,svanda2012,duvall2013,duvall2014}, the determination of typical shearing flows underneath many flaring active regions \citep{komm2012}, or mean helioseismic properties of pre-emerging active regions \citep{leka2013,birch2013,barnes2014}, among others. Excellent and abundant helioseismic data make these types of studies feasible.


%
%

%
%
%

\acknowledgements
The authors gratefully acknowledge support by the NASA SDO Science Center through contract NNH09CE41C awarded to NWRA and very fruitful discussions with Doug Braun. K.D. and J.J. also acknowledge funding from NSF award NSF/AGS-1351311 and subaward AURA/NSO No. N06504C-N, as well as a NASA EPSCoR award to NMSU under contract NNX09AP76A. M. Rempel is partially supported through NASA contracts NNH09AK02I, NNH12CF68C and NASA grant NNX12AB35G. The National Center for Atmospheric Research is sponsored by the National Science Foundation. Resources supporting this work were provided by the NASA High-End Computing (HEC) Program through the NASA Advanced Supercomputing (NAS) Division at Ames Research Center under project s0925.

\bibliography{mybib}

\begin{thebibliography}{}
\expandafter\ifx\csname natexlab\endcsname\relax\def\natexlab#1{#1}\fi

\bibitem[{{Barnes} {et~al.}(2014){Barnes}, {Birch}, {Leka}, \&
  {Braun}}]{barnes2014}
{Barnes}, G., {Birch}, A.~C., {Leka}, K.~D., \& {Braun}, D.~C. 2014, \apj, 786,
  19

\bibitem[{{Basu} {et~al.}(2004){Basu}, {Antia}, \& {Bogart}}]{basu2004}
{Basu}, S., {Antia}, H.~M., \& {Bogart}, R.~S. 2004, \apj, 610, 1157

\bibitem[{{Birch}(2011)}]{birch2011}
{Birch}, A.~C. 2011, Journal of Physics Conference Series, 271, 012001

\bibitem[{{Birch} {et~al.}(2013){Birch}, {Braun}, {Leka}, {Barnes}, \&
  {Javornik}}]{birch2013}
{Birch}, A.~C., {Braun}, D.~C., {Leka}, K.~D., {Barnes}, G., \& {Javornik}, B.
  2013, \apj, 762, 131

\bibitem[{{Birch} \& {Gizon}(2007)}]{birch2007}
{Birch}, A.~C., \& {Gizon}, L. 2007, Astronomische Nachrichten, 328, 228

\bibitem[{{Bogart} {et~al.}(2008){Bogart}, {Basu}, {Rabello-Soares}, \&
  {Antia}}]{bogart2008}
{Bogart}, R.~S., {Basu}, S., {Rabello-Soares}, M.~C., \& {Antia}, H.~M. 2008,
  \solphys, 251, 439

\bibitem[{Braun \& Lindsey(2003)}]{braunlindsey2003}
Braun, D., \& Lindsey, C. 2003, in ESA Conference Proceedings, Vol. SP-517,
  Local and Global Helioseismology: The Present and Future, ed.
  H.~Sawaya-Lacoste (Noordwijk: ESA Publications Division), 15--22

\bibitem[{{Braun} \& {Birch}(2008)}]{braun2008}
{Braun}, D.~C., \& {Birch}, A.~C. 2008, \solphys, 251, 267

\bibitem[{{Braun} {et~al.}(2012){Braun}, {Birch}, {Rempel}, \&
  {Duvall}}]{braun2012}
{Braun}, D.~C., {Birch}, A.~C., {Rempel}, M., \& {Duvall}, T.~L. 2012, \apj,
  744, 77

\bibitem[{{Braun} {et~al.}(1987){Braun}, {Duvall}, \& {Labonte}}]{braun1987}
{Braun}, D.~C., {Duvall}, Jr., T.~L., \& {Labonte}, B.~J. 1987, \apjl, 319, L27

\bibitem[{{Couvidat} {et~al.}(2006){Couvidat}, {Birch}, \&
  {Kosovichev}}]{couvidat2006}
{Couvidat}, S., {Birch}, A.~C., \& {Kosovichev}, A.~G. 2006, \apj, 640, 516

\bibitem[{{Couvidat} {et~al.}(2012){Couvidat}, {Zhao}, {Birch}, {Kosovichev},
  {Duvall}, {Parchevsky}, \& {Scherrer}}]{couvidat2012}
{Couvidat}, S., {Zhao}, J., {Birch}, A.~C., {et~al.} 2012, \solphys, 275, 357

\bibitem[{{Crouch} {et~al.}(2005){Crouch}, {Cally}, {Charbonneau}, {Braun}, \&
  {Desjardins}}]{crouch2005}
{Crouch}, A.~D., {Cally}, P.~S., {Charbonneau}, P., {Braun}, D.~C., \&
  {Desjardins}, M. 2005, \mnras, 363, 1188

\bibitem[{{DeGrave} {et~al.}(2014){DeGrave}, {Jackiewicz}, \&
  {Rempel}}]{degrave2014}
{DeGrave}, K., {Jackiewicz}, J., \& {Rempel}, M. 2014, \apj, 788, 127

\bibitem[{{Dombroski} {et~al.}(2013){Dombroski}, {Birch}, {Braun}, \&
  {Hanasoge}}]{dombroski2013}
{Dombroski}, D.~E., {Birch}, A.~C., {Braun}, D.~C., \& {Hanasoge}, S.~M. 2013,
  \solphys, 282, 361

\bibitem[{{Duvall} \& {Hanasoge}(2013)}]{duvall2013}
{Duvall}, T.~L., \& {Hanasoge}, S.~M. 2013, \solphys, 287, 71

\bibitem[{{Duvall} {et~al.}(2014){Duvall}, {Hanasoge}, \&
  {Chakraborty}}]{duvall2014}
{Duvall}, T.~L., {Hanasoge}, S.~M., \& {Chakraborty}, S. 2014, \solphys,
  arXiv:1404.2533

\bibitem[{{Duvall} \& {Birch}(2010)}]{duvall2010}
{Duvall}, Jr., T.~L., \& {Birch}, A.~C. 2010, \apjl, 725, L47

\bibitem[{{Duvall} {et~al.}(1993){Duvall}, {Jefferies}, {Harvey}, \&
  {Pomerantz}}]{duvall1993}
{Duvall}, Jr., T.~L., {Jefferies}, S.~M., {Harvey}, J.~W., \& {Pomerantz},
  M.~A. 1993, \nat, 362, 430

\bibitem[{{Duvall} {et~al.}(1997){Duvall}, {Kosovichev}, {Scherrer}, {Bogart},
  {Bush}, {de Forest}, {Hoeksema}, {Schou}, {Saba}, {Tarbell}, {Title},
  {Wolfson}, \& {Milford}}]{duvall1997}
{Duvall}, Jr., T.~L., {Kosovichev}, A.~G., {Scherrer}, P.~H., {et~al.} 1997,
  \solphys, 170, 63

\bibitem[{{Fan} {et~al.}(1995){Fan}, {Braun}, \& {Chou}}]{fan1995}
{Fan}, Y., {Braun}, D.~C., \& {Chou}, D.-Y. 1995, \apj, 451, 877

\bibitem[{{Gizon} \& {Birch}(2002)}]{gb02}
{Gizon}, L., \& {Birch}, A.~C. 2002, \apj, 571, 966

\bibitem[{{Gizon} \& {Birch}(2004)}]{gb04}
---. 2004, \apj, 614, 472

\bibitem[{Gizon {et~al.}(2000)Gizon, Duvall~Jr, \& Larsen}]{gizon2000}
Gizon, L., Duvall~Jr, T., \& Larsen, R. 2000, J. Astrophys. Astron., 21, 339

\bibitem[{Gizon {et~al.}(2001)Gizon, Duvall~Jr, \& Larsen}]{gizon2001}
Gizon, L., Duvall~Jr, T., \& Larsen, R. 2001, in IAU Symposia, Vol. 203, Recent
  insights into the Physics of the Sun and Heliosphere: Highlights from SOHO
  and Other Space Missions, ed. P.~Brekke, B.~Fleck, \& J.~Gurman (San
  Francisco, U.S.A.: Astronomical Society of the Pacific), 189--191

\bibitem[{{Gizon} {et~al.}(2009){Gizon}, {Schunker}, {Baldner}, {Basu},
  {Birch}, {Bogart}, {Braun}, {Cameron}, {Duvall}, {Hanasoge}, {Jackiewicz},
  {Roth}, {Stahn}, {Thompson}, \& {Zharkov}}]{gizon2009}
{Gizon}, L., {Schunker}, H., {Baldner}, C.~S., {et~al.} 2009, \ssr, 144, 249

\bibitem[{Haber {et~al.}(2002)Haber, Hindman, Toomre, Bogart, Larsen, \&
  Hill}]{haber2001}
Haber, D., Hindman, B., Toomre, J., {et~al.} 2002, Astrophys. J., 570, 855

\bibitem[{{Haber} {et~al.}(2004){Haber}, {Hindman}, {Toomre}, \&
  {Thompson}}]{haber2004}
{Haber}, D.~A., {Hindman}, B.~W., {Toomre}, J., \& {Thompson}, M.~J. 2004,
  \solphys, 220, 371

\bibitem[{{Hindman} {et~al.}(2004){Hindman}, {Gizon}, {Duvall}, {Haber}, \&
  {Toomre}}]{hindman2004}
{Hindman}, B.~W., {Gizon}, L., {Duvall}, Jr., T.~L., {Haber}, D.~A., \&
  {Toomre}, J. 2004, \apj, 613, 1253

\bibitem[{{Hindman} {et~al.}(2009){Hindman}, {Haber}, \&
  {Toomre}}]{hindman2009}
{Hindman}, B.~W., {Haber}, D.~A., \& {Toomre}, J. 2009, \apj, 698, 1749

\bibitem[{{Jackiewicz} {et~al.}(2012){Jackiewicz}, {Birch}, {Gizon},
  {Hanasoge}, {Hohage}, {Ruffio}, \& {{\v S}vanda}}]{jackiewicz2012}
{Jackiewicz}, J., {Birch}, A.~C., {Gizon}, L., {et~al.} 2012, \solphys, 276, 19

\bibitem[{{Komm} {et~al.}(2012){Komm}, {Howe}, \& {Hill}}]{komm2012}
{Komm}, R., {Howe}, R., \& {Hill}, F. 2012, \solphys, 277, 205

\bibitem[{{Korzennik}(2006)}]{korzennik2006}
{Korzennik}, S.~G. 2006, in ESA Special Publication, Vol. 624, Proceedings of
  SOHO 18/GONG 2006/HELAS I, Beyond the spherical Sun

\bibitem[{{Kosovichev}(1996)}]{kosovichev1996}
{Kosovichev}, A.~G. 1996, \apjl, 461, L55

\bibitem[{{Kosovichev}(2012)}]{kosovichev2012}
---. 2012, \solphys, 279, 323

\bibitem[{{Leka} {et~al.}(2013){Leka}, {Barnes}, {Birch}, {Gonzalez-Hernandez},
  {Dunn}, {Javornik}, \& {Braun}}]{leka2013}
{Leka}, K.~D., {Barnes}, G., {Birch}, A.~C., {et~al.} 2013, \apj, 762, 130

\bibitem[{{Moradi} {et~al.}(2010){Moradi}, {Baldner}, {Birch}, {Braun},
  {Cameron}, {Duvall}, {Gizon}, {Haber}, {Hanasoge}, {Hindman}, {Jackiewicz},
  {Khomenko}, {Komm}, {Rajaguru}, {Rempel}, {Roth}, {Schlichenmaier},
  {Schunker}, {Spruit}, {Strassmeier}, {Thompson}, \& {Zharkov}}]{moradi2010}
{Moradi}, H., {Baldner}, C., {Birch}, A.~C., {et~al.} 2010, \solphys, 267, 1

\bibitem[{{Pijpers} \& {Thompson}(1992)}]{pijpers1992}
{Pijpers}, F.~P., \& {Thompson}, M.~J. 1992, \aap, 262, L33

\bibitem[{{Rajaguru} {et~al.}(2006){Rajaguru}, {Birch}, {Duvall}, {Thompson},
  \& {Zhao}}]{rajaguru2006}
{Rajaguru}, S.~P., {Birch}, A.~C., {Duvall}, Jr., T.~L., {Thompson}, M.~J., \&
  {Zhao}, J. 2006, \apj, 646, 543

\bibitem[{{Rempel}(2012)}]{rempel2012}
{Rempel}, M. 2012, \apj, 750, 62

\bibitem[{{Rempel} {et~al.}(2009{\natexlab{a}}){Rempel}, {Sch{\"u}ssler},
  {Cameron}, \& {Kn{\"o}lker}}]{rempel2009}
{Rempel}, M., {Sch{\"u}ssler}, M., {Cameron}, R.~H., \& {Kn{\"o}lker}, M.
  2009{\natexlab{a}}, Science, 325, 171

\bibitem[{{Rempel} {et~al.}(2009{\natexlab{b}}){Rempel}, {Sch{\"u}ssler},
  {Cameron}, \& {Kn{\"o}lker}}]{rempel2009a}
---. 2009{\natexlab{b}}, Science, 325, 171

\bibitem[{{Rempel} {et~al.}(2009{\natexlab{c}}){Rempel}, {Sch{\"u}ssler}, \&
  {Kn{\"o}lker}}]{rempel2009b}
{Rempel}, M., {Sch{\"u}ssler}, M., \& {Kn{\"o}lker}, M. 2009{\natexlab{c}},
  \apj, 691, 640

\bibitem[{{Sheeley}(1969)}]{sheeley1969}
{Sheeley}, Jr., N.~R. 1969, \solphys, 9, 347

\bibitem[{{Sheeley}(1972)}]{sheeley1972}
---. 1972, \solphys, 25, 98

\bibitem[{{{\v S}vanda}(2012)}]{svanda2012}
{{\v S}vanda}, M. 2012, \apjl, 759, L29

\bibitem[{{{\v S}vanda} {et~al.}(2011){{\v S}vanda}, {Gizon}, {Hanasoge}, \&
  {Ustyugov}}]{svanda2011}
{{\v S}vanda}, M., {Gizon}, L., {Hanasoge}, S.~M., \& {Ustyugov}, S.~D. 2011,
  \aap, 530, A148

\bibitem[{Zhao \& Kosovichev(2004)}]{zhaokosovichev2004}
Zhao, J., \& Kosovichev, A. 2004, Astrophys. J., 603, 776

\bibitem[{{Zhao} \& {Kosovichev}(2003)}]{zhao2003}
{Zhao}, J., \& {Kosovichev}, A.~G. 2003, in ESA Special Publication, Vol. 517,
  GONG+ 2002. Local and Global Helioseismology: the Present and Future, ed.
  H.~{Sawaya-Lacoste}, 417--420

\bibitem[{{Zhao} {et~al.}(2001{\natexlab{a}}){Zhao}, {Kosovichev}, \&
  {Duvall}}]{zhaokosovichev2001}
{Zhao}, J., {Kosovichev}, A.~G., \& {Duvall}, Jr., T.~L. 2001{\natexlab{a}},
  \apj, 557, 384

\bibitem[{{Zhao} {et~al.}(2001{\natexlab{b}}){Zhao}, {Kosovichev}, \&
  {Duvall}}]{zhao2001}
---. 2001{\natexlab{b}}, \apj, 557, 384

\end{thebibliography}

\end{document}